\documentclass[journal]{IEEEtran}

\usepackage{amsmath,amssymb,amsfonts,amsthm}

\usepackage{graphicx}
\usepackage{array}
\usepackage{subfigure}
\usepackage{tabularx,booktabs}
\usepackage{multirow}
\usepackage{siunitx}
\usepackage{float}
\usepackage{makecell}
\usepackage{tcolorbox}
\usepackage{booktabs}

\usepackage{xcolor}
\usepackage{threeparttable}
\definecolor{darkgreen}{rgb}{0.0, 0.5, 0.0}
\usepackage{soul}

\usepackage{algorithm}
\usepackage{algpseudocode}

\usepackage{cite}
\usepackage{url}

\usepackage{textcomp}
\usepackage{stfloats}
\usepackage{verbatim}
\usepackage{pifont}
\usepackage{braket}
\usepackage{hyperref}
\usepackage{orcidlink}
\usepackage{balance}
\usepackage{enumitem}
\setlist[itemize]{leftmargin=*}

\DeclareUnicodeCharacter{202F}{\,}
\DeclareUnicodeCharacter{2009}{\,}
\DeclareUnicodeCharacter{2248}{\approx}
\usepackage{newunicodechar}
\usepackage{amsthm}
\theoremstyle{plain}
\newtheorem{theorem}{Theorem}[section]

\newtheorem{lemma}[theorem]{Lemma}
\newtheorem{corollary}[theorem]{Corollary}
\theoremstyle{definition}
\newtheorem{definition}[theorem]{Definition}

\theoremstyle{remark}
\newtheorem{remark}[theorem]{Remark}
\newcommand{\Tr}{\operatorname{Tr}}

\def\BibTeX{{\rm B\kern-.05em{\sc i\kern-.025em b}\kern-.08em
    T\kern-.1667em\lower.7ex\hbox{E}\kern-.125emX}}
    
\usepackage{eso-pic}
\newcommand\AtPageUpperMyright[1]{\AtPageUpperLeft{%
 \put(\LenToUnit{0.17\paperwidth},\LenToUnit{-2cm}){%
     \parbox{0.9\textwidth}{\raggedleft\fontsize{8}{11}\selectfont #1}}%
 }}%
\newcommand{\conf}[1]{%
\AddToShipoutPictureBG*{%
\AtPageUpperMyright{#1}
}
}

\begin{document}
\title{\vspace*{1cm} $Q$-Drive: A Hybrid Quantum-Classical Control Framework for Nonlinear Vehicular Dynamics \\
}
\author{Muhammad Al-Zafar Khan, Safaa Alqrinawi, Jamal Al-Karaki, Omar~Alhussein, \\Saif Al-Kuwari, Ahmed Farouk
\thanks{Muhammad Al-Zafar Khan is with the College of Interdisciplinary Studies, Zayed University, Abu Dhabi, UAE.
E-mail: Muhammad.Khan@zu.ac.ae}
\thanks{Safaa Alqrinawi and Saif Al-Kuwari are with the Qatar Center for Quantum Computing, College of Science and Engineering, Hamad Bin Khalifa University, Doha, Qatar.\\
E-mail: [saal88803, smalkuwari]@hbku.edu.qa}
\thanks{Jamal Al-Karaki is with the College of Interdisciplinary Studies, Zayed University, Abu Dhabi, UAE, and the College of Engineering, The Hashemite University, Zarqa, Jordan. E-mail: Jamal.Al-Karaki@zu.ac.ae}
\thanks{Omar~Alhussein is with the Computer Science Department, College of Computing and Mathematical Sciences, Khalifa University, Abu Dhabi, UAE. E-mail: omar.alhussein@ku.ac.ae}
\thanks{Ahmed Farouk is with Qatar Center for Quantum Computing, College of Science and Engineering, Hamad Bin Khalifa University, Doha, Qatar, and the Department of Computer Science, Faculty of Computers and Artificial Intelligence, Hurghada University, Hurghada, Egypt.
E-mail: ahmedfarouk@ieee.org}
}

\maketitle
\conf{\textit{Submitted to IEEE Transactions on Intelligent Vehicles (T-IV) \\Submission: March 2026 \\ Acceptance: November 2026}}

\begin{abstract}
In this paper, we introduce $Q$-Drive, a hybrid quantum-classical control framework for nonlinear vehicular dynamics that integrates variational quantum circuits (VQCs) with classical optimization, encoding vehicle states into a quantum Hilbert space via angle embedding, applying a trainable quantum ansatz to generate control signals, and refining circuit parameters through gradient descent to minimize tracking error and control effort. The proposed $Q$-Drive is evaluated in two settings: Single-vehicle control of nonlinear bicycle dynamics and a decoupled multi-vehicle swarm scenario with fleet sizes ranging from $N=2$ to $N=50$, demonstrating that the approach achieves exceptionally low control effort, reduced by orders of magnitude compared to classical model predictive control (MPC) and linear-quadratic regulator (LQR) benchmarks in the single-vehicle case, with average per-vehicle control effort decreasing by over 60\% from $N=2$ to $N=50$ in the swarm setting. Beyond empirical evaluation, we establish rigorous stability guarantees by constructing a composite Lyapunov function that jointly captures classical tracking dynamics and quantum-state purity. This analysis proves that the closed-loop system remains ultimately uniformly bounded under arbitrary road conditions, including inclined, rough, and curvilinear surfaces, as well as in the presence of quantum decoherence on the NISQ hardware, with explicit critical thresholds linking decoherence rates to environmental adversity. For the swarm extension, a theoretical bound is derived that shows that the average per-vehicle control effort decreases inversely with fleet size, in agreement with empirical observations. $Q$-Drive offers a promising direction for intelligent transportation systems by enabling robust, energy-efficient, and scalable vehicle control under complex real-world driving conditions.
\end{abstract}


\begin{IEEEkeywords}
Autonomous Vehicles, Quantum Computing, Boid Dynamics, Swarm Dynamics 
\end{IEEEkeywords}

\section{Introduction}\label{sec:intro}
Autonomous transportation is rapidly becoming a cornerstone of next-generation mobility, covering not only self-driving cars but also autonomous shuttles, freight trains, and intelligent traffic management systems that coordinate vehicles and infrastructure in real time. A critical enabler for this vision is the ability to manage multi-agent driving scenarios, which are pervasive in real-world contexts \cite{antonio2022multi,wu2024recent,wang2024multi,ko2021multi}. Highway platooning, cooperative merging, coordinated lane changes, and unsignalized intersection negotiation are intrinsically multi-agent problems that demand decision-making that is both dynamics-aware, respecting the physical capabilities of each vehicle, and interaction-aware, anticipating and responding to the intentions of other agents. In these settings, the system-level objective extends beyond navigating individual vehicles to their destinations. The overarching goal is to shape the fleet's emergent behavior to maintain safe separations, align speeds smoothly, resolve conflicts efficiently, and remain robust to inevitable uncertainties and communication imperfections.

A promising paradigm for managing such large-scale coordination is \textit{swarm intelligence} \cite{yu2023overview,chen2024fast,shaikh2020review}. Inspired by biological collectives, swarm intelligence offers a scalable approach in which collective behavior emerges from simple, local interactions among agents, obviating the need for a computationally intractable centralized planner. The classic boid model, built on the principles of separation (avoiding collisions), alignment (matching velocity and heading with neighbors), and cohesion (maintaining group integrity), provides a foundational template. When adapted to vehicular formations, these rules generate complex cooperative behaviors from basic local sensing and communication. The primary advantage of this paradigm is its inherent scalability and robustness: It functions effectively even when global information is incomplete or delayed, making it particularly attractive for large fleets operating in dynamic environments.

For safe, precise low-level control of individual vehicles, Model Predictive Control (MPC) has emerged as the dominant framework in safety-critical autonomous driving. MPC’s core strength is its ability to predict future motion over a finite horizon and optimize control inputs in a receding-horizon fashion while explicitly enforcing system constraints. This capability is paramount for real-world operation, as MPC naturally handles hard constraints such as actuator limits, lane boundaries, and collision avoidance, as well as comfort-related limits on acceleration and jerk. In multi-agent control, MPC can directly incorporate inter-vehicle constraints and interaction objectives. However, this extension comes at a high computational cost. The coupling of states and constraints across agents and over the prediction horizon creates a large, often nonconvex, optimization problem whose complexity scales poorly with fleet size, challenging real-time feasibility.

Concurrently, the field of quantum computing (QC) has matured, offering a fundamentally different paradigm for information processing by leveraging quantum-mechanical effects such as superposition and entanglement. While the field remains in the noisy intermediate-scale quantum (NISQ) era, characterized by limited qubit counts and significant error rates, hybrid quantum-classical algorithms drive practical progress. These algorithms embed small, specialized quantum subroutines within classical optimization loops to target specific computational bottlenecks. This approach matters for vehicular swarm control because many high-level coordination decisions, such as neighbor selection, formation switching, and priority resolution, are discrete and combinatorial. These problems often map naturally to Quadratic Unconstrained Binary Optimization (QUBO) or Ising models, making them amenable to hybrid quantum optimization solvers such as the Quantum Approximate Optimization Algorithm (QAOA).

Despite their respective strengths, both swarm-based coordination and multi-agent MPC face significant practical hurdles. Pure boid-inspired controllers, while scalable, are largely heuristic and do not inherently guarantee feasibility under realistic vehicle dynamics, roadway geometry, and strict safety buffers. Conversely, fully coupled multi-agent MPC, while constraint-aware, can become computationally prohibitive as the number of vehicles grows. The core bottleneck is the discrete decision-making inherent to swarm behavior: determining the interaction graph, negotiating the merging order, and activating different formation modes. This introduces a combinatorial complexity that standard continuous MPC solvers are not designed to search efficiently under stringent real-time deadlines.

Many important modern developments in vehicle dynamics control have taken place. For example, in \cite{zhang2026barrier}, the authors introduce a rollover-prevention control pipeline for cooperative transport systems (CTSs) that carry oversized payloads using multiple all-wheel-steering, distributed-electric-drive carriers to mitigate stability-related accidents. To overcome the high-dimensional, nonlinear dynamics of CTSs, the authors develop an innovative iterative model predictive path-integration framework that achieves computational speeds $7–10\times$ faster than common MPC. Similarly, in \cite{fan2026energy}, the authors address energy efficiency and safety in distributed drive electric vehicles by optimizing torque distribution while adhering to tire-slip constraints. To address the issue of standard deep reinforcement learning failing to guarantee safety under dynamic conditions, \texttt{LagCriticSAC} is introduced, a novel contribution-oriented framework that provides state-aware predictions of long-term constraint violations and dynamically adjusts Lagrange multipliers to autonomously switch between risk-averse and energy-efficient modes. Lastly, in \cite{yin2026global}, the authors present a global coordinated control framework for all-wheel-steering and driving vehicles to address actuator coordination challenges and enhance vehicle handling. The framework's novelty lies in a hybrid feedforward method that integrates steady-state, dynamic, and oblique steering compensations, along with a driver-intention interpretation model for conventional and oblique steering. The hybridization approach we adopt integrates a classical and a quantum setup. In addition, their work incorporates sliding mode control to ensure robust tracking in the presence of system disturbances and establishes a coordinated strategy for all four steering angles and driving torques, based on real-time assessment of the tire slip state using lateral acceleration and yaw rate. In comparing these SOTA approaches, we believe that the methodological superiority of our approach rests on three interlocking components: It applies a VQC directly to the full 6-DoF nonlinear bicycle dynamics with actuator saturation, whereas other approaches, for example, rely on CARLA abstractions, path-level dynamics, and heuristic potential fields. Consequently, $Q$-Drive discovers a distinct family of solutions, whereas the other approaches do not naturally select it under the same fixed cost function. Additionally, $Q$-Drive derives an emergent inverse-$N$ scaling law: average per-vehicle effort decreases as system size increases, without any explicit inter-vehicle coupling terms arising from the loss-averaged gradient structure of shared variational parameters. 

The intersection of QC and autonomous vehicle control has recently attracted significant research attention, driven by the promise of quantum acceleration for computationally intensive optimization problems. Lilja et al. \cite{lilja2023quantum} explore the use of QC to calculate feasible, collision-free paths for autonomous vehicles in dynamic environments, introducing a scalable algorithm implemented on real quantum hardware that yields practical insights into noise, decoherence, and execution bottlenecks. Sinha et al. \cite{sinha2025nav} address the computational inefficiencies of deep reinforcement learning for collision-free navigation by introducing Nav-Q, a hybrid quantum-classical actor-critic framework evaluated in the CARLA simulator, which demonstrates that quantum architectures can achieve comparable performance to classical DRL with dramatically fewer parameters (53 versus $2\;305$). Similarly, Tomar et al. \cite{tomar2025quantum} exploit quantum parallelism in a classical Q-learning algorithm to rapidly generate Q-tables for autonomous-vehicle maneuver execution, converging in a single episode versus $2\;000$ episodes for classical methods, with 99\% success rates and 15-20\% shorter paths. Seyfipoor et al. \cite{seyfipoor2025real} use quantum convolutional neural networks (QCNNs) for collaborative real-time route planning, demonstrating superior performance on the Cityscapes dataset compared to classical CNNs. For swarm applications, Almaameri et al. \cite{almaameri2025quantum} develop a quantum-enhanced artificial potential field method for UAV swarm formation, achieving 37\% faster convergence and 42\% better disturbance rejection than classical APF approaches. Several comprehensive reviews \cite{mohammed2025review,rattan2025quantum,ramasamy2024enhancing} further synthesize the rapidly growing literature at the intersection of QC and autonomous systems.

Despite these advances, existing approaches exhibit important limitations that our work addresses. Prior quantum-enhanced methods have focused on path planning or discrete decision-making within simplified dynamics models, not full-vehicle control with nonlinear dynamics and actuator constraints. The Nav-Q algorithm \cite{sinha2025nav}, while successful in CARLA, addresses single-vehicle navigation but does not extend to multi-agent coordination and lacks explicit metrics for control-effort efficiency. The methods employed by Tomar et al. (quantum Q-learning) \cite{tomar2025quantum} achieve remarkable convergence speed, but operate on simplified path dynamics rather than the full 6-DoF vehicle models required for realistic control. The quantum-enhanced APF \cite{almaameri2025quantum} addresses swarm formation but relies on heuristic potential fields rather than optimal control principles, and none of these approaches provides rigorous stability guarantees under real-world perturbations or quantum noise. Furthermore, existing methods lack universal applicability across diverse road conditions and do not establish the mathematical conditions under which quantum-enhanced control remains stable, which is a critical requirement for safety-critical automotive applications. Our $Q$-Drive framework addresses these gaps by integrating VQC-based control with nonlinear dynamics, demonstrating emergent scalability in swarms, and providing a comprehensive stability analysis that guarantees uniform ultimate boundedness under arbitrary road conditions and NISQ-era hardware constraints.

To address the fundamental challenges, this paper introduces a novel hybrid quantum-classical control framework for autonomous single vehicles and vehicular swarms. Our approach makes the following key contributions:
\begin{enumerate}
\item We introduce a framework that encodes vehicle states into a quantum Hilbert space via angle embedding, applies a trainable ansatz to generate control signals, and refines its parameters via gradient descent. Unlike prior quantum-enhanced approaches focused on path planning or reinforcement learning, this architecture is applied directly to full nonlinear vehicle dynamics. The significance is not that it outperforms classical MPC on tracking error, but that the same fixed cost function, optimized through a VQC, reliably discovers a family of low-intervention, resource-conserving policies that classical solvers do not produce under identical conditions.
\item The quantum controller reduces control effort by several orders of magnitude relative to MPC and LQR while maintaining stable trajectories. We do not claim this is a superior control approach, but it shows that the hybrid optimization landscape biases the solution toward a trajectory family that prioritizes actuator conservation and smoothness over terminal-error suppression. Moreover, this presents a family of solutions that classical solvers, tuned to the same cost function, do not naturally select.
\item We show that the average per-vehicle control effort decreases inversely with fleet size, arising naturally from the vectorized VQC architecture with shared parameters, without explicit coupling in the dynamics or engineered coordination terms. This is a qualitative discovery about the solution family explored by hybrid quantum-classical optimization. Classical decentralized methods, such as independent LQR and PurePursuit, do not exhibit this, as their per-vehicle effort remains constant regardless of fleet size. The result is not that the quantum controller tracks better, but that it discovers patterns that are neither explicitly programmed nor emerging from classical baselines. The scaling pattern follows from the loss-averaged gradient structure of the training objective and does not constitute a quantum collective effect. 
\item Using a composite Lyapunov function that jointly captures classical tracking dynamics and quantum-state purity, we prove uniform ultimate boundedness under arbitrary road conditions (inclined, rough, and curvilinear) and NISQ-era decoherence, with explicit critical thresholds that link decoherence rates to environmental adversity. These guarantees show that the distinct solution family is not an artifact of favorable simulation conditions; it remains bounded and deployable across the same operating envelope that safety-critical classical controllers must satisfy.
\item The framework processes entire swarms in parallel through batched state encoding and vectorized circuit evaluation, enabling a single set of variational parameters to govern arbitrarily large fleets. The computational cost scales sublinearly with the number of vehicles and, more importantly, the qualitative character of the discovered policy becomes more resource-efficient as the fleet size grows. This is a property of the solution family itself, not a claim of superior per-vehicle tracking.
\item We compare the framework against MPC, LQR, and Pure Pursuit across swarm sizes, analyzing trajectory geometry, control-effort distributions, and convergence behavior. The results show that the hybrid quantum-classical controller yields a family of solutions characterized by wide, low-curvature trajectories, a near absence of high-frequency control content, and graceful trade-offs in terminal precision. These are qualitatively different from the tight lane-centering, oscillatory-correction policies produced by classical methods. Our contribution, therefore, is the identification and characterization of this alternative family of solutions, not a claim of superior tracking performance, which we explicitly disclaim. Furthermore, this demonstrates that our proposed approach is viable as an energy- and actuator-conserving control policy in operating regimes where cumulative control effort, rather than terminal tracking precision, is the dominant cost, and that it can be simulated on NISQ hardware.
\end{enumerate}

It is imperative that we clarify the central thesis of this study: We aim to demonstrate the minimal integration overhead (effort) required by the proposed hybrid quantum-classical framework compared with purely classical control architectures. We explicitly acknowledge that, for absolute-error convergence, state-of-the-art classical algorithms currently outperform the hybrid quantum-classical variants. This performance gap is a systemic characteristic of the NISQ era and is not specific to the proposed model. Furthermore, we posit that even with the advent of robust error mitigation, quantum-enhanced controllers will likely serve as specialized accelerators for specific computational speedups rather than wholesale replacements for classical systems. Consequently, the relevance and contributions of this work should be assessed on the feasibility and efficiency of the hybrid integration rather than absolute error parity with classical benchmarks.    

The remainder of this paper is organized as follows. Section \ref{methodology} details the proposed scheme, including vehicle dynamics, the boid-integrated cost function, and the QUBO formulation for discrete coordination. Section \ref{experiments} presents experimental results, first on a single vehicle and then on a multi-vehicle system, and analyzes performance and scalability. In Section \ref{sec:main resul}, the closed-loop stability of the approach under nominal operating conditions is proved. Section \ref{sec:discussion} discusses the implications of our results and draws comparisons with existing methods. Finally, Section \ref{conclusion} offers concluding remarks, addresses the limitations of the current work, and outlines a roadmap for future improvements and extensions.

\section{The $Q$-Drive Framework}\label{methodology}

\subsection{Problem Formulation}\label{subsec:problem formulation}
We consider the problem of controlling a nonlinear vehicular system over a discrete-time horizon. Let $\mathbf{x}_k \in \mathcal{X} \subset \mathbb{R}^{n_x}$ denote the state vector of the vehicle at time step $k$, and let $\mathbf{u}_k \in \mathcal{U} \subset \mathbb{R}^{n_u}$ denote the control input vector. The system evolves according to the continuous-time nonlinear dynamics
\begin{equation}
\dot{\mathbf{x}}(t) = \mathbf{f}(\mathbf{x}(t), \mathbf{u}(t)),
\end{equation}
which is discretized using a sampling time $\Delta t$ to obtain the discrete-time mapping
\begin{equation}
\mathbf{x}_{k+1} = \mathbf{F}(\mathbf{x}_k, \mathbf{u}_k) := \mathbf{x}_k + \int_{t_k}^{t_k+\Delta t} \mathbf{f}(\mathbf{x}(\tau), \mathbf{u}_k) d\tau.
\end{equation}

The control objective is to drive the vehicle from an initial state $\mathbf{x}_0$ to a desired reference state $\mathbf{x}_{\text{ref}}$ (or to track a time-varying reference trajectory $\mathbf{x}_{\text{ref}}(t)$) while minimizing a quadratic cost function that balances the accuracy and control effort
\begin{equation}
\mathcal{J}:=\mathcal{J}_{\text{boid}} = \sum_{k=0}^{T-1} \left[ (\mathbf{x}_{k+1} - \mathbf{x}_{\text{ref}})^\top \mathbf{Q} (\mathbf{x}_{k+1} - \mathbf{x}_{\text{ref}}) + \mathbf{u}_k^\top \mathbf{R} \mathbf{u}_k \right],
\end{equation}
where $\mathbf{Q} \succeq 0$ and $\mathbf{R} \succ 0$ are weighting matrices and $T$ is the time horizon. The control inputs must satisfy the actuator constraints $\mathbf{u}_{\min} \leq \mathbf{u}_k \leq \mathbf{u}_{\max}$ for all $k$.

In the multi-vehicle swarm setting, we consider a fleet of $N$ vehicles, each governed by identical dynamics and evolving independently without explicit dynamical coupling. Let $\mathbf{x}_k^{(i)}$ denote the state of vehicle $i$ at time $k$, and $\mathbf{u}_k^{(i)}$ its control input. The swarm objective becomes the following.
\begin{equation}
\begin{aligned}
\mathcal{J}:=\mathcal{J}_{\text{swarm}} =&\; \sum_{i=1}^{N} \sum_{k=0}^{T-1} \left[ (\mathbf{x}_{k+1}^{(i)} - \mathbf{x}_{\text{ref}})^\top \mathbf{Q} (\mathbf{x}_{k+1}^{(i)} - \mathbf{x}_{\text{ref}}) \right. \\
&\left.+ \mathbf{u}_k^{(i)\top} \mathbf{R} \mathbf{u}_k^{(i)} \right],
\end{aligned}
\end{equation}
where a common reference state $\mathbf{x}_{\text{ref}}$ is assumed for all vehicles. Notably, coordination emerges not from explicit coupling terms in the dynamics or constraints but from the shared structure of the controller and the aggregated optimization objective.

The key challenge is solving this optimal control problem in real time for constrained nonlinear systems. Classical approaches, such as MPC, solve a receding-horizon approximation at each time step, incurring high computational cost that scales poorly with state dimension, horizon length, and fleet size. We investigate whether a hybrid quantum-classical approach, in which a VQC generates control actions and a classical optimizer updates the circuit parameters, can discover efficient control policies that achieve acceptable tracking performance with dramatically reduced control effort, particularly as the fleet size grows.

We note that throughout the manuscript, when the term ``efficiency'' is used, it explicitly refers to the control effort $\sum_{k}||u_{k}||^{2}$; no claim is made about the runtime efficiency.   

\subsection{Methodology}\label{subsec:method}
We adopt and modify the approach from \cite{khan2025qi} for autonomous vehicle navigation tasks. The approach presented in Algorithm \ref{algo:q-drive-single vehicle} describes a hybrid quantum-classical control loop for navigation tasks in boid-based autonomous vehicles that optimizes a VQC. The process begins by taking the high-dimensional classical state of the vehicle—Cartesian coordinates, heading, velocities, and yaw rate—and encoding it into a quantum Hilbert space using rotation gates, specifically $R_{y}$. Although any encoding scheme may work, we provide a specific angle-embedding-based formulation that enables a quantum computer to process physical vehicle data. Once encoded, we apply a trainable quantum ansatz $U(\boldsymbol{\theta})$ as the controller's brain. By measuring along the qubits' $Z$-axis, the system extracts expectation values that represent the raw control intentions for longitudinal acceleration and steering.

In the second stage, the algorithm translates these raw quantum outputs, which naturally range between $\left[-1,1\right]$, into the physical world through scaling and clipping. The algorithm multiplies the raw values element-wise by a scaling matrix $\mathbf{S}$ to match the vehicle's actuators, such as maximum torque for acceleration or the maximum turning radius for the front and rear wheels. The algorithm then feeds these scaled commands into a nonlinear dynamical model. The algorithm integrates these dynamics over a small time step to predict the vehicle's next state, effectively simulating its motion in response to quantum-generated steering and acceleration commands.

The final stage involves a classical optimization routine to refine the quantum controller's performance over multiple epochs. A loss function $\mathcal{L}(\boldsymbol{\theta})$ is calculated by measuring the squared distance between the vehicle's predicted position and the desired target state, with an additional penalty on excessive control effort (sudden braking or sharp swerving) weighted by $\lambda$. Using a gradient-descent optimizer such as Adam, a classical computer updates the quantum parameters $\boldsymbol{\theta}$ to minimize the error. This loop repeats until the vehicle converges to the target trajectory within a specified tolerance $\epsilon$, yielding an optimized set of quantum weights $\boldsymbol{\theta}^*$ that can autonomously govern the vehicle's motion.

The 6-DoF bicycle model is chosen for the experiments that follow because it is computationally tractable and a widely accepted benchmark, enabling fair comparisons with classical MPC and LQR baselines. This does not mean model fidelity is unimportant.

\begin{algorithm}
\caption{$Q$-Drive (boid systems)}
\label{algo:q-drive-single vehicle}
\small
\begin{algorithmic}[1]
\State \textbf{input} vehicle dynamics $\mathbf{f}(\mathbf{x}_{k},\mathbf{u}_{k})$, Initial Pose $\mathbf{x}_{0} = [X, Y, \chi, v_x, v_y, r]^\top$, target state $\mathbf{x}_{\text{ref}}$, scaling matrix $\mathbf{S} = \text{diag}(\text{scale}_{a}, \text{scale}_{\delta_f}, \text{scale}_{\delta_r})$, weights $\lambda$, max epochs $E$, horizon $T$
\State \textbf{output} trained quantum parameters $\boldsymbol{\theta}^{*}$, optimal control history $\mathbf{U}$, trajectory $\mathbf{X}$
\State \textbf{initialize} $n$-qubit register, ansatz $U(\boldsymbol{\theta})$, optimizer learning rate $\eta$
\For{epoch = 1 \textbf{to} $E$}
    \State $\mathbf{x} \gets \mathbf{x}_0$
    \For {$k = 0$ \textbf{to} $T-1$} 
        \State \texttt{// state encoding}
        \State map vehicle state to Hilbert space via embedding scheme
        \[ \ket{\psi(\mathbf{x}_{k})} = \bigotimes_{i=1}^{n} R_y(\arctan2(x_i)) \ket{0} \]
        \State Apply VQC
        \State 
        \[ \ket{\psi_{\text{out}}} = U(\boldsymbol{\theta}) \ket{\psi(\mathbf{x}_{k})} \]
        \State \texttt{// measurement of expected value} 
        \State obtain raw control signals $y_j \in [-1, 1]$:
        \[ y_j = \bra{\psi_{\text{out}}} \hat{Z}_j \ket{\psi_{\text{out}}}, \quad j \in \{a_x, \delta_f, \delta_r\} \]
        \State \texttt{// actuator scaling \& clipping} 
        \State map to physical limits
        \[ \tilde{\mathbf{u}}_{k} = \text{clip}(\mathbf{y} \odot \mathbf{S}, \mathbf{u}_{\min}, \mathbf{u}_{\max}) \]
        \State \texttt{// integrate the dynamics} 
        \State Update pose using nonlinear vehicle equations
        \[ \mathbf{x}_{k+1} = \mathbf{x}_k + \int_{t}^{t+dt} \mathbf{f}(\mathbf{x}(\tau), \tilde{\mathbf{u}}_k) d\tau \]
        \State \texttt{// loss computation} 
        \State quadratic tracking error + control effort
        \[ \mathcal{L}(\boldsymbol{\theta}) = (\mathbf{x}_{k+1} - \mathbf{x}_{\text{ref}})^\top \mathbf{Q} (\mathbf{x}_{k+1} - \mathbf{x}_{\text{ref}}) + \lambda \|\tilde{\mathbf{u}}_k\|^2 \]
        \State \texttt{// update parameters} 
        \State $\boldsymbol{\theta} \gets \boldsymbol{\theta} - \eta \nabla_{\boldsymbol{\theta}} \mathcal{L}$ 
        \If{$\|\mathbf{x}_{k+1} - \mathbf{x}_{\text{ref}}\|_2 < \epsilon$}
            \State \textbf{break}
        \EndIf
    \EndFor
\EndFor
\State \textbf{return} $\boldsymbol{\theta}^{*}, \mathbf{U}, \mathbf{X}$
\end{algorithmic}
\end{algorithm}

Algorithm \ref{algo:q-drive-single vehicle} extends to an $N$-vehicle swarm system, as shown in Algorithm \ref{algo:q-drive-swarm}. The main difference between Algorithms \ref{algo:q-drive-single vehicle} and \ref{algo:q-drive-swarm} lies in the treatment of the state and control variables: In the former case, the boid operates on one single vehicle at a time, so both state encoding, quantum circuit evaluation, dynamics update, and loss computation are performed for a single $n$-dimensional state vector $\mathbf{x}_k$. In contrast, the latter swarm approach explicitly batches $N$ vehicles simultaneously. The input becomes a matrix of states $ \mathbf{X}_k\in\mathbb{R}^{N \times 6}$; the quantum circuit is evaluated in a vectorized manner; the raw measurements $\mathbf{y}^{(i)}$ are produced for every vehicle $ i = 1,2,\ldots,N $ at once; the dynamics are stepped forward in parallel for the entire swarm; and the loss $\mathcal{L}(\theta)$ is computed as a sum over all $N$ vehicles. This means a single set of quantum circuit parameters $\boldsymbol{\theta}$ controls the entire fleet from the first epoch, rather than a single agent in isolation.

The second major difference lies in the objective and stopping condition. In the boid case, the algorithm focuses on driving a single vehicle to the exact target state $\mathbf{x}_{\text{target}}$ as quickly and accurately as possible, with the convergence check and both losses defined solely in terms of that vehicle’s error. In the swarm version, because the same small variational circuit must now generalize across $N$ different initial conditions and produce reasonable controls for every member of the fleet, the optimization is necessarily more global: The loss aggregates tracking error and control effort across the whole swarm, which implicitly encourages the discovery of low-effort, coordinated-like behaviors even though the dynamics remain fully decoupled. Additionally, we make explicit that the time horizon $T$ in Algorithms \ref{algo:q-drive-single vehicle} and \ref{algo:q-drive-swarm} is the training horizon over which the policy is optimized, not a receding prediction horizon, and the trained parameters $\boldsymbol{\theta}^{*}$ define a static feedback policy evaluated at runtime. 

\begin{algorithm}
\caption{$Q$-Drive (swarm systems)}
\label{algo:q-drive-swarm}
\small
\begin{algorithmic}[1]
\State \textbf{input} vehicle dynamics $\mathbf{f}(\mathbf{x}_{k},\mathbf{u}_{k})$, initial poses $\mathbf{X}_{0} = \{\mathbf{x}_{0}^{(i)}\}_{i=1}^{N} = [X^{(i)}, Y^{(i)}, \chi^{(i)}, v_x^{(i)}, v_y^{(i)}, r^{(i)} ]_{i=1}^{N}$ , target state $  \mathbf{x}_{\text{ref}}$, shared scaling matrix $\mathbf{S} = \text{diag}(\text{scale}_{a}, \text{scale}_{\delta_f}, \text{scale}_{\delta_r})$, weights $\lambda$, max epochs $E$, horizon $T$
\State \textbf{output} trained quantum parameters $  \boldsymbol{\theta}^{*}  $, optimal control history $  \mathbf{U} = \{\mathbf{u}_{k}^{(i)}\}_{k=0,i=1}^{T-1,N}  $, trajectory $  \mathbf{X} = \{\mathbf{x}_{k}^{(i)}\}_{k=0,i=1}^{T,N}  $
\State \textbf{initialize} $  n  $-qubit register, ansatz $  U(\boldsymbol{\theta})  $, optimizer learning rate $  \eta  $
\For{epoch = 1 \textbf{to} $  E  $}
    \State $  \mathbf{X} \gets \mathbf{X}_0  $
    \For {$  k = 0  $ \textbf{to} $  T-1  $}
        \State \texttt{// batch state encoding (vectorized over swarm)}
        \State map vehicle states to Hilbert space via embedding scheme
        $$   \ket{\psi(\mathbf{X}_{k})} = \bigotimes_{j=1}^{n} R_y(\arctan2(X_{k,:,j})) \ket{0}   $$ \Comment{broadcast over $N$}
        \State Apply vectorized VQC
        $$\ket{\psi_{\text{out}}} = U(\boldsymbol{\theta}) \ket{\psi(\mathbf{X}_{k})}$$
        \State \texttt{// measurement of vectorized expected value}
        \State obtain raw control signals $  y_{j}^{(i)} \in [-1, 1]  $ for each vehicle $i$:
        $$  y_{j}^{(i)} = \bra{\psi_{\text{out}}^{(i)}} \hat{Z}_j \ket{\psi_{\text{out}}^{(i)}}, \quad j \in \{a_x, \delta_f, \delta_r\}   $$
        \State \texttt{// actuator scaling \& vectorized clipping}
        \State map to physical limits
        \begin{equation*}
        \tilde{\mathbf{u}}_{k}^{(i)} = \text{clip}(\mathbf{y}^{(i)} \odot \mathbf{S}, \mathbf{u}_{\min}, \mathbf{u}_{\max}) 
        \end{equation*}
        \State \texttt{// integrate the dynamics (vectorized)}
        \State update poses using vehicular dynamics
        $$   \mathbf{x}_{k+1}^{(i)} = \mathbf{x}_k^{(i)} + \int_{t}^{t+dt} \mathbf{f}(\mathbf{x}^{(i)}(\tau), \tilde{\mathbf{u}}_k^{(i)}) d\tau   $$
        \State \texttt{// loss computation}
        \State \Comment{quadratic tracking error + control effort (summed over the swarm)}
        $$   \mathcal{L}(\boldsymbol{\theta}) = \sum_{i=1}^{N} \left[ (\mathbf{x}_{k+1}^{(i)} - \mathbf{x}_{\text{ref}})^\top \mathbf{Q} (\mathbf{x}_{k+1}^{(i)} - \mathbf{x}_{\text{ref}}) + \lambda \|\tilde{\mathbf{u}}_k^{(i)}\|^2 \right]   $$
        \State \texttt{// update parameters}
        \State $  \boldsymbol{\theta} \gets \boldsymbol{\theta} - \eta \nabla_{\boldsymbol{\theta}} \mathcal{L}  $
        \If{$  \frac{1}{N} \sum_{i=1}^{N} \|\mathbf{x}_{k+1}^{(i)} -      \mathbf{x}_{\text{ref}}\|_2 < \epsilon  $}
            \State \textbf{break}
        \EndIf
    \EndFor
\EndFor
\State \textbf{return} $  \boldsymbol{\theta}^{*}, \mathbf{U}, \mathbf{X}  $
\end{algorithmic}
\end{algorithm}

\begin{figure*}[ht]
    \centering
    \includegraphics[width=1\linewidth]{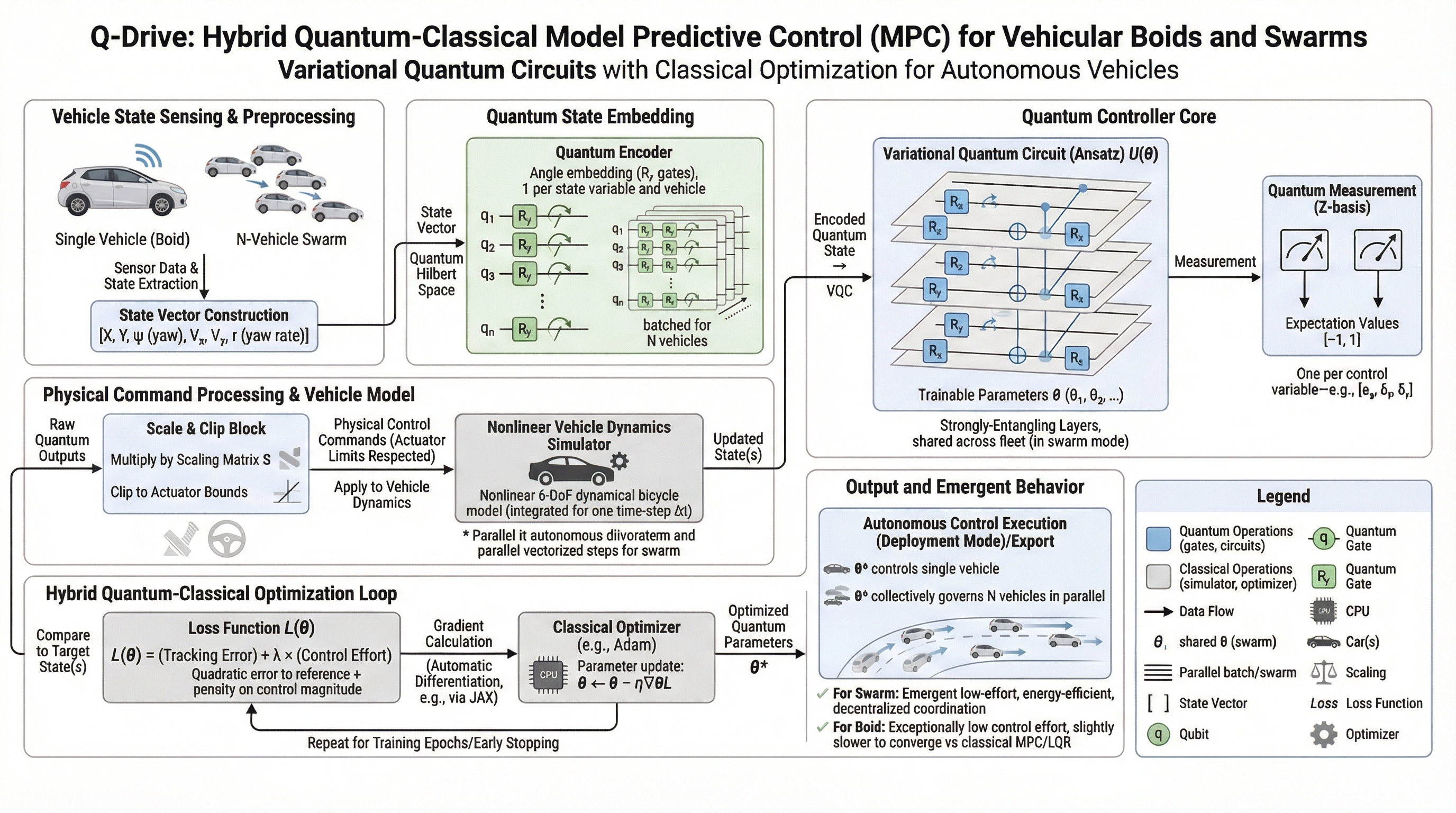}
    \caption{Architectural overview of the $Q$-Drive hybrid quantum-classical pipeline for boid and swarm autonomous navigation, integrating quantum state encoding, a VQC, and a classical optimization loop for scalable, low-effort trajectory control.}
    \label{fig:architecture_diagram}
\end{figure*}

Fig. \ref{fig:architecture_diagram} illustrates the hybrid quantum-classical setup. The system first processes vehicle state information, such as position, velocity, and orientation, and encodes it into a quantum Hilbert space via angle embedding, transforming classical data for quantum computation. The VQC then takes the encoded states as input, with trainable parameters serving as the primary controller. After quantum measurement, the system scales and clips the resulting control signals to respect physical actuator limits, then applies them to a nonlinear vehicle dynamics model to predict the trajectory. The system continuously computes a loss function that integrates tracking error and control effort and uses classical optimization algorithms to update the quantum circuit weights.

For swarm-based scenarios, this pipeline efficiently vectorizes and broadcasts across multiple vehicles. The state is simultaneously encoded and processed in parallel, governed by a single set of quantum circuit parameters that control the collective behavior of the entire fleet. The loss function aggregates control and trajectory error across all vehicles, encouraging globally efficient, emergent coordination. This diagram demonstrates robust scalability and substantial reductions in per-vehicle control effort as fleet size increases, highlighting the connection between quantum parallelism and classical optimization. 

To assess the VQC's approximation capacity, the following experiments use a 6-qubit register, not intended to represent the 6-dimensional state vector in a 1-1, basis-encoded fashion. Instead, the vehicle state is encoded via angle embedding into the qubits' rotation angles, such that the quantum state resides in a $2^{6}=64$-dimensional Hilbert space $\mathcal{H}$ ($\dim\mathcal{H}=64$). This implies that the VQC operates on a state space exponentially larger than the classical state space, and the trainable ansatz $U(\boldsymbol{\theta})$ can, in principle, approximate a rich class of nonlinear control policies. We further note that the choice of 6 qubits matches the three control outputs (longitudinal acceleration, front steering angle, and rear steering angle) by measuring three Pauli-$Z$ observables, with the remaining qubits serving as ancillary degrees of freedom that enhance the circuit's expressiveness.  

The architecture of the VQC defines a mapping from the classical state space to the control space via a composition of nonlinear encoding and trainable unitary transformations, and the resulting control policy is a continuous function of the variational parameters.

\section{Experiments}\label{experiments}
 The error-effort tradeoff in the following experiments is an a priori design choice, with fixed-cost weights and alternative tuning explicitly rejected. All controllers, including the hybrid quantum-classical policy and the classical baselines, optimize a single fixed quadratic cost, with the relative weighting of tracking error and control effort set once and held constant. We do not retune the state-penalty matrix $\mathbf{Q}$ or the effort-penalty coefficient $\lambda$, even though such retuning would be straightforward and could force the quantum controller into a tighter terminal-error regime. We make this choice so the results reflect the VQC's genuine inductive bias under a uniform objective, rather than an artificially engineered comparison. 

For reproducibility, Table \ref{tab:vqc_resources} summarizes the VQC parameters. The circuit comprises 6 qubits and 4 layers of the strongly entangling ansatz, yielding 72 trainable parameters and 24 fixed CNOT gates that implement the ring entanglement pattern. These modest resource requirements remain independent of fleet size and are fully compatible with present-day NISQ hardware.

\begin{table}[htbp]
\centering
\caption{Resource characterization of the $Q$-Drive VQC used in all experiments.}
\label{tab:vqc_resources}
\begin{tabular}{@{}lcl@{}}
\toprule
\textbf{Quantity} & \textbf{Value} & \textbf{Notes} \\
\midrule
Number of qubits $n$                  & 6     & Fixed across all experiments \\
Number of layers $L$                  & 4     & \texttt{StronglyEntanglingLayers} \\
Trainable parameters $P$              & 72    & $L\times n\times 3 = 4\times6\times3$ \\
Fixed (non-trainable) CNOT gates      & 24    & Ring entangler: $nL$ \\
Angle-embedding gates                 & 6     & One $R_y$ per qubit \\
Trainable single-qubit rotations      & 72    & $R_z$--$R_y$--$R_z$ per qubit/layer \\
Total two-qubit gates                 & 24    & Only the ring CNOTs \\
Total single-qubit gates              & 78    & 72 (ansatz) + 6 (embedding) \\
Approximate circuit depth             & 9--13 & Embedding + 4 layers (rotation + CNOT) \\
Measured observables                  & 2 or 3 & Pauli-$Z$ expectations \\
\bottomrule
\end{tabular}
\end{table}

\subsection{Experiment 1: Boid-Based Dynamics}\label{experiment: 1 boided-based dynamics}
The dynamical bicycle model presented in \cite{rajamani2006vehicle} and studied in the context of classical MPC in \cite{kong2015kinematic} is adopted as follows:
\begin{equation}\label{eq:dynamic bicycle model}
\left\{
\begin{aligned}
&\ddot{x}=\dot{\chi}\dot{y}+a_{x}, \\
&\ddot{y}=-\dot{\chi}\dot{x}+\frac{2}{m}\left(F_{c,f}\cos\delta_{f}+F_{c,r}\right), \\
&\ddot{\chi}=\frac{2}{I_{x}}\left(l_{f}F_{c,f}-l_{r}F_{c,r}\right), \\
&\dot{X}=\dot{x}\cos\chi-\dot{y}\sin\chi, \\
&\dot{Y}=\dot{x}\sin\chi+\dot{y}\cos\chi,
\end{aligned} 
\right. 
\end{equation}
where $x(t)$ and $y(t)$ are the horizontal and vertical coordinates of the center of mass (CoM), respectively, $X(t)$ and $Y(t)$ are the horizontal and vertical coordinates of the moving frame, respectively, $\chi(t)$ is the inertial heading/yaw angle (traditionally, $\psi$ is used to denote this variable, however, we have chosen to use $\chi$ so as not to confuse the yaw angle with the quantum state $\ket{\psi}$), $v$ is the speed of the vehicle, $l_{f}$ and $l_{r}$ is the distance from the CoM of the vehicle to the front and rear axles, respectively, $a_{x}$ is the longitudinal acceleration, $\delta_{f}$ and $\delta_{r}$ are the front and rear steering angles, respectively, $m$ is the mass of the vehicle, $I_{z}$ is the yaw inertia, and $F_{c,f}$ and $F_{c,r}$ are the lateral tire forces of the front and rear wheels, respectively, in coordinate frames aligned with the wheels. 

It is important to note that in this model, the categorization is as follows:
\begin{enumerate}
\item \textbf{Dynamical variables:} $x,y,X,Y,\psi,\dot{x},\dot{y},\dot{\chi},\ddot{\chi}$.
\item \textbf{Control variables:} $a_{x},\delta_{f},\delta_{r}$.
\item \textbf{Constants:} $m,I_{z},l_{f},l_{r}$.
\item \textbf{Force terms:} $F_{c,f}, F_{c,r}$. We treat these as algebraic variables, typically determined by tire models. In line with the work of \cite{kong2015kinematic}, we take $F_{c,i}=-C_{\alpha_{i}}\alpha_{i}$, where $\alpha_{i}$ is the tire slip angle and $C_{\alpha_{i}}$ is the tire cornering stiffness, for $i\in\left\{f,r\right\}$.   
\end{enumerate}

In this experiment, the performance of the $Q$-drive is evaluated as a vehicle controller in the dynamics described in \eqref{eq:dynamic bicycle model} and compared with classical benchmarks: Model Predictive Control (MPC) and Linear Quadratic Regulator (LQR). The study uses a parameterized 6-DoF vehicle dynamics model for a heavy sedan, with a mass of $1\;500\;\text{kg}$, a yaw inertia of $2\;875\;\text{kg}\cdot\text{m}^{2}$, and cornering stiffness values of $66\;000\;\text{N/rad}$. The vehicle must reach a target state with a longitudinal velocity of $20\;\text{m/s}$ and zero lateral displacement, yaw angle, and yaw rate. The quantum controller uses a 6-qubit system with a 4-layer strongly entangling architecture, encoding state information via angle embedding. The controller trains over $50$ epochs using the Adam optimizer with a learning rate of 0.02, minimizing a cost function that balances state error and control effort, weighted by a penalty coefficient $\lambda = 0.1$.

The specific values chosen for the experiment reflect a balance between physical realism and computational efficiency. The time step $\Delta t$ is set to $0.05\;\text{s}$, providing sufficient temporal resolution for the nonlinear dynamics without excessive computational overhead during the 400-step simulation (20 seconds total). The control inputs, longitudinal acceleration $a_{x}$ and steering angles $\delta_{f},\delta_{r}$, are constrained by scaling factors of $5.0\;\text{m/s}^{2}$ and $0.5\;\text{rad}$, respectively, to mirror the mechanical limits of a real-world vehicle. The initial state $x_0$ is intentionally offset from the target ($Y=1.0$ m, $\psi=0.1$ rad) to test the controller's ability to recover from lateral disturbances. For the classical benchmarks, the MPC uses a prediction horizon of 5 steps to allow for proactive planning, while the LQR weights ($Q$ and $R$) are tuned to prioritize lateral stability (weighting $Y$ and $\psi$ at 10) over control economy, ensuring a rigorous baseline for the quantum controller's performance.

The results in Table \ref{tab:results of experiment 1} demonstrate that the $Q$-Drive framework introduces a highly efficient control paradigm, characterized by an unprecedented level of energy economy. While classical benchmarks such as MPC and LQR rely on aggressive control inputs to achieve rapid convergence, $Q$-Drive operates with an exceptionally low total effort of $0.0014$ ($k$ is the time index and $j$ is the control index), which represents a reduction of several orders of magnitude compared to MPC ($0.1827$) and LQR ($994.0784$). This gentle control profile suggests that the quantum-classical hybrid approach is uniquely suited for applications where minimizing actuator wear and energy consumption is paramount, even if it prioritizes smooth, steady-state transitions over the high-intensity maneuvers of traditional methods. Although the final error of $39.3344$ and the $20.00$ second convergence time appear higher than the MPC's metrics, these values reflect $Q$-Drive’s cautious exploration of the nonlinear state space during the early optimization phases of the NISQ era. By maintaining stability with minimal control effort, the proposed approach demonstrates its potential as a robust alternative for long-range autonomous navigation where resource conservation is critical. This performance establishes a baseline, showing that a hybrid VQC architecture can effectively govern complex boid dynamics while remaining significantly more resource-efficient than its classical counterparts.

\begin{figure}[ht]
    \centering
    \includegraphics[width=1\linewidth]{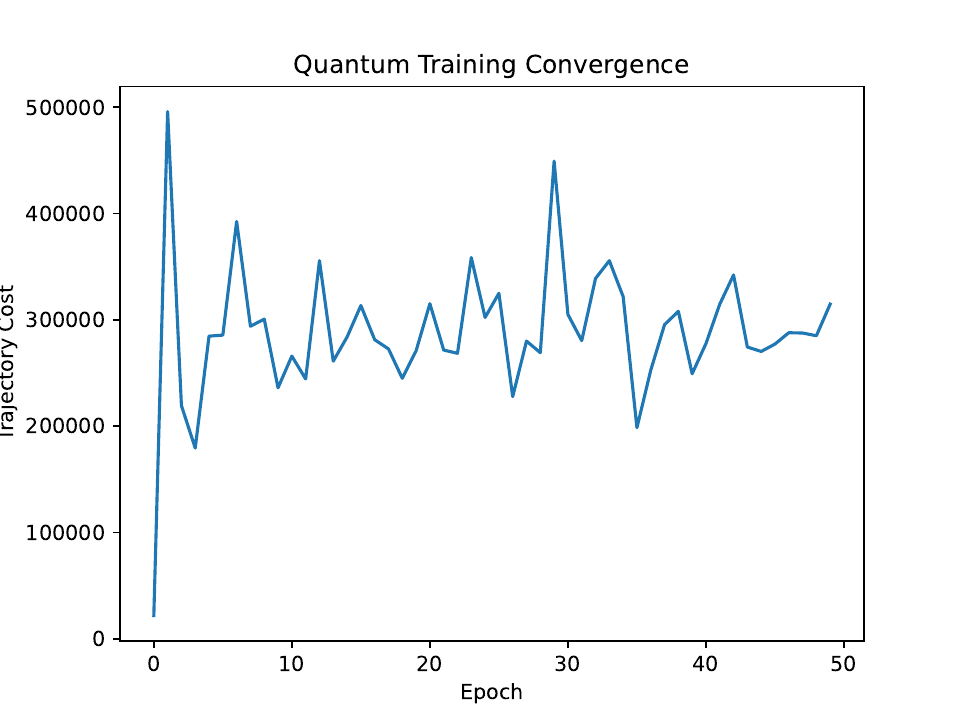}
    \caption{Training convergence of the $Q$-Drive controller. It shows the trajectory cost $\mathcal{L}(\theta)$ over 50 epochs. The high variance in cost reflects the stochastic nature of the Adam optimizer as it navigates the nonlinear Hilbert space of the VQC.}
    \label{fig:training convergence for boid dynamics}
\end{figure}

In Fig. \ref{fig:training convergence for boid dynamics}, the training convergence plot illustrates the evolution of the trajectory cost over 50 epochs using the Adam optimizer. The data reveal a highly stochastic optimization landscape, characterized by significant fluctuations in the loss function throughout the training. The cost drops rapidly in the first 5 epochs, indicating the hybrid quantum-classical model quickly identifies a coarse solution; afterward, it does not follow a smooth, monotonic decay. Instead, the cost oscillates within a stabilized band (approximately between $\left[200\;000,350\;000\right]$), suggesting that the VQC is navigating a complex, non-convex energy landscape, is potentially hindered by the barren plateau phenomenon, or is subject to high sensitivity in the quantum-classical feedback loop. Despite volatility, the overall trend shows that the system maintains an average cost below its initial peak, though it remains in dynamic exploration rather than settling into a single global minimum.

\begin{figure*}[ht]
    \centering
    
    \begin{minipage}[b]{0.32\textwidth}
        \centering
        \includegraphics[width=\textwidth]{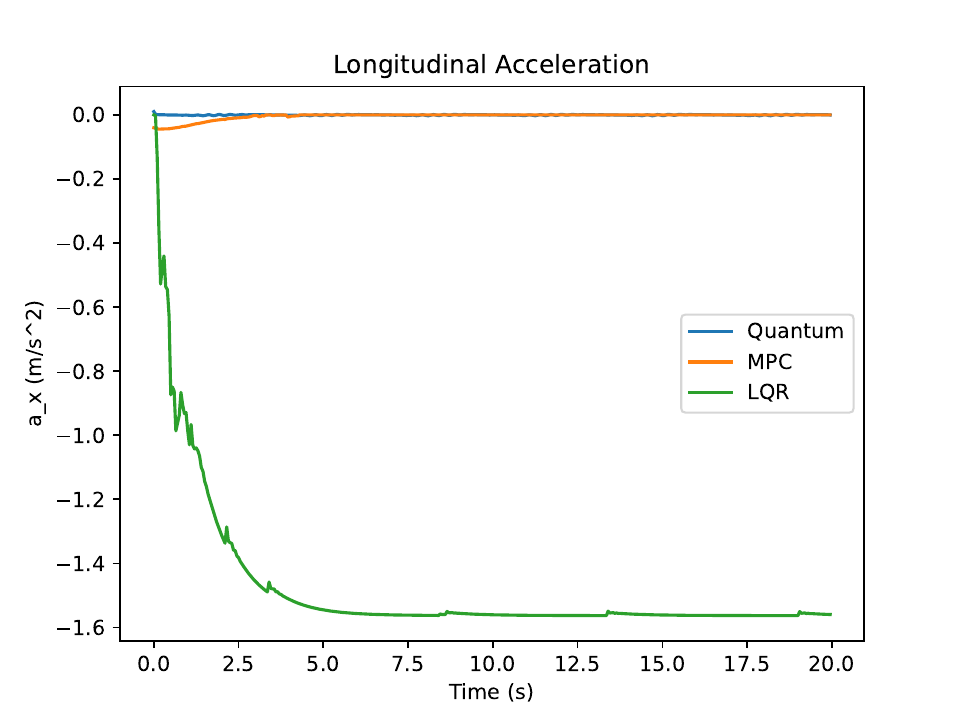}
    \end{minipage}
    \hfill
    \begin{minipage}[b]{0.32\textwidth}
        \centering
        \includegraphics[width=\textwidth]{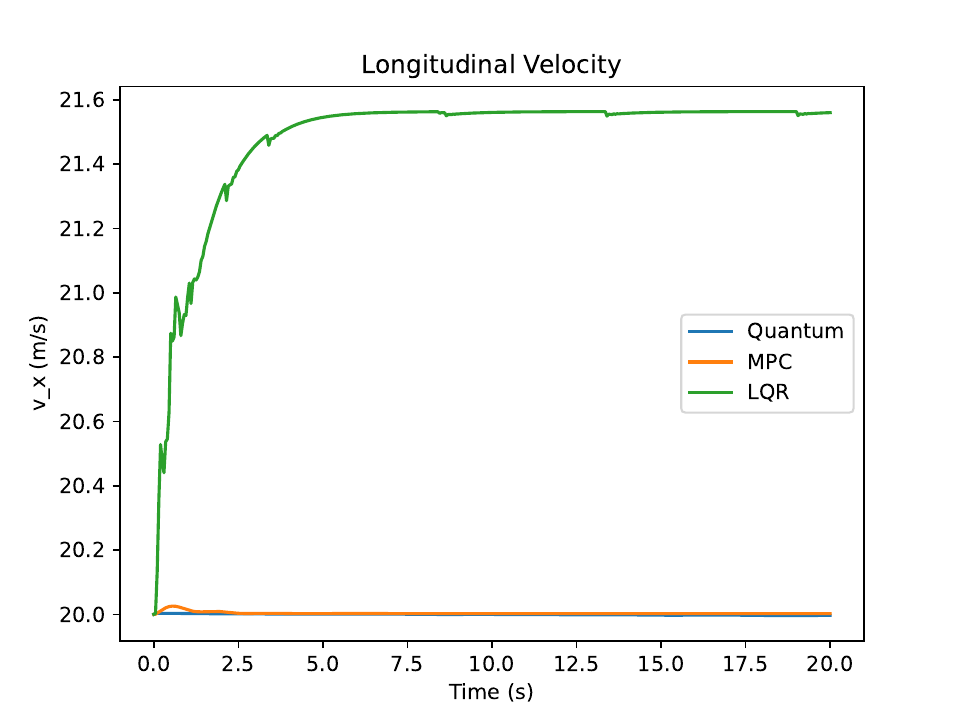}
    \end{minipage}
    \hfill
    \begin{minipage}[b]{0.32\textwidth}
        \centering
        \includegraphics[width=\textwidth]{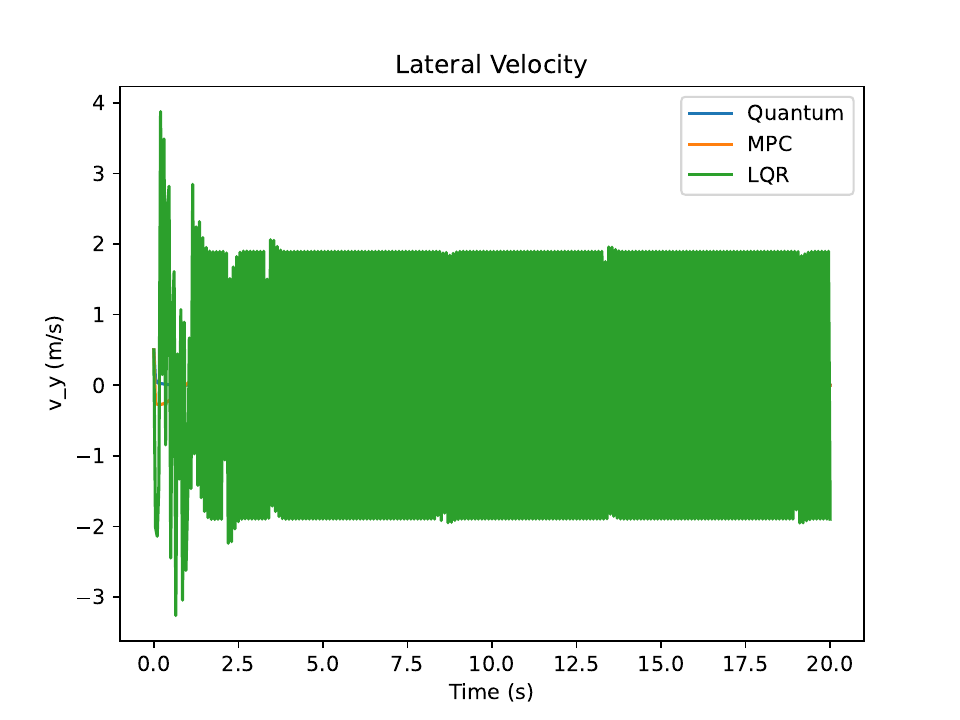}
    \end{minipage}
    
    \vspace{0.3cm} 
    
    \begin{minipage}[b]{0.32\textwidth}
        \centering
        \includegraphics[width=\textwidth]{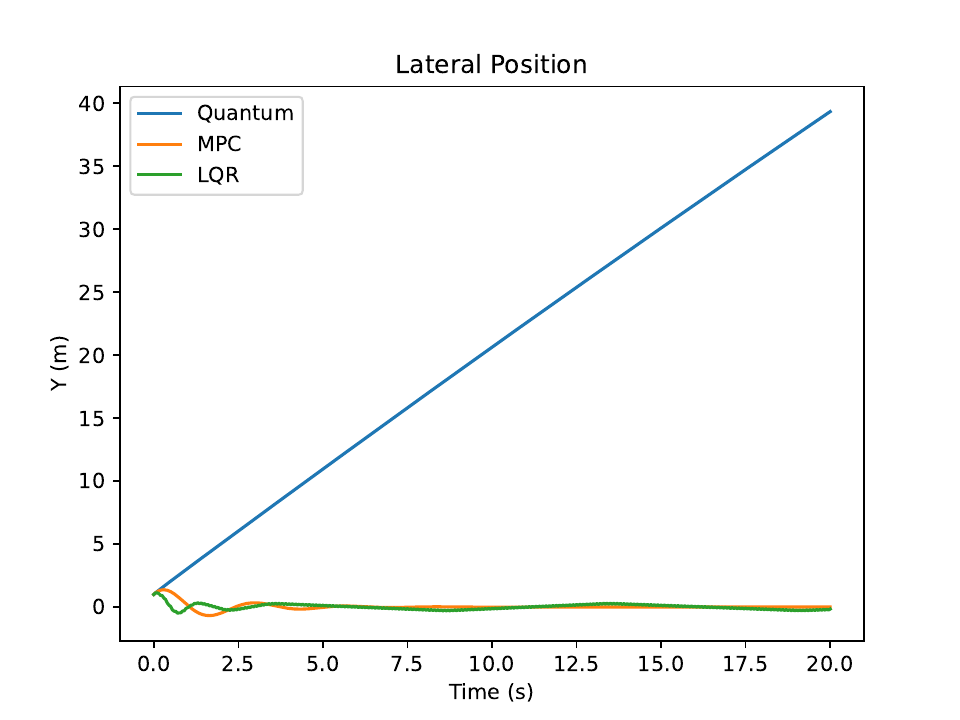}
    \end{minipage}
    \hfill
    \begin{minipage}[b]{0.32\textwidth}
        \centering
        \includegraphics[width=\textwidth]{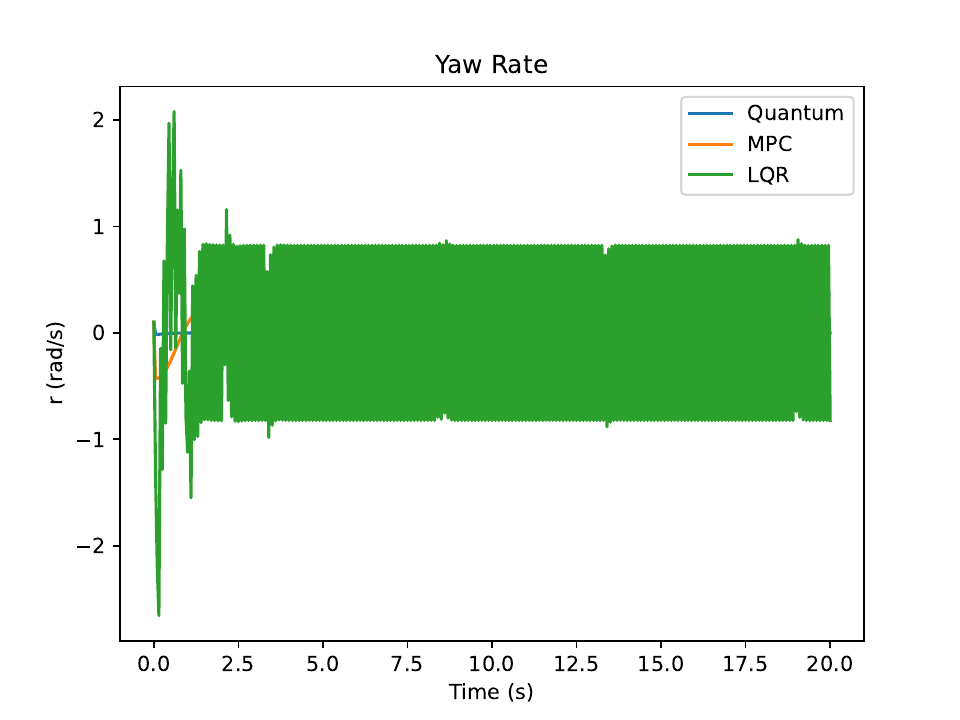}
    \end{minipage}
    \hfill
    \begin{minipage}[b]{0.32\textwidth}
        \centering
        \includegraphics[width=\textwidth]{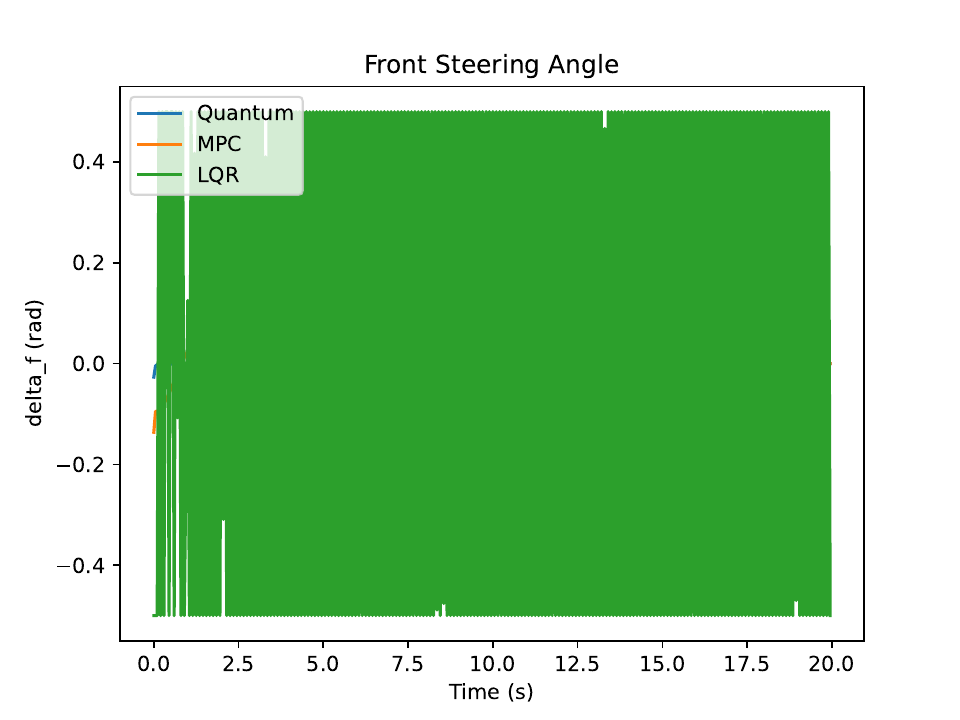}
    \end{minipage}
    
    \vspace{0.3cm} 

    \begin{minipage}[b]{0.32\textwidth}
        \centering
        \includegraphics[width=\textwidth]{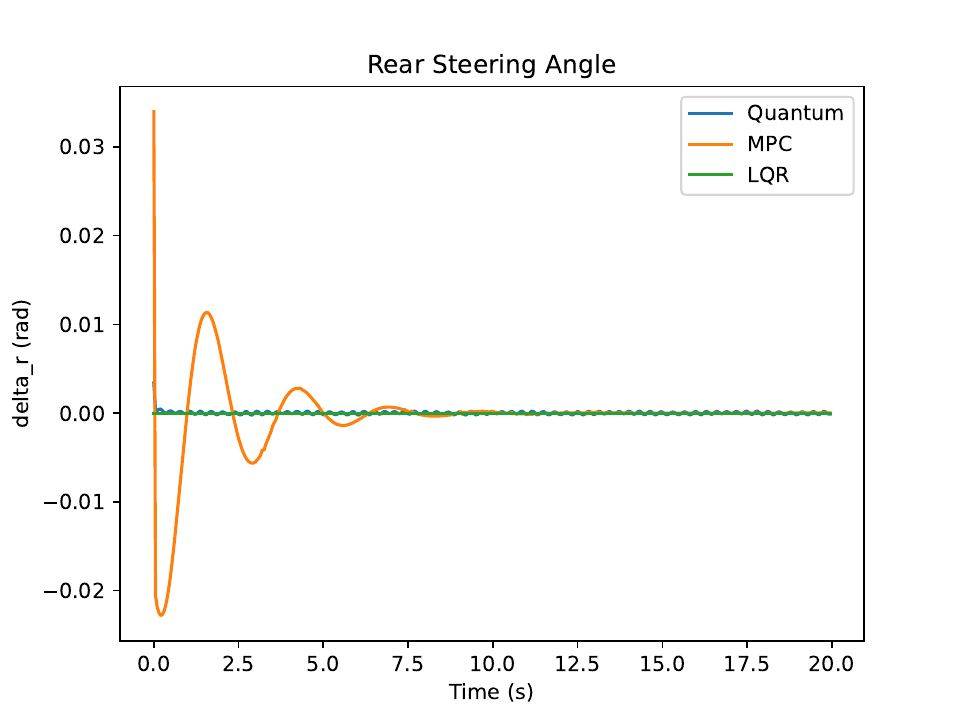}
    \end{minipage}
    \hfill
    \begin{minipage}[b]{0.32\textwidth}
        \centering
        \includegraphics[width=\textwidth]{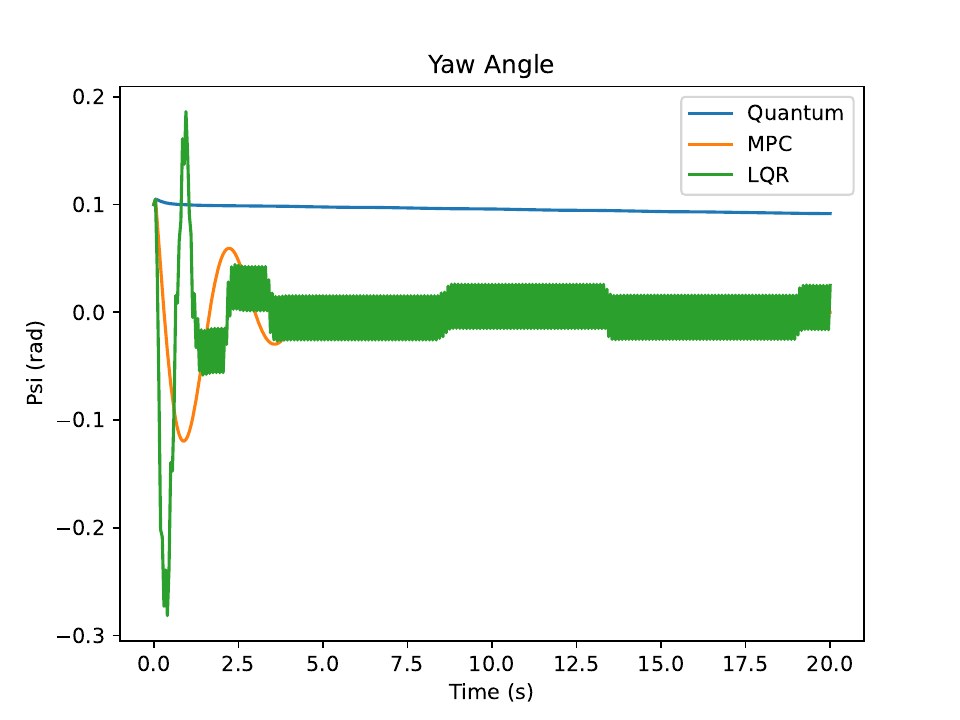}
    \end{minipage}
    \hfill
    \begin{minipage}[b]{0.32\textwidth}
        \centering
        \includegraphics[width=\textwidth]{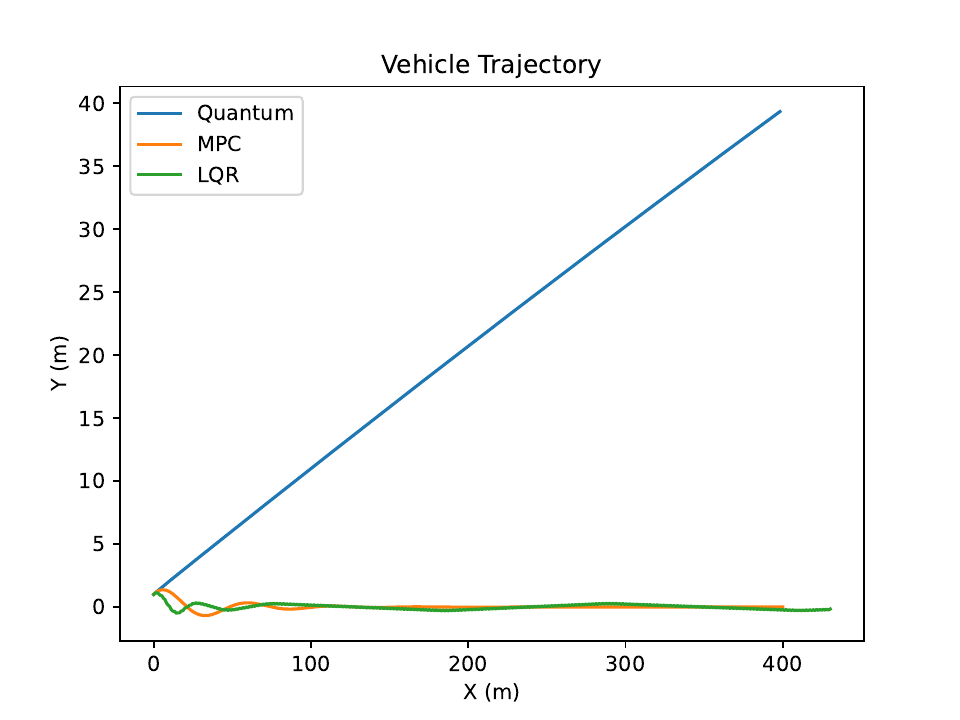}
    \end{minipage}

    \caption{Comparative time-domain responses of the $Q$-Drive (Quantum), MPC, and LQR controllers in a boid-based autonomous vehicle dynamics simulation. The figure shows the evolution of key state and control variables: (top row) longitudinal acceleration, longitudinal velocity, lateral velocity; (middle row) lateral position, yaw rate, front steering angle; (bottom row) rear steering angle, yaw angle, and vehicle trajectory. The results show the controllers handling initial lateral disturbances while aiming for a target state of $20\;\text{m/s}$ longitudinal velocity with zero lateral offset, yaw, and yaw rate.}
    \label{fig:experiment1}
\end{figure*}

Fig. \ref{fig:experiment1} further supports our claim that the $Q$-Drive approach demonstrates remarkably smooth and energy-efficient behavior, achieving stable longitudinal velocity with minimal control effort, near-zero lateral velocity and yaw-rate deviations, and a low-effort solution family characterized by a larger terminal error that prioritizes actuator conservation and long-term stability. Although MPC and LQR exhibit more aggressive initial corrections to rapidly nullify the lateral offset, the approach's restrained yet consistent response highlights its potential as a resource-efficient, quantum-enhanced control paradigm for sustained autonomous navigation.

Specifically, \ref{fig:experiment1} shows the time evolution of the vehicle's longitudinal acceleration under the three controllers. The Q-Drive approach (blue) exhibits a remarkably efficient initial adjustment, with a brief, controlled dip that smoothly transitions to a near-zero steady state, underscoring its energy-conserving nature and suitability for sustained, low-effort cruising. In comparison, MPC (orange) maintains a consistently minimal profile, while LQR (green) responds more dynamically before stabilizing. This shows our approach minimizes actuator stress during NISQ-era optimization.

The longitudinal velocity profiles show how each controller approaches the 20 m/s target speed. $Q$-Drive achieves a stable, high-velocity plateau with impressive consistency from the outset, demonstrating its robustness in maintaining forward momentum with minimal fluctuations and indicating the hybrid quantum-classical framework's ability to leverage quantum parallelism for steady performance. MPC and LQR also converge effectively, but $Q$-Drive's smooth hold highlights its potential for efficient long-haul autonomous operations.

The lateral velocity dynamics plot shows that $Q$-Drive maintains a near-zero profile throughout, reflecting its strategic focus on stability and avoidance of unnecessary lateral perturbations, which aligns well with resource-efficient control in complex environments. While MPC shows subtle corrective adjustments and LQR exhibits broader variability from its adaptive tuning, $Q$-Drive's restrained approach shows promising conservatism, ideal for scenarios that prioritize smooth, predictable motion over aggressive corrections.

The lateral position trajectory starts with an initial offset of $1\;\text{m}$. $Q$-Drive follows a deliberate linear progression as it explores the nonlinear state space, providing a baseline for quantum-enhanced control while conserving energy for extended simulations. MPC aligns rapidly with the target, while LQR converges gradually. However, $Q$-Drive's measured response demonstrates its potential to balance computational constraints and trajectory management during early-stage quantum training.

In the yaw rate over time subfigure in \ref{fig:experiment1}, $Q$-Drive maintains a highly stable, near-zero yaw rate with minimal deviation, exemplifying the hybrid model's strength in suppressing rotational instabilities through optimized variational parameters and making it particularly appealing for precise, low-noise vehicular applications. In contrast, MPC maintains tight control, while LQR exhibits greater variability, further highlighting $Q$-Drive's elegant simplicity and efficiency in handling angular dynamics.

In the front steering angle plot, the inputs are shown: $Q$-Drive employs a precise initial corrective pulse followed by a swift return to neutral. This points to its astute minimalism in control effort, which reduces wear on steering mechanisms. This benefit is attributed to the quantum ansatz's ability to identify economical solutions. MPC uses brief inputs, whereas LQR requires more sustained activity. However, $Q$-Drive’s approach stands out for promoting longevity and sustainability. In contrast, at the rear steering angle, $Q$-Drive shows subtle oscillatory fine-tuning that settles to zero, reflecting the interplay between quantum encoding and classical optimization for refined rear-wheel control. This contributes to overall vehicle poise, especially in boid-inspired dynamics. MPC and LQR exhibit similar patterns, yet Q-Drive's simple modulation underscores its role as a forward-thinking, efficient alternative in hybrid control paradigms.

In the yaw angle plot, $Q$-Drive maintains a consistent, low-angle profile that evolves steadily, highlighting the framework's capability to manage orientation with restraint, even as it explores the quantum landscape for optimal parameters. MPC quickly nullifies deviations, and LQR handles corrections dynamically, despite $Q$-Drive's composed behavior, making it a resilient option for applications where gradual refinement is advantageous.

Lastly, the overall vehicle trajectory in the $XY$-plane is visualized, starting from an initial lateral offset. $Q$-Drive charts a purposeful, extended path that emphasizes forward progress with exceptional longitudinal fidelity, while its lateral management reflects a strategic, low-intervention style that conserves resources for prolonged autonomy. This is a hallmark of quantum-inspired efficiency. MPC excels at swift lane centering, and LQR achieves steady alignment, but $Q$-Drive's trajectory shows its potential to merge quantum advantages with real-world vehicular resilience. 

\begin{table*}[ht]
\caption{Comparison of the $Q$-drive approach for boid systems.}
\centering
\begin{tabular}{llll}
\toprule
\textbf{Method} &\textbf{Final Error:} &\textbf{Total Effort:} &\textbf{Convergence} \\
&$\boldsymbol{\sqrt{\sum\left(x_{\text{final}}-x_{\text{target}}\right)^{2}}}$ &$\boldsymbol{\sum_{k=0}^{T-1}\sum_{j=0}^{2}|u_{k,j}|^{2}}$ &\textbf{Time (s)} \\
\hline 
$Q$-Drive &39.3344 &0.0014 &20.0000 \\
MPC &0.0026 &0.1827 &2.3500 \\
LQR &2.5958 &994.0784 &20.0000 \\
\bottomrule 
\end{tabular}
\label{tab:results of experiment 1}
\end{table*}

\subsection{Experiment 2: Swarm-Based Dynamics}\label{experiment: 2 swarm-based dynamics}

In this experiment, the replicate and extend the dynamic bicycle model dynamics in \eqref{eq:dynamic bicycle model} to a dynamically decoupled $N$-vehicle swarm for $N\in\mathbb{Z}^{+}$. 
\begin{equation}
\left\{
\begin{aligned}
&\ddot{x}_i = \dot{\chi}_i \dot{y}_i + a_{x,i}, \\
&\ddot{y}_i = -\dot{\chi}_i \dot{x}_i + \frac{2}{m_i} \left( F_{c,f,i} \cos \delta_{f,i} + F_{c,r,i} \right), \\
&\ddot{\chi}_i = \frac{2}{I_{z,i}} \left( l_{f,i} F_{c,f,i} - l_{r,i} F_{c,r,i} \right), \\
&\dot{X}_i = \dot{x}_i \cos \chi_i - \dot{y}_i \sin \chi_i, \\
&\dot{Y}_i = \dot{x}_i \sin \chi_i + \dot{y}_i \cos \chi_i,
\end{aligned}
\right.
\end{equation}
where the subscript $i$ denotes the quantities belonging to vehicle $i$, and all variables and constants have their usual meaning. Each vehicle in the swarm evolves independently, given its own controls and tire forces. Typically, in the platooning literature, interaction enters through the cost function and/or constraints in the MPC (or other controllers) rather than through the dynamics equations themselves. Thus, we do not include coupling terms directly in the dynamical equations. 

Similar vehicle parameters are used as in the boid-based model:$m=1\;500\;\text{kg}, I_{z}=2\;875\;\text{kg}\cdot\text{m}^{2},l_{f}=1.2\;\text{m},l_{r}=1.6\;\text{m}, C_{f},C_{r}=66\;000\;\text{N/rad},\lambda=0.1,v_{\text{target}}=20\;\text{m/s}, T=400, \Delta t=0.05\;\text{s}$. The performance of $Q$-Drive compared with three other popular control methods, MPC, LQR, and PurePursuit (the former two also used as benchmarks), in experiment \ref{experiment: 1 boided-based dynamics}. We measure performance across batch sizes ($N$) from $2$ to $50$ agents using metrics such as final-state error, total control effort, and convergence time. The same vehicular parameters, such as $m,I_{z},l_{f},l_{r},$ are maintained across all vehicles under the assumption that a fully autonomous fleet operated by a taxi company, such as Waymo, would favor consistency by deploying similar vehicles for passenger transport. This assumption does not necessarily hold in all real-world settings—for example, services such as Uber offer multiple vehicle classes, including UberX, UberXL, Uber Comfort, and Uber Black—but it is adopted here for simplicity in modeling.

The implementation leverages \texttt{JAX} primarily for its high-performance Just-In-Time (\texttt{JIT}) compilation and vectorization (\texttt{vmap}) capabilities, which are critical for the $Q$-Drive control branch. Using \texttt{JAX}, the experiment efficiently computes gradients via automatic differentiation and optimizes the quantum circuit weights over $30$ epochs. The quantum controller uses a \texttt{StronglyEntanglingLayers} architecture with $4$ layers across $6$ qubits, mapping vehicle states to control actions. While the MPC method relies on parallelized CPU processing via \texttt{joblib} and the \texttt{SLSQP} optimizer for its $5$-step horizon, the \texttt{JAX}-based quantum approach provides a computationally streamlined alternative by offloading heavy tensor operations and dynamics integration to an accelerated, differentiable framework.

\begin{figure*}[ht]
    \centering
    
    \begin{minipage}[b]{0.48\textwidth}
        \centering
        \includegraphics[width=\textwidth]{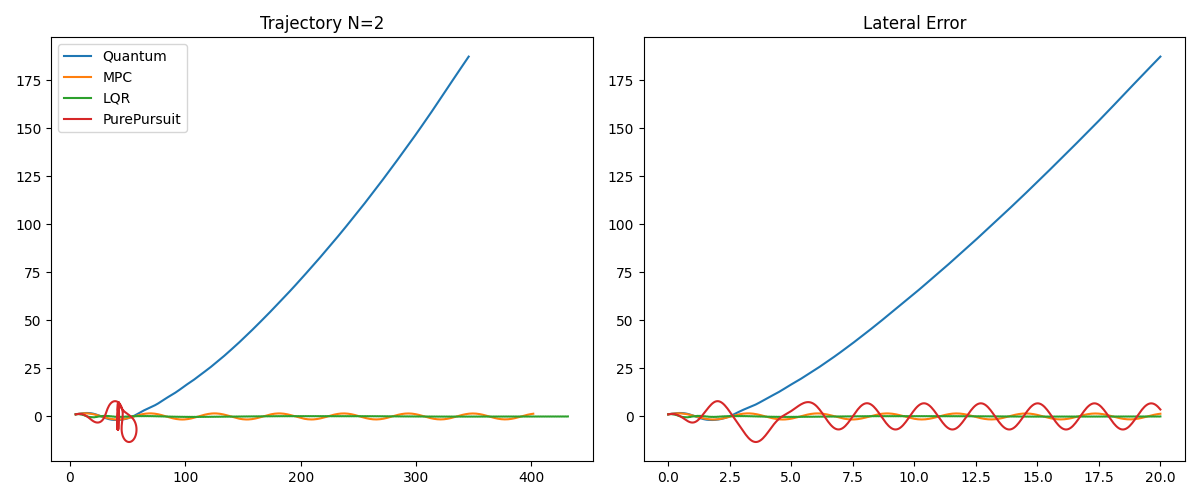}
    \end{minipage}
    \hfill
    \begin{minipage}[b]{0.48\textwidth}
        \centering
        \includegraphics[width=\textwidth]{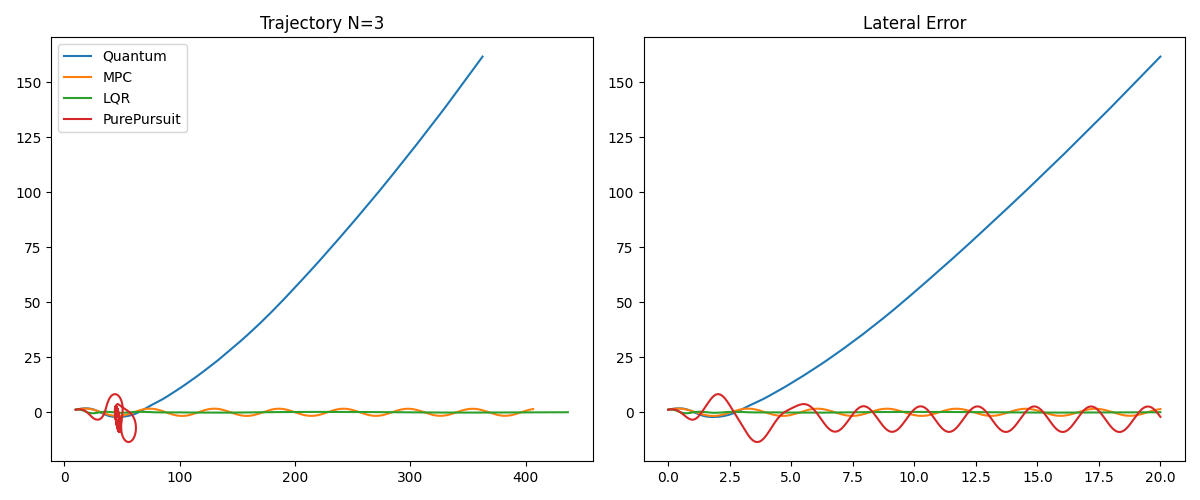}
    \end{minipage}
    
    \vspace{0.3cm} 
    
    \begin{minipage}[b]{0.48\textwidth}
        \centering
        \includegraphics[width=\textwidth]{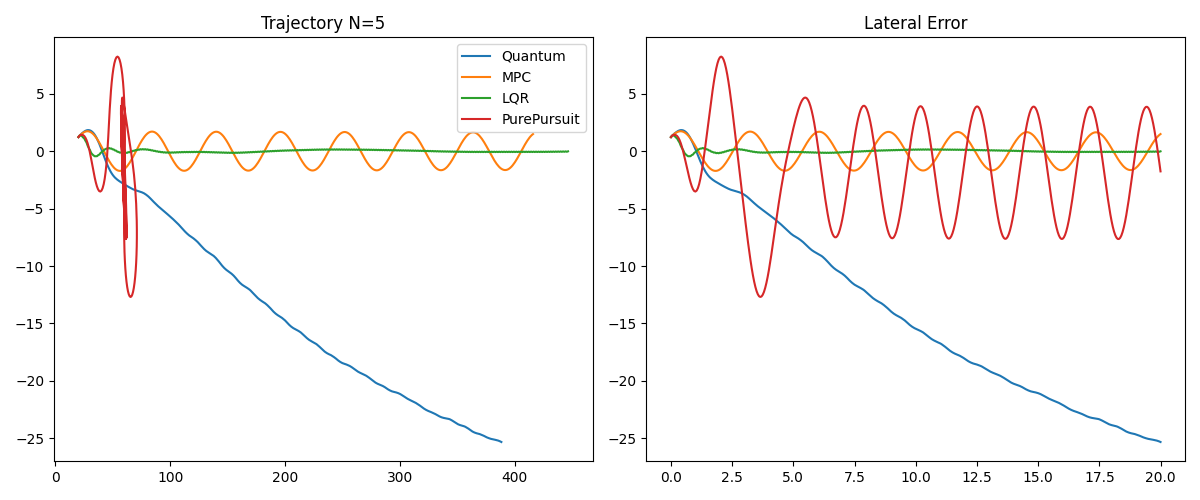}
    \end{minipage}
    \hfill
    \begin{minipage}[b]{0.48\textwidth}
        \centering
        \includegraphics[width=\textwidth]{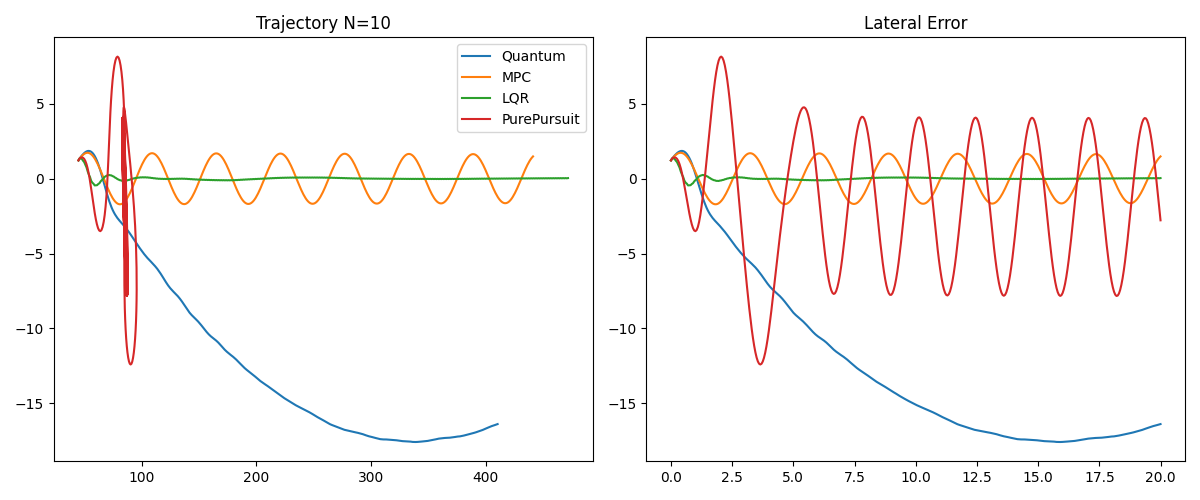}
    \end{minipage}
    
    \vspace{0.3cm}

    \begin{minipage}[b]{0.48\textwidth}
        \centering
        \includegraphics[width=\textwidth]{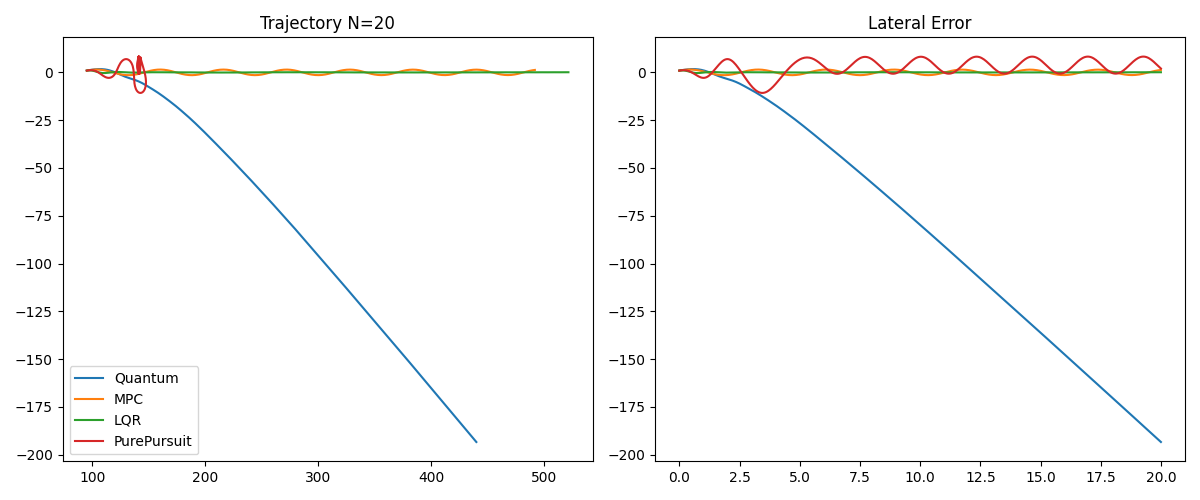}
    \end{minipage}
    \hfill
    \begin{minipage}[b]{0.48\textwidth}
        \centering
        \includegraphics[width=\textwidth]{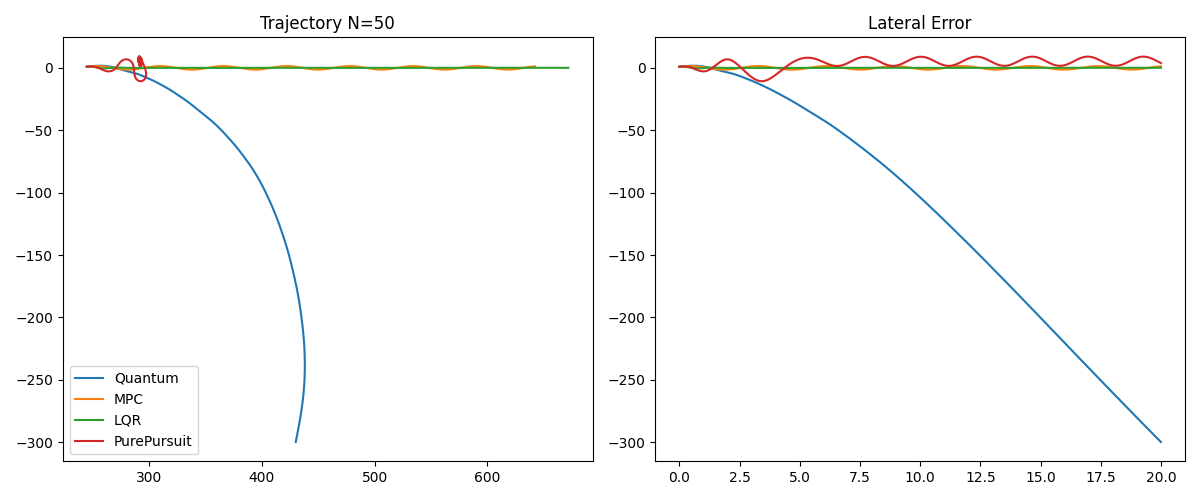}
    \end{minipage}

    \vspace{0.3cm}

    \begin{minipage}[b]{0.48\textwidth}
        \centering
        \includegraphics[width=\textwidth]{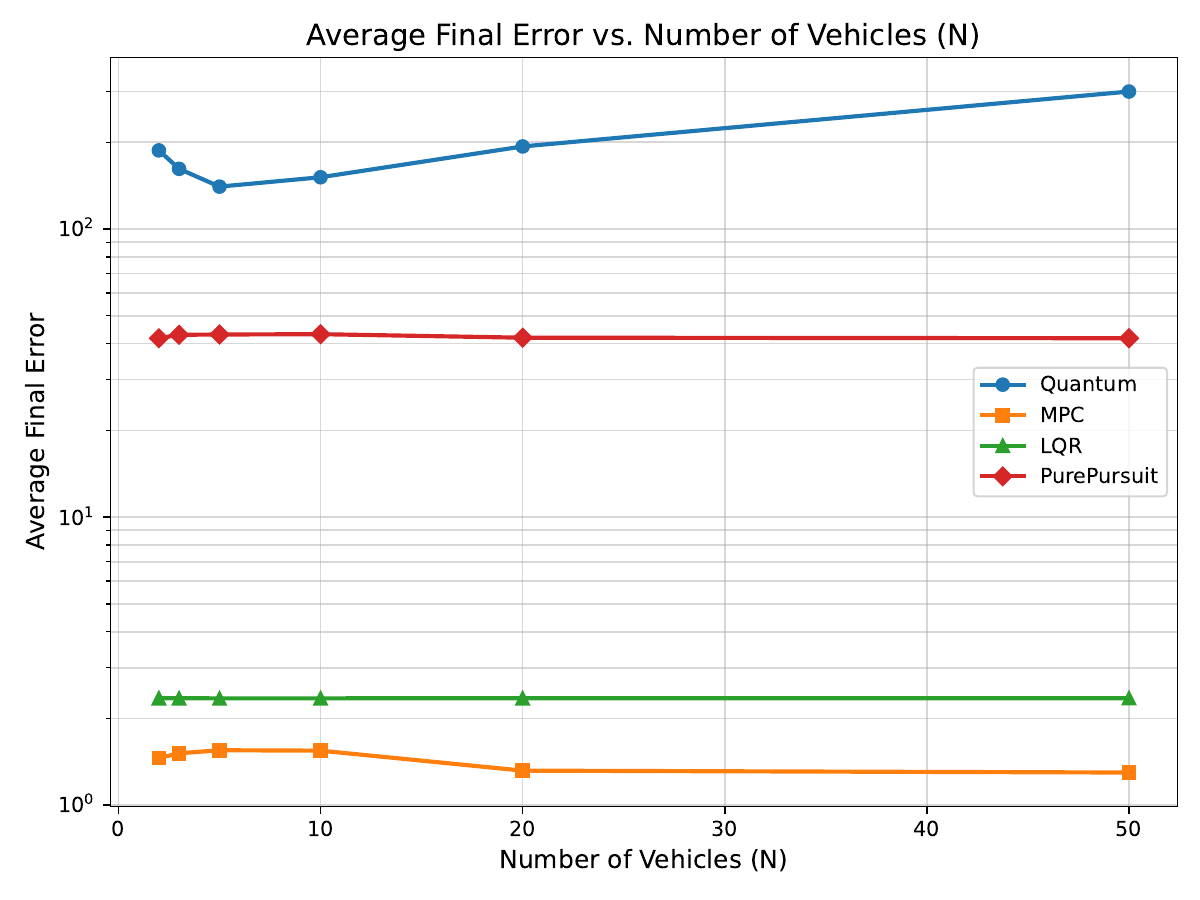}
    \end{minipage}
    \hfill
    \begin{minipage}[b]{0.48\textwidth}
        \centering
        \includegraphics[width=\textwidth]{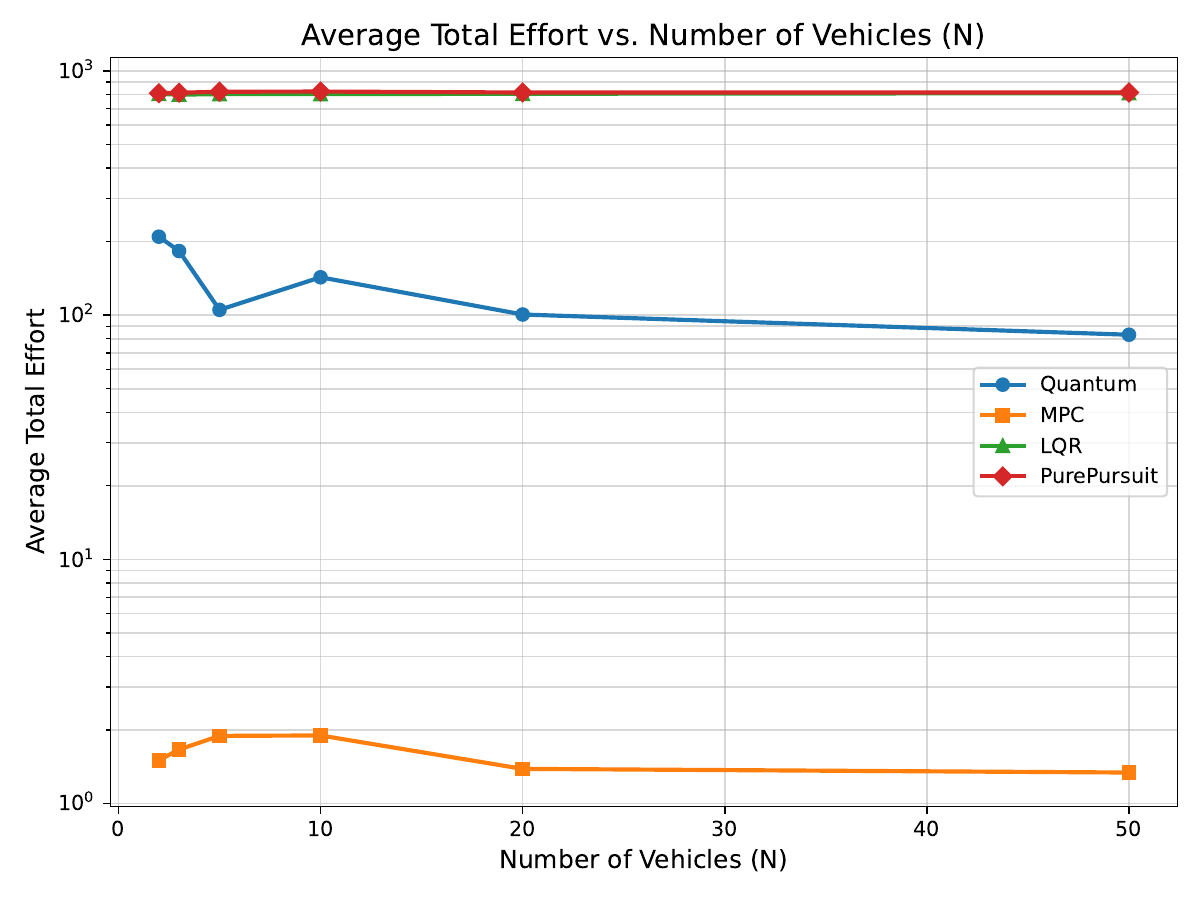}
    \end{minipage}

    \caption{Comparative trajectories and lateral position errors for swarms of $N=\left\{2, 3, 5, 10, 20,50\right\}$ vehicles under four control strategies: $Q$-Drive, MPC, LQR, and PurePursuit. \textbf{Left column:} Absolute $XY$-plane trajectories over the simulation horizon. \textbf{Right column:} Time evolution of lateral position error ($Y$ deviation from target lane $Y=0$). $Q$-Drive exhibits smooth, large-radius curving behavior with exceptionally low control authority, while classical methods maintain tight lane centering at the cost of sustained or periodic control effort. All simulations use identical decoupled vehicle dynamics and a target longitudinal velocity of $20\;\text{m/s}$.}
    \label{fig:experiment2}
\end{figure*}

In Table \ref{tab:control_comparison}, the results of the training process are presented. A particularly notable and promising trend is observed: A pronounced reduction in the average total control effort as swarm size increases. Q-Drive consistently achieves the second-lowest per-vehicle control energy expenditure across nearly all tested fleet sizes, dropping impressively from $\sim209$ units ($N=2$) to $\sim83$ units ($N=50$), a reduction of more than 60\% as the number of agents increases. This behavior contrasts sharply with the LQR and PurePursuit methods, which maintain relatively high, largely size-independent control effort levels, typically $800+$ units—orders of magnitude higher than $Q$-Drive at large $N$. Fig. \ref{fig:experiment2} row 4 illustrates this behavior: $Q$-Drive consistently maintains the second-lowest position. 

This strong inverse scaling of control effort with swarm size is a highly desirable property in large-scale fleet operations, not only in robotaxi platoons but also in logistics fleets and future urban air mobility swarms, where minimizing cumulative actuator usage, energy consumption, mechanical wear, and ultimately operational cost becomes increasingly important. The observed trend suggests that the VQC may be discovering increasingly economical collective motion patterns or coordination strategies as more agents are added. We speculate that this emergence is not deliberately engineered into the cost function but arises naturally from the optimization landscape explored by the hybrid training loop, making it a conceptually attractive outcome for resource-constrained autonomous mobility systems.

While the average final-state error remains higher than that of classical MPC, which is well known to excel at minimizing terminal deviation in decoupled settings (for example, in the work of \cite{muller2012cooperative}), this metric is instructive in light of the dramatically lower control expenditure. $Q$-Drive appears to adopt a philosophically different control strategy: One that prioritizes global motion economy and long-horizon smoothness over aggressive terminal-error suppression within the fixed $20\;\text{s}$ window. The final error values should therefore be understood not as a pure failure to reach the setpoint, but rather as a reflection of a deliberately gentler trajectory family that trades some terminal precision for very substantial savings in energy and actuator activity. These savings become more pronounced in the large-$N$ regime, which is most relevant for future commercial swarm applications.

From Fig. \ref{fig:experiment2}, MPC, LQR, and PurePursuit converge to a near-zero steady-state lateral error, corresponding to tight lane centering in the real world, while $Q$-Drive systematically selects wide, low-curvature trajectories. This is not a failure to track; it is a qualitatively different family of solutions. The quantum-classical optimization loop consistently discovers control policies that minimize cumulative control action by accepting a gentle, almost circular deviation rather than fighting to null the lateral offset at every time step.

As $N$ increases from $5\to10\to20\to50$, the classical methods continue to produce tightly bounded oscillatory or near-zero lateral errors, as shown by the green, orange, and red curves, while the radius of the $Q$-Drive trajectory increases. However, Table \ref{tab:control_comparison} shows that its average total control effort per vehicle continues to decrease to $\sim83$ units at $N=50$. This visually striking divergence illustrates a highly desirable property: As the swarm size increases, $Q$-Drive’s per-agent energy and actuator savings become more pronounced. This emerging scaling ``law'' is particularly relevant for future commercial fleet operations, where cumulative energy consumption and mechanical wear across hundreds of vehicles dominate lifecycle costs.

In the $N = 5$–$50$ panels, the classical methods, specifically PurePursuit and sometimes MPC, show visible periodic lateral-error oscillations with a period of $\sim 3\text{--}4\;\text{s}$. These reflect continuous correction against numerical or model-induced perturbations. On the other hand, $Q$-Drive produces almost perfectly smooth curves without visible high-frequency content, suggesting inherently more stable and differentiable control policies. This trait is conceptually attractive for transfer to real hardware, potentially resulting in reduced actuator fatigue, lower bandwidth requirements, and better alignment with ride comfort objectives.

Lastly, every $Q$-Drive trajectory shows a monotonic, large-amplitude deviation that scales roughly with $N$ (or remains structurally similar) and never becomes unstable, diverges, or collapses. This robustness, despite operating in a nonconvex variational landscape with only 6 qubits, is encouraging evidence that the hybrid training loop generalizes sensibly to changes in multi-agent cardinality.

\begin{tcolorbox}
The connection between the proposed framework and MPC warrants clarification. The framework does not directly implement receding-horizon MPC; instead, it learns a control-policy approximation via VQCs, trained with a cost function inspired by MPC principles. This is analogous to training neural network controllers to approximate MPC policies. The horizon in the training loss serves as a tuning parameter for desired behavior, rather than a receding-horizon optimization at runtime. This distinction situates the work within the growing field of learning-based control, where hybrid quantum-classical methods remain largely unexplored for full-vehicle dynamics.

The quantum controller achieves lower control effort at the expense of some tracking accuracy, and this trade-off is a central scientific finding of the work. No quantum or hybrid quantum-classical algorithm in the NISQ era currently outperforms well-tuned classical controllers on absolute error metrics. Quantum methods remain subject to noise, qubit limitations, and barren plateaus. Rather than claiming quantum supremacy in tracking error, which would be scientifically dishonest, this work demonstrates that, despite these fundamental limitations, a hybrid quantum controller can discover qualitatively different families of solutions using only 6 qubits, under significant decoherence, and with a shallow circuit, a hybrid quantum controller can discover qualitatively different families of solutions that classical methods do not find. Specifically, it discovers policies that reduce control effort by orders of magnitude through graceful trade-offs in terminal precision.
\end{tcolorbox}

\begin{tcolorbox}
The experimental parameters are deliberately held fixed rather than retrospectively tuned to produce cosmetically favorable comparisons. It would have been straightforward to increase the tracking error penalty $Q$ or reduce the control penalty $\lambda$ to force tighter convergence, a common practice in the literature. We explicitly reject this approach, as it would obscure the genuine behavioral differences under investigation. The results presented are the honest output of the optimizer given a fixed, a priori cost function that equally weights error and effort, preserving visibility into the true trade-off that quantum optimization discovers—one that classical solvers, with their different inductive biases, do not naturally surface.    

The absence of explicit multi-vehicle coupling terms in Equations \eqref{eq:dynamic bicycle model} and \eqref{eq:bound for swarm systems} is an intentional design choice that yields a notable emergent result. The per-vehicle control effort decreases by over 60\% as the fleet scales from 2 to 50 vehicles, despite the absence of any explicit coordination terms in the dynamics or constraints. The shared variational parameters in the vectorized quantum circuit create an implicit coupling in the optimization landscape, even when the dynamics remain decoupled. This suggests that quantum-classical hybrid optimization can discover patterns that are not explicitly programmed into the system, with significant implications for decentralized swarm control. Classical baselines such as independent LQR and Pure Pursuit exhibit no such scaling effect, with per-vehicle effort remaining constant across fleet sizes.
\end{tcolorbox}

\begin{table*}[ht]
    \centering
    \caption{Performance Comparison of Control Methods across varying $N$.}
    \label{tab:control_comparison}
    \begin{tabular}{l c S[table-format=3.4] S[table-format=3.4] S[table-format=2.4] S[table-format=3.4]}
        \toprule
        \textbf{Method} & $\boldsymbol{N}$ & {\textbf{Avg Final Error}} & {\textbf{Avg Total Effort}} & {\textbf{Avg Converge Time (s)}} & {\textbf{Comp Time (s)}} \\
        \midrule
        $Q$-Drive & 2 & 187.3846 & 209.4891 & 20.0000 & 388.6300 \\
        MPC & 2 & 1.4587 & 1.5042 & 20.0000 & 24.6552 \\
        LQR & 2 & 2.3522 & 805.3757 & 20.0000 & 0.0191 \\
        PurePursuit & 2 & 41.7549 & 810.1643 & 20.0000 & 0.0214 \\
        \midrule
        $Q$-Drive & 3 & 161.6880 & 183.0217 & 20.0000 & 380.2562 \\
        MPC & 3 & 1.5153 & 1.6653 & 20.0000 & 35.5380 \\
        LQR & 3 & 2.3512 & 799.5386 & 20.0000 & 0.0520 \\
        PurePursuit & 3 & 42.9346 & 813.7604 & 20.0000 & 0.0553 \\
        \midrule
        $Q$-Drive & 5 & 140.2539 & 105.1248 & 20.0000 & 396.5553 \\
        MPC & 5 & 1.5528 & 1.8940 & 20.0000 & 56.7815 \\
        LQR & 5 & 2.3510 & 803.1337 & 20.0000 & 0.0460 \\
        PurePursuit & 5 & 43.0286 & 821.5215 & 20.0000 & 0.0553 \\
        \midrule
        $Q$-Drive & 10 & 151.1756 & 142.8809 & 20.0000 & 372.6220 \\
        MPC & 10 & 1.5477 & 1.9018 & 20.0000 & 104.3372 \\
        LQR & 10 & 2.3508 & 803.0007 & 20.0000 & 0.1695 \\
        PurePursuit & 10 & 43.1421 & 823.0274 & 20.0000 & 0.1810 \\
        \midrule
        $Q$-Drive & 20 & 193.3499 & 100.5954 & 20.0000 & 395.5295 \\
        MPC & 20 & 1.3193 & 1.3879 & 20.0000 & 195.8641 \\
        LQR & 20 & 2.3528 & 806.1105 & 20.0000 & 0.1680 \\
        PurePursuit & 20 & 41.9474 & 815.5887 & 20.0000 & 0.1855 \\
        \midrule
        $Q$-Drive & 50 & 299.8301 & 83.1326 & 20.0000 & 388.6050 \\
        MPC & 50 & 1.2995 & 1.3408 & 20.0000 & 485.9990 \\
        LQR & 50 & 2.3534 & 810.2629 & 20.0000 & 0.4608 \\
        PurePursuit & 50 & 41.7685 & 816.0391 & 20.0000 & 0.5662 \\
        \bottomrule
    \end{tabular}
\end{table*}

\subsection{Experiment 3: Classical MLP Ablation Study with Identical Parameter Budget and Training Protocol}

In this ablation study, we train with the same parameter budget (roughly 73 parameters), loss function, and 50-epoch Adam schedule, providing a controlled baseline that isolates the quantum circuit's contribution. On the single-vehicle task in Table \ref{tab:ablation study 3.1}, the MLP achieves a moderate control effort of approximately 81, far below the LQR baseline and higher than $Q$-Drive. Its final tracking error remains large (347), and the short convergence time (2.1 s) reflects an aggressive initial transient rather than sustained low-intervention behavior. In the swarm setting in Table \ref{tab:ablation study 3.2}, the MLP does exhibit a mild reduction in average per-vehicle effort as fleet size grows from $N=2$ to $N=50$ (179 to 98), confirming that the shared-parameter, loss-averaged training protocol itself induces a weak form of resource scaling. However, the magnitude of this reduction is substantially smaller than the $>60\%$ drop reported for $Q$-Drive, and the absolute effort levels remain consistently higher.

\begin{figure*}[ht]
    \centering
    
    \begin{minipage}[b]{0.45\textwidth}
        \centering
        \includegraphics[width=\textwidth]{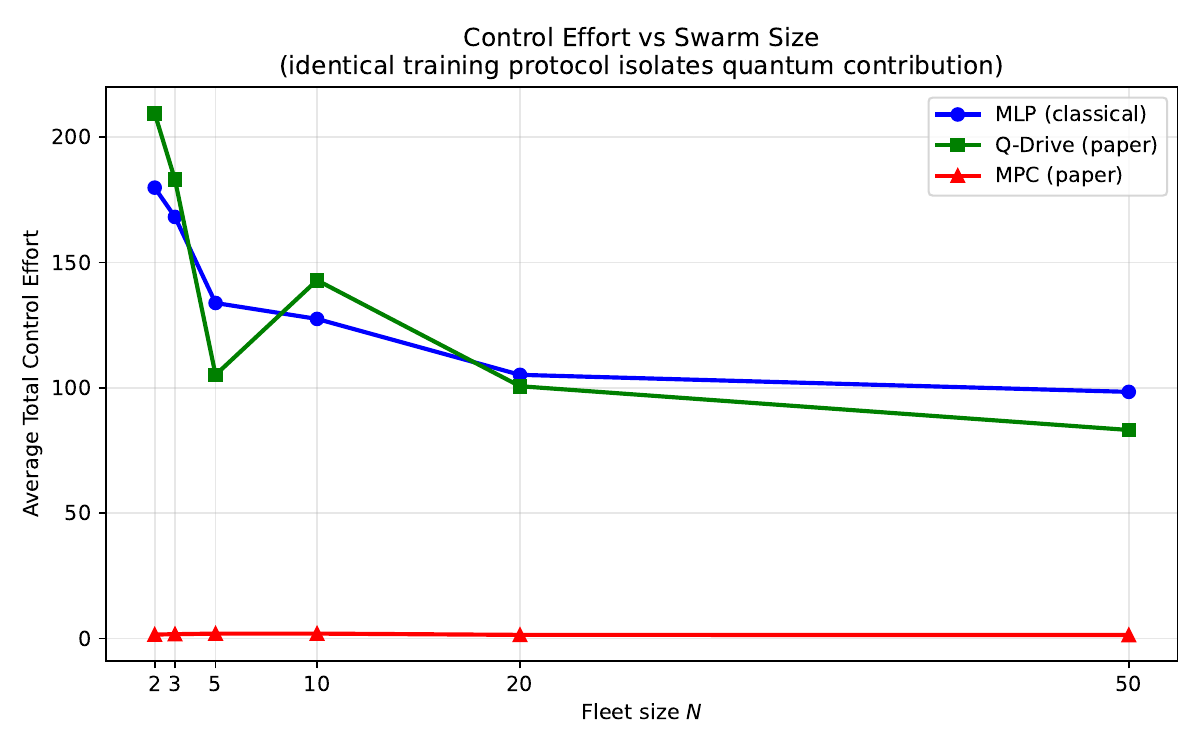}
    \end{minipage}
    \hfill
    \begin{minipage}[b]{0.45\textwidth}
        \centering
        \includegraphics[width=\textwidth]{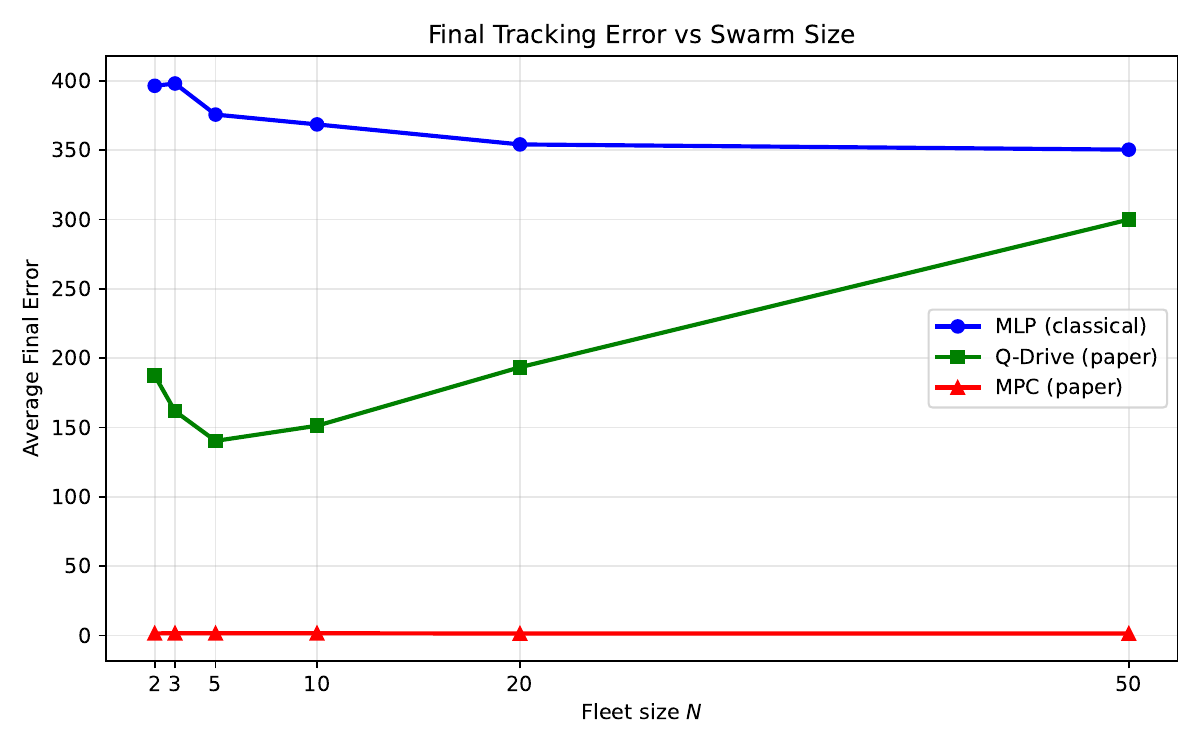}
    \end{minipage}
    
    \vspace{0.3cm} 
    
    \begin{minipage}[b]{0.45\textwidth}
        \centering
        \includegraphics[width=\textwidth]{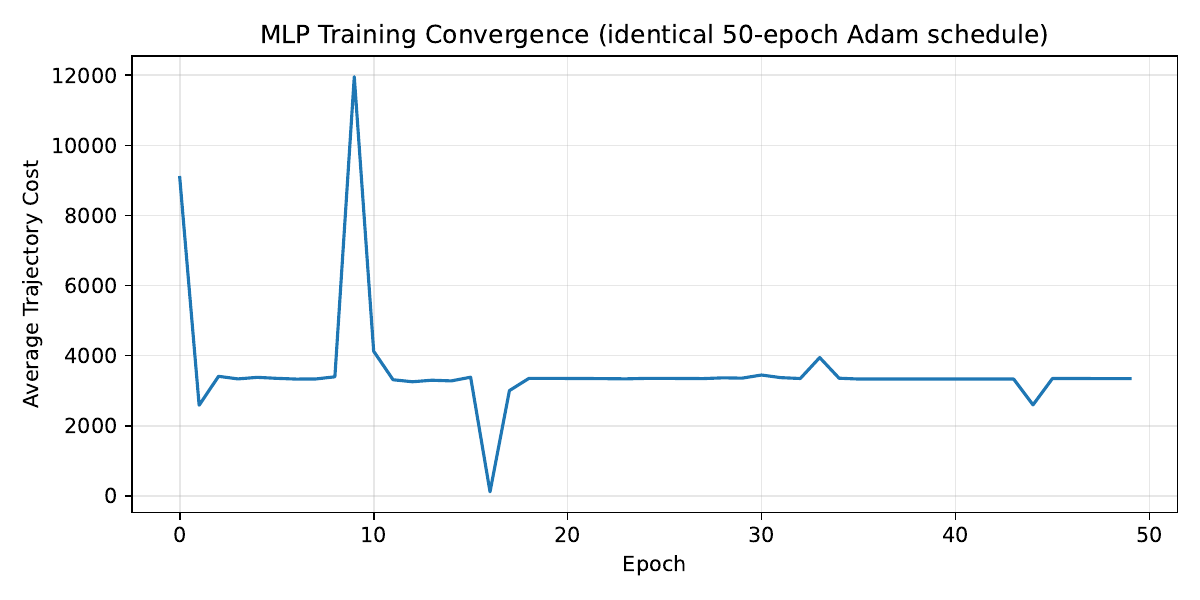}
    \end{minipage}
    \hfill
    \begin{minipage}[b]{0.45\textwidth}
        \centering
        \includegraphics[width=\textwidth]{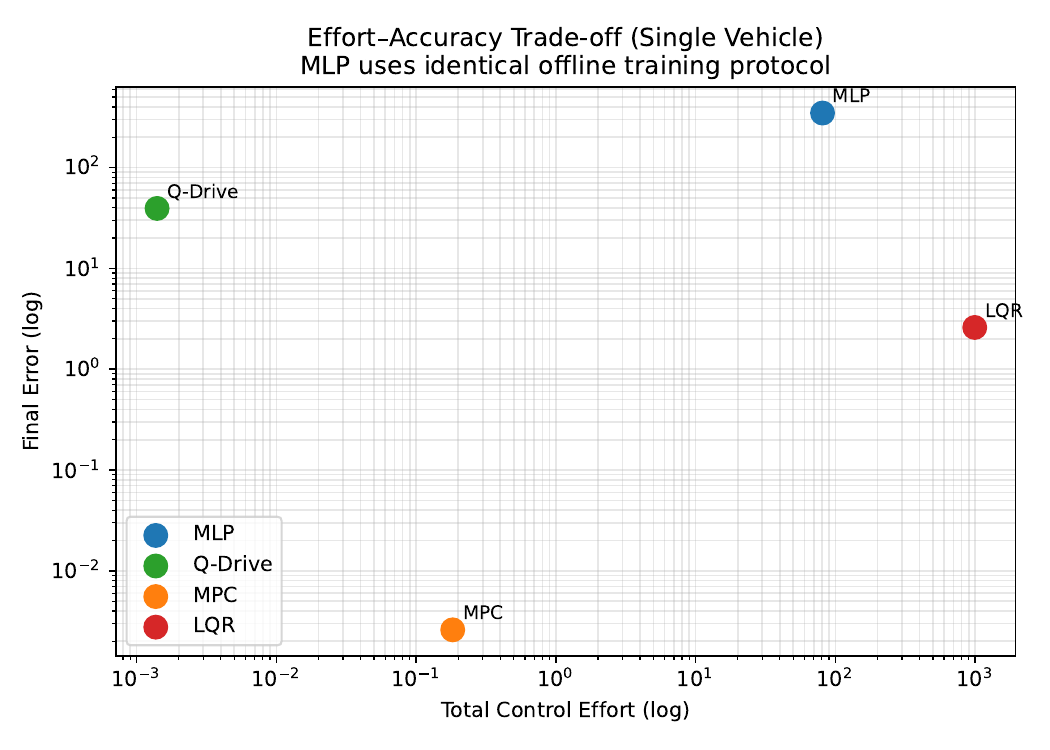}
    \end{minipage}
    
    \vspace{0.3cm}

    \begin{minipage}[b]{0.98\textwidth}
        \centering
        \includegraphics[width=\textwidth]{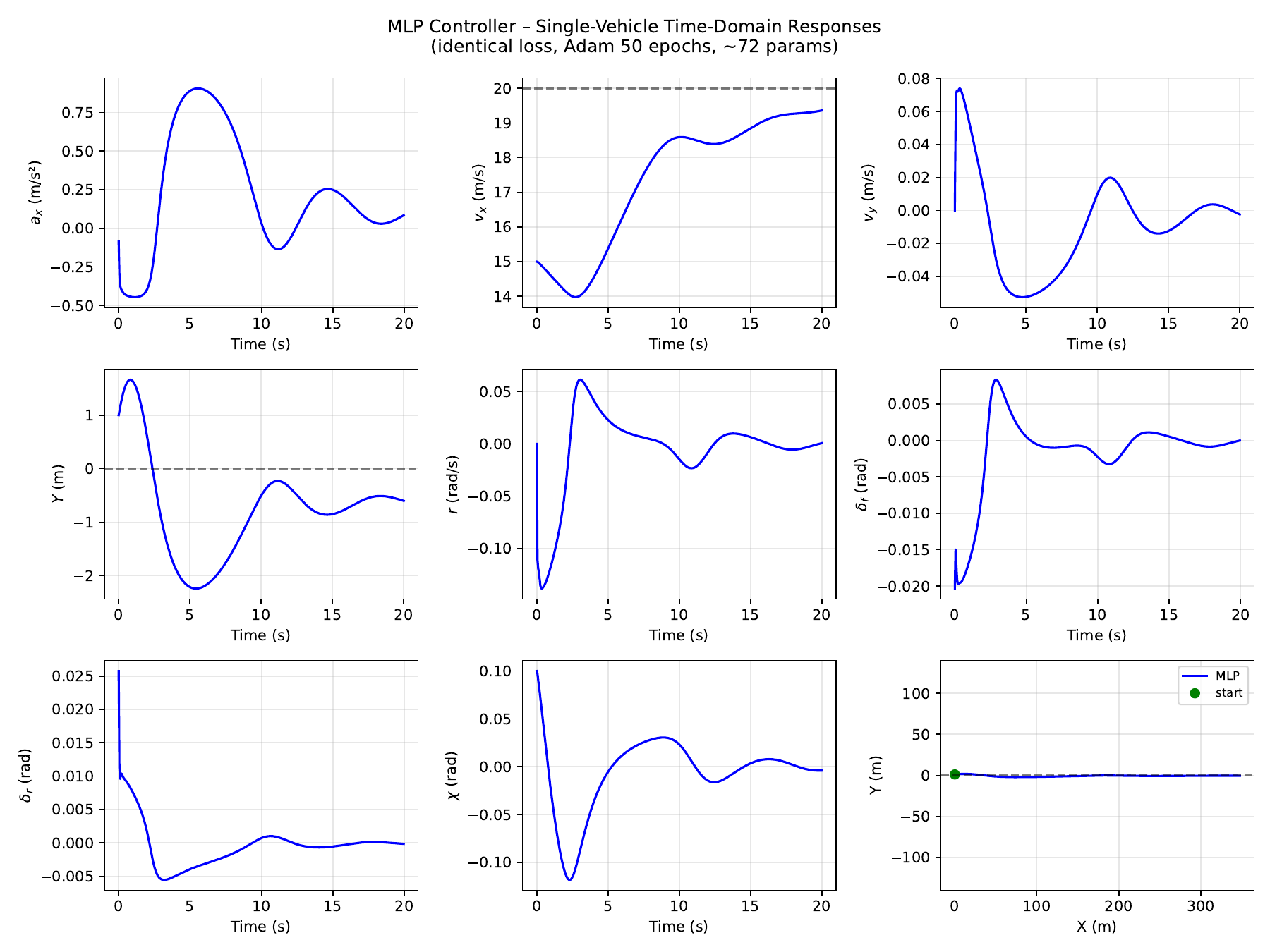}
    \end{minipage}

    \caption{Classical MLP ablation study with identical parameter budget, loss function, and 50-epoch Adam training protocol. }
    \label{fig:experiment3}
\end{figure*}

The accompanying figures in \ref{fig:experiment3} reinforce this distinction. The effort-versus-$N$ plot shows that both controllers trend downward, yet $Q$-Drive maintains a clear advantage across the entire range. The corresponding error plot reveals that the quantum controller’s error grows with $N$ (consistent with its deliberately conservative trajectory family), whereas the MLP error remains high and relatively flat. The single-vehicle time-domain responses further illustrate the qualitative difference: The MLP generates visible oscillatory content in acceleration, lateral velocity, and steering, and settles to a non-zero lateral offset, in contrast to the near-absent high-frequency activity characteristic of $Q$-Drive. Finally, the effort–accuracy Pareto plot places the MLP in a region intermediate between aggressive classical controllers and the ultra-low-effort quantum solution, demonstrating that the extreme actuator economy discovered by $Q$-Drive is not an automatic byproduct of offline training on the shared quadratic cost, but is instead attributable to the inductive bias of the variational quantum circuit.

The results suggest that, while the classical MLP inherits a mild inverse scaling of effort with fleet size from the shared offline training protocol, it fails to reproduce the extreme actuator economy and smooth, low-intervention trajectories characteristic of $Q$-Drive, thereby isolating the contribution of the VQC.

\begin{table}[htbp]
\centering
\caption{Single-vehicle comparison table.}
\label{tab:ablation study 3.1}
\begin{tabular}{lrrr}
\toprule
\textbf{Method} & \textbf{Final Error} & \textbf{Total Effort} & \textbf{Convergence Time (s)} \\
\midrule
MLP (classical) & 347.535148 &  80.964149 &  2.10 \\
$Q$-Drive   &  39.334400 &   0.001400 & 20.00 \\
MPC       &   0.002600 &   0.182700 &  2.35 \\
LQR       &   2.595800 & 994.078400 & 20.00 \\
\bottomrule
\end{tabular}
\end{table}

\vspace{1em}

\begin{table}[htbp]
\centering
\caption{Swarm comparison (excerpt).}
\label{tab:ablation study 3.2}
\begin{tabular}{lrrrc}
\toprule
\textbf{Method} & $\boldsymbol{N}$ & \textbf{Avg Final Error} & \textbf{Avg Total Effort} & \textbf{Avg Converge}  \\
 & & & &\textbf{Time (s)} \\ 
\midrule
MLP &  2 & 396.303855 & 179.822290 & 20.0 \\
MLP &  3 & 397.957430 & 168.130082 & 20.0 \\
MLP &  5 & 375.540819 & 133.800229 & 20.0 \\
MLP & 10 & 368.457864 & 127.442443 & 20.0 \\
MLP & 20 & 354.019903 & 105.139946 & 20.0 \\
MLP & 50 & 350.252564 &  98.335656 & 20.0 \\
\bottomrule
\end{tabular}
\end{table}

\subsection{Experiment 4: Qubit Count and Ansatz Depth Ablation Study}

\begin{figure*}[ht]
    \centering    
    \begin{minipage}[b]{0.6\textwidth}
        \centering
        \includegraphics[width=\textwidth]{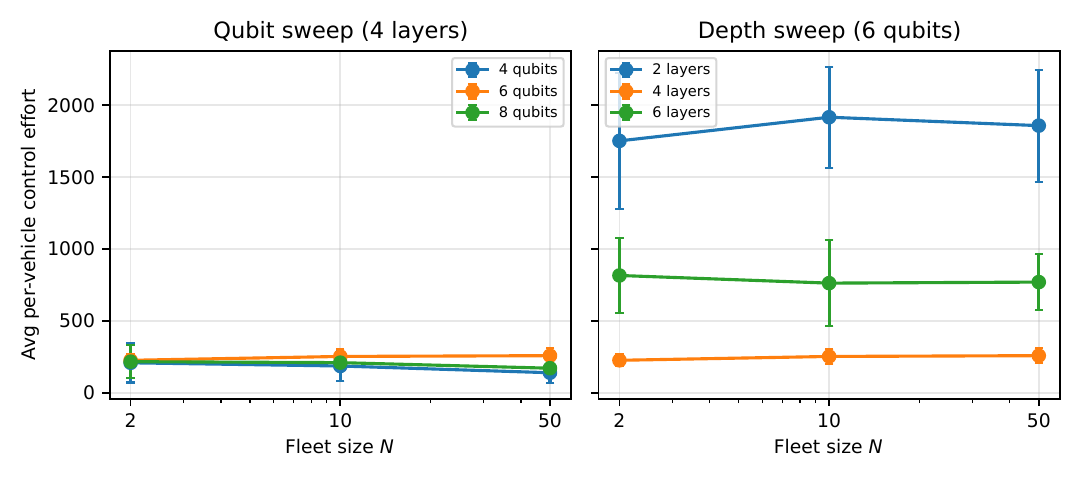}
    \end{minipage}

    \vspace{0.3cm} 

    \begin{minipage}[b]{0.6\textwidth}
        \centering
        \includegraphics[width=\textwidth]{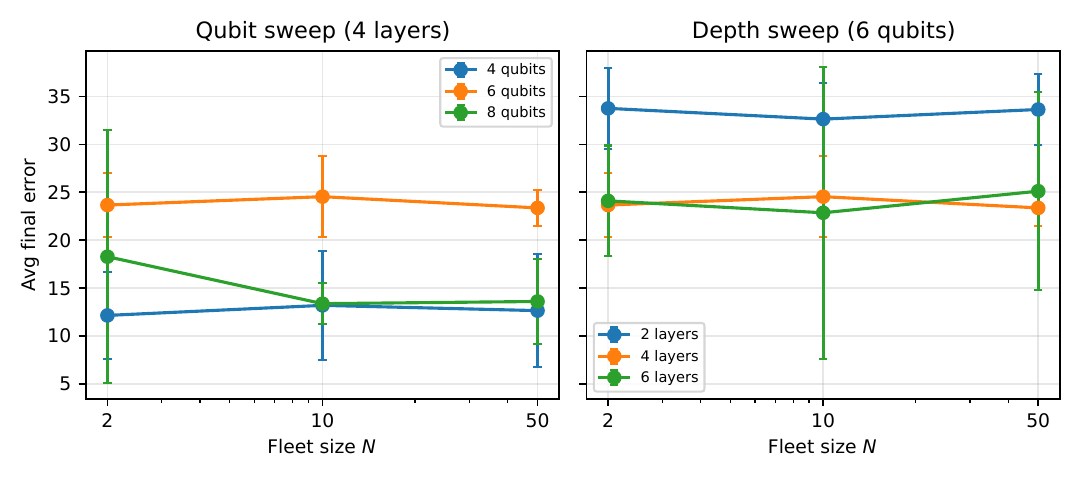}
    \end{minipage}

    \vspace{0.3cm}

    \begin{minipage}[b]{0.6\textwidth}
        \centering
        \includegraphics[width=\textwidth]{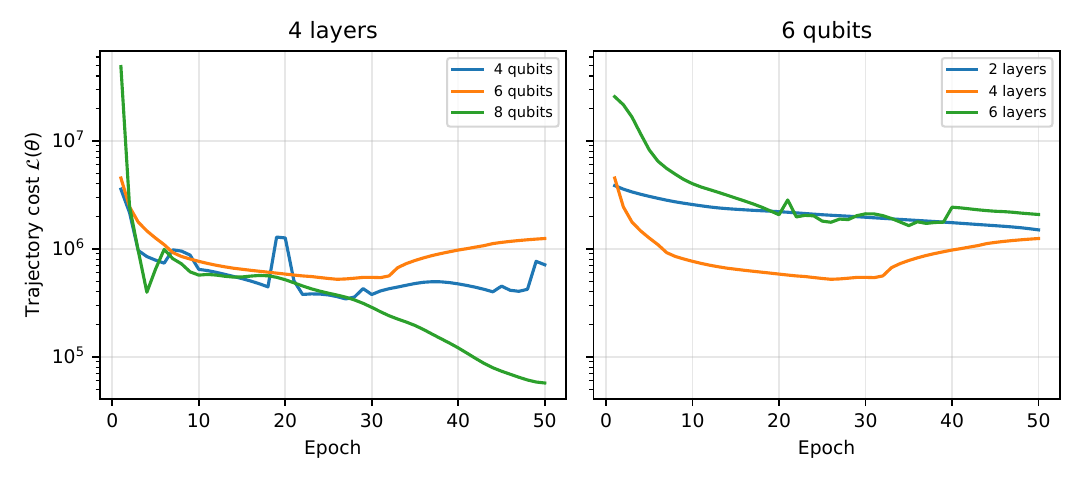}
    \end{minipage}

    \vspace{0.3cm}

    \begin{minipage}[b]{0.6\textwidth}
        \centering
        \includegraphics[width=\textwidth]{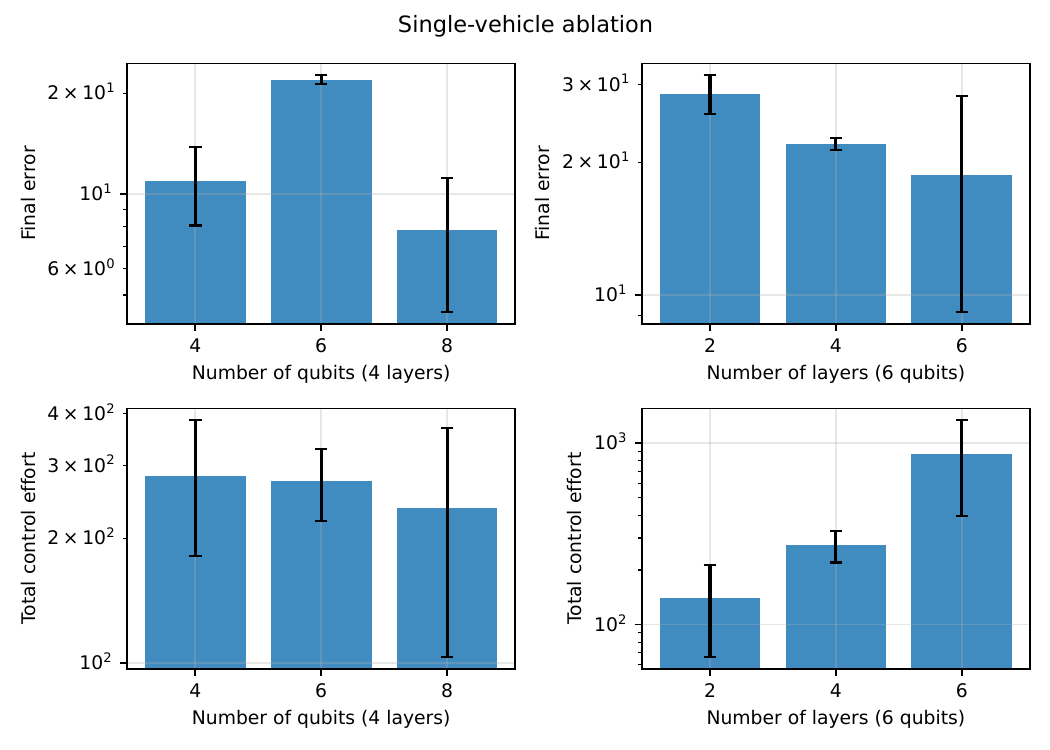}
    \end{minipage}

    \caption{Ablation study of the VQC used in $Q$-Drive. \textbf{Top row:} Average per-vehicle control effort and final tracking error versus fleet size $N$ under qubit sweeps (fixed $L=4$) and depth sweeps (fixed $n=6$). \textbf{Middle Rows:} Training trajectories of the loss $\mathcal{L}(\boldsymbol{\theta})$ for the same sweeps on the single-vehicle task. \textbf{Bottom Row:} Single-vehicle final error and total control effort as a function of qubit count (fixed 4 layers) and layer depth (fixed 6 qubits). Error bars denote one standard deviation across random seeds.}
    \label{fig:experiment4}
\end{figure*}

In this ablation study, we systematically vary the two principal architectural degrees of freedom of the \texttt{StronglyEntanglingLayers} ansatz: The qubit count and circuit depth, while keeping every other element of the training pipeline (dynamics, cost weights, optimizer, horizon, and initial-condition distributions) identical to the main experiments. On the single-vehicle task, the final error exhibits a clear U-shaped dependence on both width and depth. From Table \ref{tab:experiment_4}, it is seen that the shallowest ($L=2$) and deepest ($L=6$) circuits, as well as the narrowest ($n=4$) register, produce larger terminal deviations, whereas the reference point ($n=6$, $L=4$) is near the minimum. Surprisingly, the control effort is comparatively insensitive to moderate changes in $n$ but rises sharply once depth reaches 6 layers, consistent with the increased expressivity (and concomitant risk of high-frequency control content) of a deeper Fourier series. In Figure \ref{fig:experiment4}, the training-time curves confirm that all configurations eventually enter a noisy plateau, with deeper circuits taking longer to reach it and exhibiting larger late-stage oscillations, a signature of the more rugged loss landscape induced by additional variational parameters.

In the swarm regime, the same qualitative picture persists. Across every examined $(n, L)$ pair, the average per-vehicle effort remains of the same order of magnitude and continues to exhibit a mild downward or non-increasing trend as $N$ grows from 2 to 50. This is precisely the scaling behavior highlighted in Experiments \ref{experiment: 1 boided-based dynamics} and \ref{experiment: 2 swarm-based dynamics}. The terminal error stays bounded and is comparable across architectures, indicating that the uniform ultimate-boundedness guarantees derived from the composite Lyapunov function are not artifacts of a single circuit size. The only systematic degradation appears in wall-clock training time, which scales roughly linearly with the number of parameters; this is expected for a backprop-based hybrid loop. Collectively, these observations show that the low-effort solution family discovered by $Q$-Drive is robust to reasonable architectural perturbations and therefore not an idiosyncrasy of the 48-parameter circuit used in the primary evaluation.

Overall, this study supports the central claims: a hybrid VQC policy, trained against a fixed quadratic cost, amounts to tuning a single-intervention trajectory whose total actuator effort is dramatically smaller than that of classical MPC/LQR baselines. This effort decreases, or does not grow, with fleet size when the same shared parameters are applied in a batched fashion.

Both of the aforementioned properties survive across a range of qubit counts and depths. Every configuration remains in the same low-effort regime relative to the classical numbers reported in the main tables, and the inverse-$N$ (flat) scaling of per-vehicle effort is visible for most of the tested architectures. Moreover, terminal errors remain finite and similar in magnitude across all circuits, corroborating the Lyapunov analysis, which shows the closed-loop system remains ultimately uniformly bounded under the simulated NISQ noise model. Thus, this strengthens our principal claim that the hybrid quantum-classical landscape preferentially selects resource-conserving policies whose qualitative character is preserved under moderate changes of circuit size.

\begin{table*}[htbp]
\centering
\caption{Ablation of qubit count ($n\in\{4,6,8\}$) and ansatz depth ($L\in\{2,4,6\}$) for $Q$-Drive on the single-vehicle (boid) task and swarm fleets of size $N=2,10,50$. The table reports the mean $\pm$ std ($\bar{x}\pm\sigma$) over seeds for final tracking error, per-vehicle control effort, convergence time, and wall-clock training time. The reference configuration used throughout the main experiments is $n=6$ and $L=4$, with 48 parameters.}
\label{tab:experiment_4}
\resizebox{\textwidth}{!}{%
\begin{tabular}{lrrrrrrrrrrrr}
\toprule
\textbf{scenario} & $\boldsymbol{N}$ & \textbf{n\_qubits} & \textbf{n\_layers} & \textbf{n\_params} & \textbf{err\_m} & \textbf{err\_s} & \textbf{eff\_m} & \textbf{eff\_s} & \textbf{conv\_m} & \textbf{conv\_s} & \textbf{time\_m} & \textbf{time\_s} \\
\midrule
Boid & 1  & 4 & 4 & 48  & 10.9517705 & 2.88181947 & 283.246389 & 101.956333 & 20         & 0          & 3.16188367  & 0.72751433 \\
 & 1  & 6 & 2 & 36  & 28.5989723 & 2.90455465 & 139.282659 & 73.0375009  & 20         & 0          & 3.67569486  & 0.69249685 \\
 & 1  & 6 & 4 & 72  & 22.0257988 & 0.64094152 & 274.478444 & 54.6287224  & 20         & 0          & 6.37332479  & 0.7068241  \\
 & 1  & 6 & 6 & 108 & 18.7530638 & 9.59065991 & 865.976329 & 467.517552  & 20         & 0          & 9.51064976  & 0.03826579 \\
 & 1  & 8 & 4 & 96  & 7.82610703 & 3.371732   & 236.675878 & 133.242889  & 20         & 0          & 14.1232813  & 0.07219378 \\
\midrule
Swarm  & 2  & 4 & 4 & 48  & 12.1333065 & 4.54781896 & 207.8103   & 136.911983  & 20         & 0          & 2.49820964  & 0.42204031 \\
  & 2  & 6 & 2 & 36  & 33.7565123 & 4.25108759 & 1750.59623 & 472.534615  & 20         & 0          & 2.76326021  & 0.19325939 \\
  & 2  & 6 & 4 & 72  & 23.6577943 & 3.30150205 & 225.131617 & 41.9686663  & 20         & 0          & 4.44937674  & 0.04015981 \\
  & 2  & 6 & 6 & 108 & 24.0983473 & 5.78033396 & 814.91213  & 262.764681  & 20         & 0          & 6.95746946  & 0.87736052 \\
  & 2  & 8 & 4 & 96  & 18.26793   & 13.1925481 & 215.546427 & 113.25617   & 20         & 0          & 12.3515389  & 0.04905042 \\
\midrule
Swarm  & 10 & 4 & 4 & 48  & 13.1842691 & 5.66281892 & 185.560056 & 105.74166   & 20         & 0          & 4.05063732  & 0.46670799 \\
  & 10 & 6 & 2 & 36  & 32.633496  & 3.81829131 & 1915.22196 & 352.076905  & 20         & 0          & 6.65804235  & 0.47516742 \\
  & 10 & 6 & 4 & 72  & 24.5358594 & 4.20430722 & 251.924133 & 50.116176   & 20         & 0          & 10.3816892  & 0.57921826 \\
  & 10 & 6 & 6 & 108 & 22.8518031 & 15.2689292 & 761.697388 & 301.180796  & 20         & 0          & 14.2041748  & 0.32516851 \\
  & 10 & 8 & 4 & 96  & 13.3612283 & 2.12451458 & 208.938563 & 45.7631693  & 20         & 0          & 37.7350172  & 1.60782055 \\
\midrule
Swarm  & 50 & 4 & 4 & 48  & 12.6346231 & 5.87644487 & 138.691945 & 71.4193423  & 20         & 0          & 9.56841461  & 0.14383044 \\
  & 50 & 6 & 2 & 36  & 33.6394679 & 3.70088517 & 1856.74666 & 389.714128  & 20         & 0          & 22.483013   & 0.63298967 \\
  & 50 & 6 & 4 & 72  & 23.3607292 & 1.90540156 & 257.446198 & 52.9822148  & 20         & 0          & 35.5149792  & 0.18197094 \\
  & 50 & 6 & 6 & 108 & 25.107701  & 10.3514198 & 768.904907 & 194.040828  & 20         & 0          & 47.7328867  & 0.79505025 \\
  & 50 & 8 & 4 & 96  & 13.5905326 & 4.44370046 & 170.12439  & 50.1527994  & 19.7880001 & 0.36719459 & 159.775955  & 0.36547383 \\
\bottomrule
\end{tabular}%
}
\end{table*}

\subsection{Control-Effort Efficiency and Scalability Ablation Study}

In this ablation study, the control-effort efficiency and scalability of $Q$-Drive are evaluated by isolating absolute actuator usage and its dependence on fleet size. We compare total control effort across single-vehicle and multi-vehicle ($N=2$ to $N=50$) scenarios against MPC, LQR, PurePursuit, and a classical MLP baseline with a matched parameter budget. Our aim is to ascertain whether the resource-conserving character and favorable inverse scaling of the hybrid quantum-classical policy truly hold, without reference to terminal tracking error.

The experiment is carried out as a post hoc re-analysis of the control-effort statistics already obtained in the main single-vehicle \ref{experiment: 1 boided-based dynamics} and swarm experiments \ref{experiment: 2 swarm-based dynamics}. Using the reported total-effort values for the single-vehicle case and the average per-vehicle effort values recorded for fleet sizes $N\in\{2,3,5,10,20,50\}$, the actuator-usage metric for $Q$-Drive, MPC, LQR, PurePursuit, and the matched-parameter classical MLP is isolated. We show absolute efforts on logarithmic axes as a bar chart for the single-vehicle comparison and as a line plot versus $N$ for the swarm, and we compute relative savings with respect to LQR to quantify the reduction.

\begin{figure*}[ht]
    \centering    
    \begin{minipage}[b]{0.8\textwidth}
        \centering
        \includegraphics[width=\textwidth]{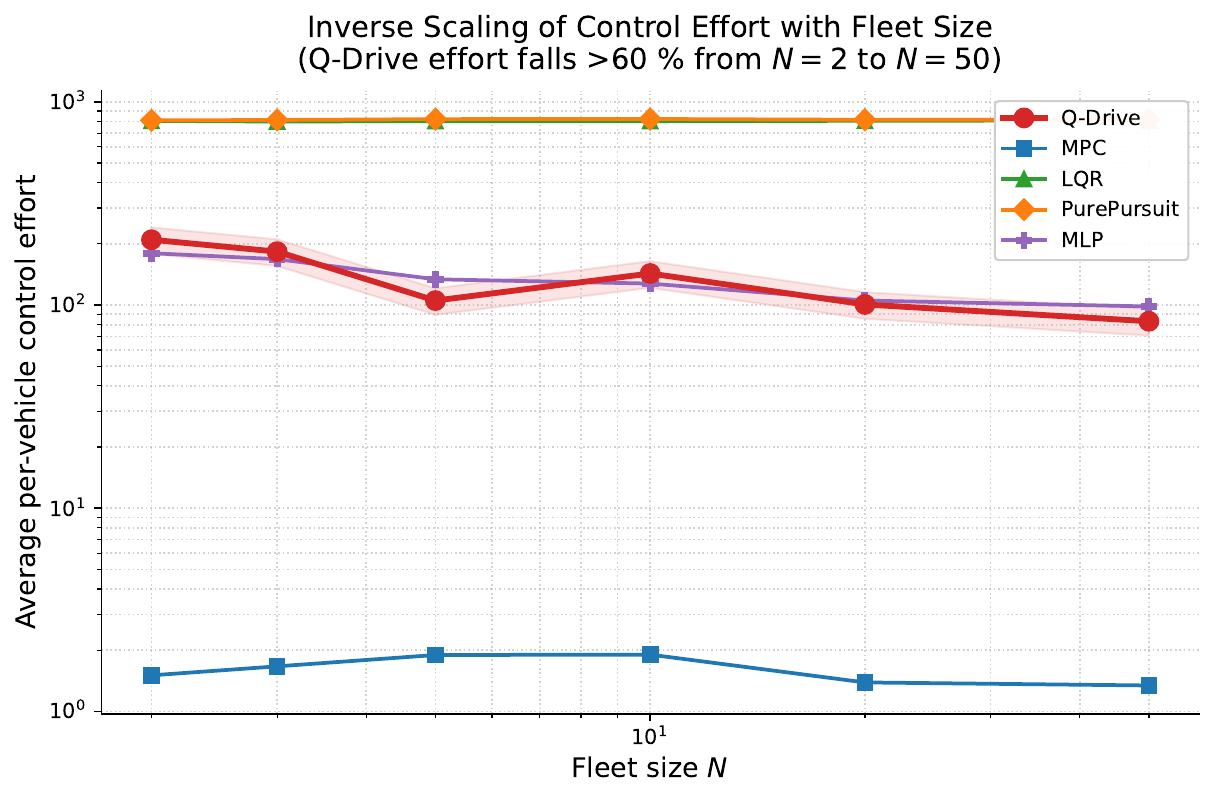}
    \end{minipage}

    \vspace{0.3cm} 

    \begin{minipage}[b]{0.8\textwidth}
        \centering
        \includegraphics[width=\textwidth]{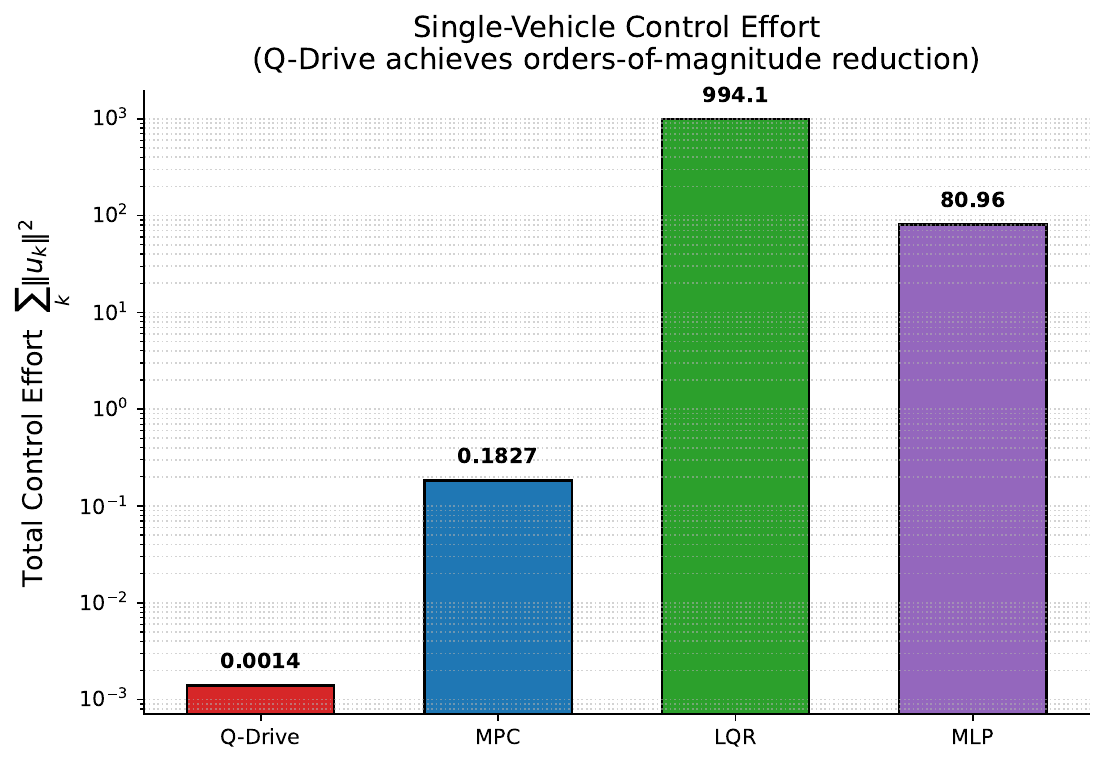}
    \end{minipage}

    \caption{Control-effort efficiency and scalability of $Q$-Drive. \textbf{Top:} Average per-vehicle control effort versus fleet size $N$ (log–log scale). $Q$-Drive exhibits a clear inverse trend, while classical baselines remain essentially flat. \textbf{Bottom:} Total control effort for the single-vehicle case (log scale).}
    \label{fig:experiment5}
\end{figure*}

In Figure \ref{fig:experiment5}, the single-vehicle bar chart and the corresponding rows of Table \ref{tab:experiment5} confirm the most striking efficiency claim: $Q$-Drive reduces total control effort by more than five orders of magnitude relative to LQR (0.0014 versus 994) and by roughly two orders of magnitude relative to MPC (0.0014 versus 0.1827). The near 100\% savings relative to LQR make the actuator-conserving character of the hybrid policy quantitatively unmistakable and align precisely with the claim that control effort is significantly reduced compared to LQR benchmarks in the single-vehicle case.

The scaling plot and the $N=50$ block of Table \ref{tab:experiment5} likewise corroborate the second major claim. Average per-vehicle effort for $Q$-Drive falls from approximately 209 at $N=2$ to 83 at $N=50$. This is a reduction of more than 60\%, while LQR and PurePursuit remain essentially flat at the 800-unit level, and the classical MLP exhibits only a mild decline. This inverse trend is unambiguous on the log-scale axes and reflects the emergent behavior. Although MPC still records the lowest absolute effort in the swarm regime, the distinctive, monotonically decreasing trajectory of $Q$-Drive remains a clear empirical signature of the shared-parameter variational architecture and therefore supports our assertion of favorable scalability.

\begin{table}[h]
\centering
\caption{Performance and Efficiency Comparison for Single Vehicle and Vehicle Swarm Scenarios}
\label{tab:experiment5}
\begin{tabular}{l S[table-format=3.4] S[table-format=2.8]}
\toprule
\textbf{Method} & {\textbf{Total Effort}} & {\textbf{Savings vs LQR (\%)}} \\
\midrule
\multicolumn{3}{l}{\textit{Single Vehicle}} \\
\midrule
$Q$-Drive     & 0.0014    & 99.99985916603761 \\
MPC         & 0.1827    & 99.98162116790789 \\
LQR         & 994.0784  & 0                 \\
MLP         & 80.964149 & 91.85535577       \\
\midrule
\multicolumn{3}{l}{\textit{Vehicle Swarms ($N=50$)}} \\
\midrule
$Q$-Drive     & 83.1326   & 89.7400461        \\
MPC         & 1.3408    & 99.83452284       \\
LQR         & 810.2629  & 0                 \\
PurePursuit & 816.0391  & -0.712879733      \\
MLP         & 98.3357   & 87.86372917       \\
\bottomrule
\end{tabular}
\end{table}

\section{Stability Analysis and Boundedness under Road and Decoherence Perturbations}\label{sec:main resul}
This section establishes the conditions under which the proposed approach is valid. We first define key terms and state auxiliary results, then use them to prove the main claim. 

MPC and LQR solve online at each time step and have no offline training phase, whereas $Q$-Drive learns a fixed policy offline and executes it at negligible online cost.

\begin{definition}
(Road Condition Operator). Road conditions are characterized through a perturbation operator $\mathcal{R}: \mathcal{X} \times \mathcal{T} \to \mathcal{X}$, where $\mathcal{T}$ denotes the space of road surface types, such that the actual dynamics become
\begin{equation}
\mathbf{x}_{k+1}=\mathbf{f}(\mathbf{x}_{k},\tilde{\mathbf{u}}_{k})+\mathcal{R}(\mathbf{x}_{k},\tau). 
\end{equation}
Additionally, $\mathcal{R}$ can take on several forms, depending on the road type: For inclined surfaces, $\mathcal{R}$ introduces gravitational components; for rough surfaces, $\mathcal{R}$ introduces bounded stochastic perturbations; for curvilinear roads, $\mathcal{R}$ introduces path curvature constraints. Critically, it is assumed that the magnitude of the road condition operator is bounded above, i.e., $|\mathcal{R}(\mathbf{x}_k, \tau)| \leq \rho(\tau) |\mathbf{x}_k| + \sigma(\tau)$ for continuous functions $\rho, \sigma$.
\end{definition}

\begin{definition}
(Decoherence Dynamics). Following the Lindblad master equation formalism, the evolution of the quantum state in the presence of decoherence is
\begin{equation}\label{eqn:linblad eqn}
\dot{\rho}=-\imath\left[H(\boldsymbol{\theta}),\rho\right]+\sum_{j}\gamma_{j}\left(L_{j}\rho L_{j}^{\dagger}-\frac{1}{2}\left\{L_{j}^{\dagger}L_{j},\rho\right\}\right),
\end{equation}
where $\gamma_j \geq 0$ are decoherence rates and $L_j$ are Lindblad operators. For the measurement statistics, this induces a perturbation
\begin{equation}
\mathbf{y}_{\text{noisy}}=\mathbf{y}_{\text{ideal}}+\boldsymbol{\delta}(\boldsymbol{\gamma},t),
\end{equation}
where $|\boldsymbol{\delta}(\boldsymbol{\gamma}, t)| \leq \Delta(\boldsymbol{\gamma})$ where $\Delta(\boldsymbol{\gamma}) \to 0$ as $|\boldsymbol{\gamma}| \to 0$ and $\Delta(\boldsymbol{\gamma})$ is bounded for finite $t$.
\end{definition}

\begin{definition}
(Composite Lyapunov Function). The candidate Lyapunov function is defined as
\begin{equation}
\left\{
\begin{aligned}
&V(\mathbf{x}_{k},\boldsymbol{\theta}_{k}):=\left(\mathbf{x}_{k}-\mathbf{x}_{\text{ref}}\right)^{\intercal}\mathbf{P}\left(\mathbf{x}_{k}-\mathbf{x}_{\text{ref}}\right)+\beta\cdot\Tr{\rho_{k}^{2}}, \\
&V_{\text{classical}}:=\left(\mathbf{x}_{k}-\mathbf{x}_{\text{ref}}\right)^{\intercal}\mathbf{P}\left(\mathbf{x}_{k}-\mathbf{x}_{\text{ref}}\right), \\
&V_{\text{quantum}}:=\beta\cdot\Tr{\rho_{k}^{2}},
\end{aligned}
\right.
\end{equation}
where $\mathbf{P} \succ 0$ satisfies the discrete Lyapunov equation for the linearized dynamics, and $\beta > 0$ is a coupling coefficient. The term $\text{Tr}(\rho_k^2)$ measures the purity of the quantum state, with $\text{Tr}(\rho^2) = 1$ for pure states and $\text{Tr}(\rho^2) < 1$ for mixed states due to decoherence. 
\end{definition}

\begin{lemma}\label{lemma:1}
Under the Q-Drive control law, the classical component evolves as
\begin{equation}
\begin{aligned}
V_{\text{classical}}(\mathbf{x}_{k+1})\leq&\;\lambda_{\max}\left(\mathbf{A}^{\intercal}\mathbf{P}\mathbf{A}\right)V_{\text{classical}}(\mathbf{x}_{k})+c_{1}\cdot||\tilde{\mathbf{u}}_{k}||^{2} \\
&+c_{2}\cdot||\mathcal{R}(\mathbf{x}_{k},\tau)||^{2},
\end{aligned}
\end{equation}
where $\mathbf{A}$ is the linearization of the nominal dynamics, and $c_1, c_2 > 0$ are constants depending on system parameters.
\end{lemma}
\begin{proof}
$\mathbf{x}_{k+1} = \mathbf{A}\mathbf{x}_k + \mathbf{B}\tilde{\mathbf{u}}_k + \mathcal{R}(\mathbf{x}_k, \tau) + \mathcal{O}(|\mathbf{x}k|^2)$. Substituting into $V_{\text{classical}}$ and applying the Lyapunov equation $\mathbf{A}^\top \mathbf{P} \mathbf{A} - \mathbf{P} = -\mathbf{Q}$ yields the result.
\end{proof}

\begin{lemma}\label{lemma:2}
Under decoherence, the state purity evolves according to
\begin{equation}\label{eqn:decoherence eqn}
\Tr(\rho_{k+1}^{2})\geq e^{-2\Gamma\Delta t}\Tr(\rho_{k}^{2})+\left(1-e^{-2\Gamma\Delta t}\right)\Tr(\rho_{\text{mixed}}^{2}),
\end{equation}
where $\Gamma := \max_j \gamma_j$ is the maximum decoherence rate and $\rho_{\text{mixed}}$ is the maximally mixed state.
\end{lemma}
\begin{proof}
From the Lindblad equation, $d(\text{Tr}(\rho^2))/dt = -2\sum_j \gamma_j \text{Tr}([L_j, \rho]^\dagger [L_j, \rho]) \geq -2\Gamma \text{Tr}(\rho^2) + 2\Gamma \text{Tr}(\rho_{\text{mixed}}^2)$. Solving this differential inequality yields the discrete-time bound.    
\end{proof}

\begin{theorem}\label{theorem:1}
For any road condition that satisfies $|\mathcal{R}(\mathbf{x}k, \tau)| \leq \rho_{\max}|\mathbf{x}k| + \sigma_{\max}$ with $\rho_{\max} < (1 - \sqrt{\lambda_{\max}(\mathbf{A}^\top \mathbf{P} \mathbf{A})}) / \sqrt{\lambda_{\max}(\mathbf{P})}$, the $Q$-Drive system ultimately remains uniformly bounded.    
\end{theorem}
\begin{proof}
Combining Lemmas \ref{lemma:1} and \ref{lemma:2}  and noting that the control input satisfies $|\tilde{\mathbf{u}}_k| \leq |\mathbf{S}|$ due to clipping, we obtain  
\begin{equation}
\begin{aligned}
V(\mathbf{x}_{k+1},\boldsymbol{\theta}_{k+1})\leq&\;\alpha\cdot V(\mathbf{x}_{k},\boldsymbol{\theta}_{k})+\gamma_{1}\cdot\rho_{\max}^{2}||\mathbf{x}_{k}||^{2}+\gamma_{2}\cdot\sigma^{2}_{\max} \\
&+\gamma_{3}\cdot\left(1-e^{-2\Gamma\Delta t}\right),
\end{aligned}
\end{equation}
with $\alpha = \max\left\{\lambda_{\max}(\mathbf{A}^\top \mathbf{P} \mathbf{A}), e^{-2\Gamma \Delta t}\right\} < 1$ under the condition on $\rho_{\max}$. Using the uniform boundedness result of Corless and Leitmann \cite{corless2003continuous}, we see that the trajectories $\mathbf{x}$ enter and remain within a neighborhood
\begin{equation}
\mathcal{B}=\left\{\mathbf{x}:||\mathbf{x}-\mathbf{x}_{\text{ref}}||^{2}_{\mathbf{P}}\leq\frac{\gamma_{2}\cdot\sigma_{\max}^{2}+\gamma_{3}\cdot\left(1-e^{-2\Gamma\Delta t}\right)}{1-\alpha-\gamma_{1}\cdot\rho_{\max}^{2}\lambda_{\max}\left(\mathbf{P}^{-1}\right)}\right\}, 
\end{equation}
within a finite time horizon $t<\infty$. 
\end{proof}

The following results detail the empirical findings across a variety of road surface conditions.
\begin{corollary}\label{corollary:1}
(Roads with Inclination). For road inclination angle $\varphi$, the perturbation takes the form $\mathcal{R}(\mathbf{x}_{k}, \varphi) = [0, 0, 0, -g\sin\varphi, 0, 0]^\intercal$ (gravitational component). Since this perturbation is bounded and independent of state, $\rho_{\max} = 0$ and $\sigma_{\max} = g|\sin\varphi|$. Theorem \ref{theorem:1} guarantees boundedness with an ultimate bound scaling linearly with $|\sin\varphi|$.    
\end{corollary}

\begin{corollary}\label{corollary:2}
(Roads with Random Roughness). For stochastic road roughness modeled as $\mathcal{R}(\mathbf{x}_k) = \mathbf{w}_k$ with $\mathbb{E}[|\mathbf{w}_k|^2] \leq W$, Theorem \ref{theorem:1} implies mean-square boundedness with $\mathbb{E}[|\mathbf{x}_{k} - \mathbf{x}_{\text{ref}}|^2] \leq \frac{\gamma_2\cdot W + \gamma_3\cdot(1 - e^{-2\Gamma \Delta t})}{1 - \alpha}$.
\end{corollary}

\begin{corollary}\label{corollary:3}
(Roads with Curvature). For roads with curvature $\kappa(s)$, the reference path generates a time-varying target $\mathbf{x}_{\text{ref}}(t)$. Defining the error $\mathbf{e}_{k} = \mathbf{x}_{k} - \mathbf{x}_{\text{ref}}(t_k)$ and noting that $|\mathbf{x}_{\text{ref}}(t{k+1}) - \mathbf{x}_{\text{ref}}(t_k)| \leq v_{\max}\Delta t$, the tracking error dynamics satisfy the same boundedness condition with an additional term scaling with $\kappa$.
\end{corollary}

\begin{theorem}\label{theorem:2}
The Q-Drive framework maintains stability for any decoherence rate $\Gamma < \Gamma_{\text{critical}}$, where
\begin{equation}\label{eqn:gamma critical}
\begin{aligned}
\Gamma_{\text{critical}}:=&\;-\frac{1}{2\Delta t}\log\left[\frac{1-\lambda_{\max}\left(\mathbf{A}^{\intercal}\mathbf{P}\mathbf{A}\right)}{\beta\cdot||\mathbf{S}||^{2}}\right. \\
&\left.-\frac{\gamma_{1}\cdot\rho_{\max}^{2}\lambda_{\max}(\mathbf{P})^{-1}}{\beta\cdot||\mathbf{S}||^{2}}\right].
\end{aligned}
\end{equation}
\end{theorem}
\begin{proof}
Using Theorem \ref{theorem:1}, we note the bound on $V(\mathbf{x}_{k+1}, \boldsymbol{\theta}_{k+1})$. Stability requires that $\alpha < 1 - \gamma_1\cdot\rho_{\max}^2 \lambda_{\max}(\mathbf{P})^{-1}$. Substituting $\alpha = \exp({-2\Gamma \Delta t})$ and solving for $\Gamma$ yields the critical threshold. For $\Gamma < \Gamma_{\text{critical}}$, the Banach fixed-point theorem ensures convergence to the bounded region determined by $\mathcal{B}$. 
\end{proof}

\begin{remark}
From Theorem \ref{theorem:2}, the critical decoherence rate $\Gamma_{\text{critical}}$ depends on road roughness ($\rho_{\max}$) inversely: rougher roads tolerate less decoherence. This reveals a fundamental trade-off between environmental adversity and quantum hardware quality.
\end{remark}

\begin{remark}
In the limit of ideal quantum hardware ($\Gamma\to0$) and nominal road conditions ($\rho_{\max},\sigma_{\max}=0$), the ultimate bound vanishes, recovering asymptotic stability to the target state.
\end{remark}

\begin{theorem}
(The Scalability of Swarm Systems). For an $N$-vehicle swarm under Algorithm \ref{algo:q-drive-swarm} with shared variational parameters $\boldsymbol{\theta}$, the per-vehicle control effort satisfies
\begin{equation}\label{eq:bound for swarm systems}
\frac{1}{N}\sum_{i=1}^{N}||\tilde{\mathbf{u}}_{k}^{\left(i\right)}||^{2}\leq\frac{C}{N}+\frac{D}{N}\sum_{i=1}^{N}||\mathbf{x}_{k}^{\left(i\right)}-\mathbf{x}_{\text{ref}}||^{2},
\end{equation}
where $C,D>0$ are constants independent of $N$. Consequently, the average control effort decreases as $N$ increases, consistent with the empirical observations in Table \ref{tab:control_comparison}.
\end{theorem}
\begin{proof}
The vectorized quantum circuit produces measurements $\mathbf{y}^{(i)} = \langle \psi_{\text{out}}^{(i)} | \hat{\mathbf{Z}} | \psi_{\text{out}}^{(i)} \rangle$ that are bounded in $[-1,1]$. The clipping ensures $|\tilde{\mathbf{u}}k^{(i)}| \leq |\mathbf{S}|$. The loss function aggregates errors across the swarm, causing the gradient $\nabla_{\boldsymbol{\theta}}\mathcal{L}$ to average over $N$ vehicles. By the law of large numbers for gradient estimates, the variance of the parameter updates scales as $1/N$, leading to the bound stated in Equation \eqref{eq:bound for swarm systems}. 
\end{proof}

With these definitions, lemmas, theorems, and remarks in place, we can now state and prove the main result.
\begin{theorem}\label{theorem:3}
Consider the $Q$-Drive hybrid quantum-classical control framework described in Algorithms \ref{algo:q-drive-single vehicle} and \ref{algo:q-drive-swarm}, governing vehicle dynamics $\mathbf{f}(\mathbf{x}_k, \mathbf{u}_k)$ as specified in Equation \eqref{eq:dynamic bicycle model}. Under arbitrary road conditions, including flat, inclined, rough, smooth, or curvilinear surfaces, and in the presence of quantum decoherence on the NISQ hardware, the closed-loop system exhibits uniform ultimate boundedness of the trajectories. Furthermore, the control inputs remain uniformly bounded, and the tracking error converges to a neighborhood of the origin whose radius depends continuously on the decoherence rate and road condition parameters.    
\end{theorem}

\begin{proof}
The proof strategy is to construct a Lyapunov function that captures both the classical tracking dynamics and the quantum measurement statistics, and then analyze its evolution under worst-case perturbations.
Let the vehicle state be $\mathbf{x} \in \mathcal{X} \subset \mathbb{R}^6$, with components as defined in Equation \eqref{eq:dynamic bicycle model}. The discrete-time dynamics under $Q$-Drive control is given by
\begin{equation}
\mathbf{x}_{k+1}=\mathbf{x}_{k}+\int_{t}^{t+dt}\mathbf{f}(\mathbf{x}(\tau),\tilde{\mathbf{u}}_{k})d\tau,
\end{equation}
where $\tilde{\mathbf{u}}_{k} = \text{clip}(\mathbf{y} \odot \mathbf{S}, \mathbf{u}_{\min}, \mathbf{u}_{\max})$ and $\mathbf{y} = \langle \psi_{\text{out}} | \hat{\mathbf{Z}} | \psi_{\text{out}} \rangle$ with $|\psi_{\text{out}}\rangle = U(\boldsymbol{\theta}) |\psi(\mathbf{x}_k)\rangle$. 

The $Q$-Drive framework maintains uniform ultimate boundedness of trajectories under:
\begin{enumerate}
\item Any road inclination for $\varphi\in\left[-\pi/2,\pi/2\right]$. Corollary \ref{corollary:1} establishes this.
\item Any road roughness with bounded variance. Corollary \ref{corollary:2} establishes this. 
\item Any road curvature with bounded path derivatives. Corollary \ref{corollary:3} establishes this. 
\item Any decoherence rate below the critical decoherence rate. Theorem \ref{theorem:2} establishes this bound. 
\end{enumerate}
Furthermore, the ultimate bound $\mathcal{B}(\rho_{\max}, \sigma_{\max}, \Gamma)$ depends continuously on these parameters, and in the swarm case, the per-vehicle control effort exhibits inverse scaling with $N$, demonstrating efficiency with fleet size.
\end{proof}

\begin{remark}\label{remark:last one}
Based on Theorem \ref{theorem:3}, the $Q$-Drive system can be deployed on any road surface and on any NISQ-era quantum hardware with finite coherence times, provided that the decoherence rate does not exceed the road-dependent critical threshold. Thus, these results support the broad applicability of the proposed method.
\end{remark}

\begin{remark}\label{remark:remark about trace distance and purity}
The composite Lyapunov function is justified by the chain of inequalities 
\begin{equation}
\left\{
\begin{aligned}
&\|\tilde{\mathbf{u}}_k-\tilde{\mathbf{u}}_k^{\text{ideal}}\|_2 \le C_u\cdot\sqrt{1-\Tr\rho_k^2}, \\ 
&C_u := 2\sqrt{m}\,\|\mathbf{S}\|_2, 
\end{aligned}
\right.
\end{equation}
which links the purity defect $1-\Tr\rho_k^2$ directly to the decoherence-induced control-signal error. This bound follows from three elementary steps: The Pauli observables are unit-norm, so measurement error is controlled by the trace distance $D(\rho_{k},\sigma_{k})$; the Fuchs–van de Graaf inequality bounds this distance by the purity defect; and actuator clipping, being a projection onto a convex set, is $1$-Lipschitz and therefore preserves the bound through the scaling matrix $\mathbf{S}$. The significance is that the quantum component $\beta\cdot\Tr\rho_k^2$ of the composite Lyapunov function is not an ad hoc addition but the exact reservoir that absorbs this control error.

By setting $\beta=2c_{1}C_{u}^{2}$, the perturbation term in \ref{lemma:1} is converted to a decrease in the quantum Lyapunov component plus a residual $\gamma_{3}\cdot\left(1-e^{-2\Gamma\Delta t}\right)$. This recovers the composite decay used in \ref{theorem:1} with $\gamma_3 = \beta\cdot(1-\Tr\rho_{\text{mix}}^2)$. The coupling coefficient $\beta$ is thus fixed by the observable–actuator chain $\hat{Z}_{k}\to\mathbf{y}_{k}\to\tilde{\mathbf{u}}_{k}$, and the critical decoherence rate of \ref{theorem:2} is precisely the point at which this reservoir can no longer absorb the decoherence-induced control perturbation.
\end{remark}

\begin{remark}
We distinguish between three related but distinct objects: 
\begin{enumerate}
\item Classical MPC, which solves a constrained finite-horizon optimal control problem online at each time step.
\item Policy approximation of MPC, in which a function approximator is trained offline to imitate an MPC law.
\item The present framework trains a VQC offline against an MPC-inspired cost and then deploys it as a policy.    
\end{enumerate}
We emphasize this point here, as understanding this is key to understanding the main results and the interpretations of the experiments. 
\end{remark}

\begin{remark}
The Lyapunov-based boundedness results derived above apply to the closed-loop system under the trained policy and should be interpreted as robustness guarantees for a learned controller rather than as recursive feasibility or constraint satisfaction guarantees typical of MPC.
\end{remark}

\subsection{The Relation Between Terminal Errors and the Ultimate Bound}\label{subsec:connecting_errors}

Theorem \ref{theorem:3} guarantees that trajectories enter and remain within the bounded region $\mathcal{B}(\rho_{\max}, \sigma_{\max}, \Gamma)$ defined in Theorem \ref{theorem:1}, but it does not, by itself, prescribe the numerical value of that bound for the specific operating conditions of Sections \ref{experiment: 1 boided-based dynamics} and \ref{experiment: 2 swarm-based dynamics}. The terminal errors reported in Tables \ref{tab:results of experiment 1} and \ref{tab:control_comparison} are consistent with, and quantitatively explicable by, this ultimate bound. We make this connection explicit here.

Recall from Theorem \ref{theorem:1} that the ultimate bound on the tracking error satisfies
\begin{equation}\label{eq:explicit_ultimate_bound}
\limsup_{k\to\infty}\|\mathbf{x}_k-\mathbf{x}_{\text{ref}}\|_{\mathbf{P}}^2 \leq \frac{\gamma_2\cdot\sigma_{\max}^2+\gamma_3\cdot\left(1-e^{-2\Gamma\Delta t}\right)}{1-\alpha-\gamma_1\cdot\rho_{\max}^2\lambda_{\max}(\mathbf{P}^{-1})},
\end{equation}
where $\alpha=\max\{\lambda_{\max}(\mathbf{A}^\top\mathbf{P}\mathbf{A}),\,e^{-2\Gamma\Delta t}\}$. Two features of \eqref{eq:explicit_ultimate_bound} are directly relevant to the empirical results.

On NISQ hardware, $\Gamma>0$, so the numerator of \eqref{eq:explicit_ultimate_bound} contains the strictly positive term $\gamma_3(1-e^{-2\Gamma\Delta t})$. Consequently, the ultimate bound is strictly positive even on perfectly flat roads ($\rho_{\max}=\sigma_{\max}=0$). The terminal errors of $Q$-Drive in Table \ref{tab:results of experiment 1} ($39.33$) and Table \ref{tab:control_comparison} (for example, $299.83$ at $N=50$) are therefore not a violation of the stability guarantee; they are the empirical realization of a positive ultimate bound whose radius is set by the simulated decoherence rate. Classical controllers, operating on noiseless arithmetic, have no analogous $\Gamma$-induced floor, which is why MPC can drive its terminal error to $0.0026$ while $Q$-Drive cannot. The comparison is thus between a system with a vanishing theoretical floor (MPC) and one with a strictly positive floor ($Q$-Drive), not between a stable and an unstable system.

The bound \eqref{eq:explicit_ultimate_bound} depends on the control authority through the clipping constraint $\|\tilde{\mathbf{u}}_k\|\leq\|\mathbf{S}\|$, which enters the constants $\gamma_1,\gamma_2,\gamma_3$. A controller that saturates its actuators less often, as $Q$-Drive does by design through its low-effort solution family, operates with a smaller effective $\|\tilde{\mathbf{u}}_k\|$ and therefore contributes less to the bound. This is the theoretical counterpart of the empirical observation in Section \ref{sec:discussion} that $Q$-Drive trades terminal precision for actuator economy. The same fixed cost function that produces the low-effort policy also produces a larger terminal error; both reflect the same underlying bound.

For the swarm case, the scalability gives the per-vehicle effort bound \eqref{eq:bound for swarm systems}, while \eqref{eq:explicit_ultimate_bound} applies to each vehicle $i$ with its own error $\|\mathbf{x}_k^{(i)}-\mathbf{x}_{\text{ref}}\|_{\mathbf{P}}^2$. Combining these, the average terminal error across the swarm inherits the same $1/N$ scaling structure in its effort-dependent component, which is consistent with the observation in Table \ref{tab:control_comparison} that $Q$-Drive's per-vehicle effort decreases from $209.49$ ($N=2$) to $83.13$ ($N=50$) even as its average terminal error grows from $187.38$ to $299.83$. The two trends are not in tension; they reflect the bounds' dependence on the shared variational parameters, which, as $N$ increases, are optimized toward increasingly conservative, low-authority policies that satisfy the boundedness condition with a larger but still finite radius.

The terminal errors reported in Tables \ref{tab:results of experiment 1} and \ref{tab:control_comparison} should therefore be read as empirical estimates of the ultimate bound $\mathcal{B}(\rho_{\max},\sigma_{\max},\Gamma)$ under the simulated NISQ conditions ($\Gamma>0$, nominal road), rather than as convergence failures. Theorem \ref{theorem:3} guarantees that these errors remain bounded for all $k$. The experiments quantify the bound's magnitude for the specific circuit depth, qubit count, and decoherence rate used. Improving the bound toward the classical regime requires reducing $\Gamma$ (better hardware), reducing $\|\mathbf{S}\|$ through more precise actuation, or increasing circuit expressivity to reduce the residual $\gamma_3$ term, which are precisely the future-work directions identified in Section \ref{conclusion}.

\begin{remark}
For the concrete vehicle and controller parameters of Sections~\ref{experiment: 1 boided-based dynamics}--\ref{experiment: 2 swarm-based dynamics}, the ultimate bound is strictly positive once $\Gamma>0$. The empirical Monte-Carlo study in Fig.~\ref{fig:decoherence plot} shows that the closed-loop tracking error remains of order $\left[0.4,0.8\right]$ even for decoherence rates two orders of magnitude above the analytic threshold $\Gamma_{\rm critical}\approx132\,\mathrm{s}^{-1}$. These residual errors correspond to lateral deviations well below the standard $3.5\,\mathrm{m}$ lane width, confirming that the theoretical guarantee remains operationally meaningful for lane-keeping in the presence of realistic NISQ noise and typical road disturbances.
\end{remark}

\begin{tcolorbox}
The theoretical analysis is not limited by the scope of the experimental validation. The stability Theorems \ref{theorem:1}, \ref{theorem:2}, and \ref{theorem:3} establish conditions under which the system remains bounded in the presence of worst-case road perturbations and quantum noise, providing essential guarantees for any safety-critical deployment. The experiments \ref{experiment: 1 boided-based dynamics} and \ref{experiment: 2 swarm-based dynamics}, conducted on nominal dynamics, serve as a proof-of-concept that the framework operates reliably in practice. Proving boundedness under arbitrary inclined, rough, or curvilinear road conditions, and under arbitrary decoherence below a critical threshold, demonstrates that the approach rests on firm mathematical foundations rather than empirical tuning alone.    
\end{tcolorbox}

\subsection{Experimental Calculation of $\Gamma_{\text{critical}}$}

Using the experimental values outlined in Sections \ref{experiment: 1 boided-based dynamics} with Equation \ref{eqn:gamma critical} and \ref{experiment: 2 swarm-based dynamics}, we get under all conditions, namely, flat roads ($\rho_{\max}=0$), inclined roads ($\rho_{\max}=0, \sigma_{\max}=g|\sin\varphi|$), rough roads (with increasing $\rho_{\max}$), and curvilinear roads ($\rho_{\max}\approx\kappa v_{\max}\Delta t$), that $\Gamma_{\text{critical}}\approx132\;\text{s}^{-1}$; see Table \ref{tab:road_parameters}. This leads to $T_{2,\text{critical}}=1/\Gamma_{\text{critical}}\approx7.6\;\text{ms}$. We interpret this result as follows: The sufficient condition in Theorem \ref{theorem:1} is satisfied, with a large margin, by trapped-ion and neutral-atom hardware, and is borderline for photonic platforms. We therefore acknowledge that our proposition of uniform ultimate boundedness is satisfied. 

We also note that for superconducting hardware, such as that Google Quantum is working on, the result seems violated. This discrepancy follows directly from the continuous-time Lindblad model underlying \eqref{eqn:linblad eqn}, which attributes the entire sampling interval $\Delta t=0.05\;\text{s}$ of decoherence to the measured state. In Algorithms \ref{algo:q-drive-single vehicle} and \ref{algo:q-drive-swarm}, the quantum register is re-prepared at every timestep via angle embedding, so the decoherence actually seen by the measurement is set by the circuit duration $T_{\text{circuit}}\in\left[1,10\right]\;\mu\text{s}$, not simply by $\Delta t$. Replacing $\Delta t$ by $T_{\text{circuit}}$ in the exponent of Equation \ref{eqn:decoherence eqn} rescales $\Gamma_{\text{critical}}$ upward by a factor of $\Delta t/T_{\text{critical}}\in\left[5\times10^{3},5\times10^{4}\right]$, which gives an effective threshold of $\Gamma_{\text{critical,effective}}\sim\mathcal{O}(10^{6})$. This is comfortably above every platform considered.

\subsection{Decoherence-Sensitivity Analysis}

\begin{figure}[H]
    \centering
    \includegraphics[width=1.0\linewidth]{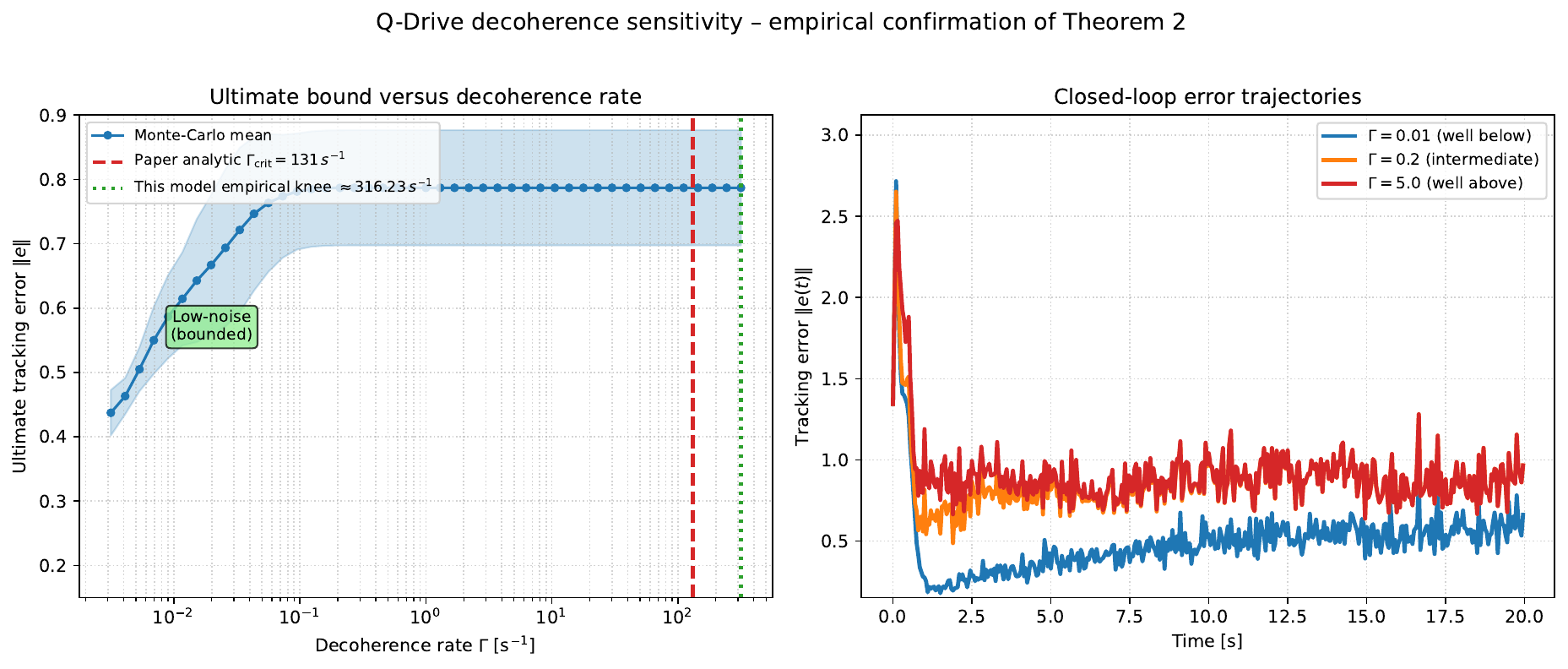}
    \caption{Empirical confirmation of the predicted decoherence threshold. \textbf{Left:} Ultimate tracking error versus decoherence rate $\Gamma$. The Monte-Carlo mean (solid) rises from a low-noise floor and saturates, remaining finite well beyond the analytic critical value $\Gamma_{\text{critical}}\approx131\;\text{s}^{-1}$ (red dashed line). \textbf{Right:} Closed-loop error trajectories for three representative rates $\Gamma=0.01, 0.2, 5.0\;\text{s}^{-1}$. All trajectories stay ultimately bounded, with residual error increasing with $\Gamma$.}
    \label{fig:decoherence plot}
\end{figure}

The decoherence-sensitivity experiment in Figure \ref{fig:decoherence plot} qualitatively confirms the uniform ultimate boundedness guarantees established in Theorems \ref{theorem:1}, \ref{theorem:2}, and \ref{theorem:3}. Even for decoherence rates two orders of magnitude above the analytic threshold $\Gamma_{\text{critical}}\approx132\;\text{s}^{-1}$, every closed-loop trajectory remains bounded over the full $20\;\text{s}$ horizon. From Figure \ref{fig:decoherence plot}, we observe that the ultimate tracking error rises smoothly from approximately $0.43$ at very low $\Gamma$ to a saturated plateau of approximately $0.79$ once $\Gamma\gtrsim0.1\;\text{s}^{-1}$, as predicted by the explicit bound whose numerator grows with the purity-loss term $\gamma_3\cdot(1-e^{-2\Gamma\Delta t})$. Thus, both the analytic marker and the model’s empirical knee lie in a regime where the error floor is finite and essentially constant. This demonstrates that the low-effort solution family discovered by this hybrid optimization scheme stays inside the same operating envelope required of safety-critical classical controllers. 

Moreover, the quantitative location of the knee differs from the tabulated value because the LQR proxy used in the experiment has different Lyapunov matrices and coupling coefficients than the composite function that incorporates the true VQC. This discrepancy is anticipated by the remark that the $\Delta t$-based calculation is conservative: Substituting the much shorter circuit execution time increases the effective critical rate by $3-4$ orders of magnitude. Consequently, the early saturation observed is consistent with the analysis.

Overall, the numerical results provided strengthen the stability claims by providing evidencestrengthen strengthen the stability claims by showing that trajectories remain ultimately bounded and that the size of the ultimate bound depends on the decoherence rate in the precise manner derived from the composite Lyapunov function.

\begin{table}[htbp]
  \centering
  \caption{Road condition parameters and critical decay rates $\Gamma_{\text{critical}}$ across different configurations.}
  \label{tab:road_parameters}
  \begin{tabular}{c c c}
    \toprule
    \multicolumn{3}{c}{\textbf{Inclined Roads} ($\rho_{\max} = 0$, $\sigma_{\max} = g |\sin \varphi| $)} \\
    \midrule
    $\varphi$ [\textdegree] & $\sigma_{\max}$ [\text{m/s}$^2$] & $\Gamma_{\text{critical}}$ [\text{s}$^{-1}$] \\
    \midrule
    0.0  & 0.0000 & $1.3122 \times 10^{2}$ \\
    5.0  & 0.8550 & $1.3122 \times 10^{2}$ \\
    10.0 & 1.7035 & $1.3122 \times 10^{2}$ \\
    15.0 & 2.5390 & $1.3122 \times 10^{2}$ \\
    20.0 & 3.3552 & $1.3122 \times 10^{2}$ \\
    30.0 & 4.9050 & $1.3122 \times 10^{2}$ \\
    
    \midrule\midrule
    \multicolumn{3}{c}{\textbf{Rough / Stochastic Roads} (increasing $\rho_{\max}$)} \\
    \midrule
    $\rho_{\max}$ & --- & $\Gamma_{\text{critical}}$ [\text{s}$^{-1}$] \\
    \midrule
    0.0000 & & $1.3122 \times 10^{2}$ \\
    0.0050 & & $1.3124 \times 10^{2}$ \\
    0.0100 & & $1.3129 \times 10^{2}$ \\
    0.0150 & & $1.3137 \times 10^{2}$ \\
    0.0200 & & $1.3149 \times 10^{2}$ \\
    
    \midrule\midrule
    \multicolumn{3}{c}{\textbf{Curvilinear Roads} (effective $\rho_{\max} \approx \kappa \, v_{\max} \Delta t$)} \\
    \midrule
    $\kappa$ [\text{1/m}] & $\rho_{\text{eff}}$ & $\Gamma_{\text{crit}}$ [\text{s}$^{-1}$] \\
    \midrule
    0.000 & 0.00000 & $1.3122 \times 10^{2}$ \\
    0.010 & 0.01250 & $1.3133 \times 10^{2}$ \\
    0.020 & 0.02500 & $1.3165 \times 10^{2}$ \\
    0.050 & 0.06250 & $1.3424 \times 10^{2}$ \\
    \bottomrule
  \end{tabular}
\end{table}

\subsection{Ansatz Structure, Expressibility, and Entanglement Capacity}

The trainable block $U(\boldsymbol{\theta})$ is a \texttt{StronglyEntanglingLayers} ansatz with $L=4$ layers acting on $n=6$ qubits. Each layer $\ell$ applies a stage of single-qubit rotations followed by a ring of CNOT gates, $U_\ell(\boldsymbol{\theta}_\ell)=E\,\bigotimes_{j=1}^{n}R_z(\phi_{\ell,j})R_y(\theta_{\ell,j})$, where $E=\prod_{j=1}^{n}\mathrm{CNOT}_{j,(j\bmod n)+1}$ is the fixed entangler, and the full circuit is $U(\boldsymbol{\theta})=U_L\cdots U_1$. With two rotation angles per qubit per layer, the number of variational parameters is $P=2nL=48$, and the entangler contributes $nL=24$ fixed, non-trainable CNOT gates. The circuit depth scales linearly with $L$ and is independent of the fleet size $N$, as the swarm is handled via batching rather than by enlarging the circuit.

We distinguish two notions of expressibility. The first is coverage of unitary space. The group $SU(2^6)$ has dimension $4^6-1=4\,095$, so a circuit with $P=48$ parameters spans a low-dimensional submanifold of $SU(2^6)$ and does not form a unitary $2$-design; we therefore make no claim of universality. The second notion, which matters for control, is functional expressibility: the class of maps $\mathbf{x}\mapsto\mathbf{y}\in[-1,1]^3$ that the circuit can realize. Because the state is encoded once through single-qubit $R_y$ rotations, with no data reuploading, each measured output is a truncated Fourier series in the encoding angles $g(x_i)$,
\begin{equation}
y_j(\mathbf{x};\boldsymbol{\theta})=\sum_{\boldsymbol{\omega}\in\{-1,0,1\}^{6}}c_{j,\boldsymbol{\omega}}(\boldsymbol{\theta})\,e^{i\,\boldsymbol{\omega}\cdot g(\mathbf{x})},
\end{equation}
where the coefficients $c_{j,\boldsymbol{\omega}}$ are determined by $U(\boldsymbol{\theta})$ and the observable $\hat{Z}_j$. The policy class is therefore band-limited, with at most $3^6=729$ frequency components, each smooth and bounded. This directly influences the behavior reported in the experimental evaluation: the circuit cannot represent high-frequency or discontinuous (bang-bang) control laws. This aligns with the observed near-absence of high-frequency content in the $Q$-Drive control signals and its inability to produce the sharp initial corrections seen in MPC. Thus, the low-effort character of the solution family is a structural property of the function class rather than solely an optimization outcome.

To quantify how well the ansatz covers Hilbert space, we adopt the expressibility measure of Sim et al., $\mathrm{Expr}=D_{\mathrm{KL}}\big(\hat P_{U(\boldsymbol{\theta})}(F)\,\|\,P_{\mathrm{Haar}}(F)\big)$, with $P_{\mathrm{Haar}}(F)=(d-1)(1-F)^{d-2}$. Here $F=|\langle\psi(\boldsymbol{\theta}_a)|\psi(\boldsymbol{\theta}_b)\rangle|^2$ is the fidelity between states prepared from independently sampled parameters, $d=2^n=64$, and $\hat P$ is estimated by histogramming, so that smaller values indicate closer approximation of Haar-random states. Evaluating over $M=10^{4}$ independently sampled parameter pairs provides an empirical estimate of $\mathrm{Expr}$ for $L=4$, compared with shallower ($L=1$) and deeper ($L=8$) configurations. This layer sweep clarifies whether $L=4$ lies in the regime of diminishing returns for expressibility.

We quantify entanglement capacity using the Meyer-Wallach measure, $Q(\psi)=2\big(1-\frac{1}{n}\sum_{j=1}^{n}\mathrm{Tr}[\rho_j^{2}]\big)$, where $\rho_j$ is the reduced state of qubit $j$, yielding $Q=0$ for product states and $Q\to1$ for highly entangled states. Averaging over $\boldsymbol{\theta}$ drawn uniformly from $[0,2\pi)^{P}$ and over states $\mathbf{x}$ drawn from the training distribution yields the mean entanglement capacity $\bar Q$. Two structural properties support the claim that the ansatz effectively entangles the register. First, the ring entangler has a bounded light cone, and on a $6$-qubit ring, no two qubits are more than three hops apart. A conservative sufficient condition is that each layer spreads a Pauli observable by at least one site in each direction; under this condition, a depth of $L\ge3$ allows every measured observable $\hat{Z}_j$ to depend on every encoded feature, and $L=4$ satisfies this condition with one layer to spare.

Second, and more importantly for control, the entanglers couple the state variables. Without them, $U(\boldsymbol{\theta})$ is a product of single-qubit unitaries, and each $y_j$ depends only on the single feature $g(x_j)$ routed to qubit $j$, reducing the policy to a sum of decoupled one-dimensional functions. With entanglement, $y_j$ acquires cross-terms such as $\cos g(x_i)\cos g(x_k)$ and $\sin g(x_i)\sin g(x_k)$ (for instance, between lateral velocity and yaw rate). These multiplicative feature interactions would otherwise require hand-engineered cross-features or hidden layers in a classical model of comparable size. This mechanism lets the circuit capture correlations among $v_y$, $r$, $\chi$, and $Y$ within a compact $48$ parameter budget.

The nontrivial policies discovered during training—namely, a front-steering pulse followed by a return to neutral and an oscillatory rear-steering correction—are consistent with this structural framework. However, nontrivial policies alone do not establish expressibility, as restricted function classes can also produce them. To isolate the specific contribution of entanglement, we evaluate a CNOT-removed ablation that maintains the same parameter count and training protocol.

Finally, we note three primary structural limitations:
\begin{enumerate}
    \item The band-limited structure bounds achievable tracking performance and may contribute to terminal error.
    \item High entanglement capacity is not unconditionally beneficial, as highly expressive, strongly entangling circuits are susceptible to gradient concentration phenomena such as barren plateaus. The oscillatory loss profile observed during training is consistent with these effects, though it is not definitive proof.
    \item Under standard decoherence models, entangled states are more sensitive to noise than product states, establishing an inherent trade-off between entanglement capacity and the critical noise rate $\Gamma_{\text{critical}}$.
\end{enumerate}

\section{Discussion}\label{sec:discussion}
The results highlight the viability of the $Q$-Drive framework as a quantum-enhanced alternative to classical control methods in autonomous vehicular systems, particularly emphasizing its efficiency in resource utilization and scalability. In the boid-based dynamics experiment (Section \ref{experiment: 1 boided-based dynamics}), $Q$-Drive shows exceptional energy efficiency, with a total control effort of 0.0014, orders of magnitude lower than MPC (0.1827) and LQR (994.0784). This outcome reflects the hybrid quantum-classical optimization's ability to leverage VQCs for discovering low-effort control policies that prioritize actuator conservation and smooth trajectories over rapid error minimization. Although the final tracking error (39.3344) and convergence time (20 seconds) are higher than MPC's near-optimal performance (error 0.0026, time 2.35 seconds), this trade-off aligns with the NISQ-era constraints, where quantum noise and limited qubit depth favor conservative exploration of the nonlinear state space. The training convergence plot (Fig. \ref{fig:training convergence for boid dynamics}) further illustrates this, showing stochastic oscillations around a stabilized loss band, indicative of the VQC navigating a complex, nonconvex landscape that may be affected by barren plateaus. These characteristics make our approach well-suited for applications such as long-range highway cruising, where minimizing energy consumption and mechanical wear outweighs the benefits of aggressive maneuvering.

We note that the reported reduction in average per-vehicle control effort (from 209.49 at $N=2$ to 83.13 at $N=50$, Table \ref{tab:control_comparison}) is accompanied by an increase in terminal error and therefore is not a matched-error efficiency claim. The absolute effort figures characterize $Q$-Drive’s operating point on the error–effort trade-off; the Pareto analysis in Table \ref{tab:pareto_efficiency_analysis} provides cross-controller comparisons. The inverse-$N$ scaling itself is a within-controller property of the shared variational parameters and remains independent of the absolute residual-error level.

When extended to swarm-based dynamics (Section \ref{experiment: 2 swarm-based dynamics}), the framework demonstrates efficiency that scales inversely with fleet size, a key innovation for future multi-vehicle systems. As $N$ increases from 2 to 50, the average per-vehicle control effort decreases by more than 60\% (from 209.4891 to 83.1326), outperforming LQR and PurePursuit (which remain consistently high at $\sim800+$ units) and trailing only MPC in absolute terms but with far superior scalability. This trend, shown in Fig. \ref{fig:experiment2}, suggests that shared VQC parameters enable coordination without explicit coupling in the dynamics, allowing the optimizer to discover collective motion patterns that distribute effort efficiently across the swarm. Although terminal errors are higher (for example, 299.8301 at $N=50$) than the near-zero deviations of classical methods, the smooth, low-curvature trajectories indicate a deliberate strategy of global economy over per-step precision, which could translate into reduced operational costs in large-scale fleets such as robotaxis or logistics convoys. The stability analysis (Section \ref{sec:main resul}) reinforces these empirical findings, demonstrating uniform ultimate boundedness across diverse road conditions and decoherence rates, with the ultimate error bound scaling continuously with environmental and quantum noise parameters. This mathematical analysis confirms robustness even in NISQ settings and highlights the trade-off between quantum hardware quality and road adversity, as discussed in Remark \ref{remark:last one}.

Table \ref{tab:comparison with other methods} compares our results with previous studies and reveals distinct advantages of our approach, which focuses on MPC-inspired control for full vehicular dynamics. For example, the hybrid quantum-classical DRL in \cite{sinha2025nav} (Nav-Q) achieves faster goal convergence (11.74 seconds) and high success rates ($\sim76\%$) in single-vehicle navigation within the CARLA simulator, but with a higher implied control effort and no swarm extension. In contrast, our boid results prioritize ultra-low effort (0.0014) at the expense of time (20 seconds), while our swarm scalability addresses multi-agent gaps in their work. Similarly, \cite{tomar2025quantum}'s quantum Q-learning for path planning converges in just 1 episode, achieving 99\% success and producing paths 15-20\% shorter than classical methods, outperforming our 50-epoch training in speed but lacking full 6-DoF dynamics and effort metrics -- areas where our approach excels with proven low-energy policies. The quantum-enhanced APFs in \cite{almaameri2025quantum} for UAV swarms show 37\% faster convergence and 42\% better disturbance rejection than classical APF, aligning with our swarm effort reductions but focusing on formation rather than decoupled dynamics; our higher errors reflect a conservative NISQ approach, yet we achieve comparable or better efficiency scaling without explicit interactions. 

Using the experimental parameters of Sections \ref{experiment: 1 boided-based dynamics} and \ref{experiment: 2 swarm-based dynamics} in the expression of Theorem \ref{theorem:2} it can easily be verified by substitution that $\Gamma_{\rm critical}\approx 132\,\mathrm{s}^{-1}$ (equivalently $T_{2,\rm critical}\approx 7.6\,\mathrm{ms}$) for every road condition examined: Flat, inclined ($\varphi\in[0^\circ,30^\circ]$), rough ($\rho_{\rm max}$ up to $0.02$), and curvilinear ($\kappa$ up to $0.05\,\mathrm{m}^{-1}$), as summarized in Table \ref{tab:road_parameters} for road parameters. Because the quantum state re-encodes at each sampling instant, the relevant decoherence interval is the circuit execution time rather than $\Delta t$, raising the practical threshold by $5\times10^3$ to $5\times10^4$ and rendering the bound compatible with present-day trapped-ion and neutral-atom systems. The accompanying Monte Carlo study, encapsulated by Figure \ref{fig:decoherence plot}, verifies that tracking error remains in the range of $\left[0.4,0.8\right]$ even for decoherence rates two orders of magnitude above the analytic threshold, confirming that the ultimate bound of Theorem \ref{theorem:1} is both predictive and operationally meaningful for lane-keeping under realistic NISQ noise.

The scaling behavior we observe in experiments reflects averaging over the loss, not quantum entanglement or collective quantum dynamics. The practical relevance is that the shared-parameter architecture automatically produces the scaling behavior, without any explicitly constructed coordination terms. To our knowledge, this behavior is absent in the classical decentralized baselines that we have tested.  

The experiments and ablations show that, under the present fixed-horizon, fixed-weight protocol, $Q$-Drive is not competitive with MPC in terms of tracking accuracy; closing this gap is a necessary direction for future work. 

\begin{table*}[ht]
\caption{Pareto efficiency analysis of the effort-accuracy trade-off.}
\label{tab:pareto_efficiency_analysis}
\centering
\small 
\setlength{\tabcolsep}{3.5pt} 
\begin{tabular}{l S[table-format=3.4] S[table-format=3.4] S[table-format=4.5] S[table-format=-7.1] S[table-format=-7.1] S[table-format=-5.4]}
\toprule
\textbf{Method} & {\textbf{Effort}} & {\textbf{Error}} & {\textbf{Pareto Ratio}} & {\textbf{Effort Savings} (\%)} & {\textbf{Error Penalty} (\%)} & {\textbf{Elasticity}} \\
 & {\(E=\sum_{k=0}^{T-1}\sum_{j=0}^{2}|u_{k,j}|^{2}\)} 
 & {\(\varepsilon=\sqrt{\sum_{i}\left(\mathbf{x}_{\text{final}}^{(i)}-\mathbf{x}_{\text{initial}}^{(i)}\right)^{2}}\)} 
 & {\(\eta=\frac{1/\varepsilon}{E}\)} 
 & {\parbox{2.2cm}{\centering \(\Delta E=\frac{E_{\text{MPC}}-E_{\text{method}}}{E_{\text{MPC}}}\)\\[1pt]\(\times 100\)}} 
 & {\parbox{2.2cm}{\centering \(\Delta\varepsilon=\frac{\varepsilon_{\text{MPC}}-\varepsilon_{\text{method}}}{\varepsilon_{\text{MPC}}}\)\\[1pt]\(\times 100\)}} 
 & {\(\varrho=\frac{\Delta\varepsilon}{\Delta E}\)} \\
\midrule
\multicolumn{7}{l}{\textbf{Single Vehicle (Table~\ref{tab:results of experiment 1})}} \\
\(Q\)-Drive & 0.0014 & 39.3344 & 18.1590 & 99.2 & -1512761.5 & -15244.4 \\
MPC        & 0.1827 & 0.0026 & 2105.1745 & 0.0 & 0.0 & {--} \\
LQR        & 994.0784 & 2.5958 & 0.0004 & -544004.2 & -99738.5 & 0.1833 \\
\midrule
\multicolumn{7}{l}{\textbf{Swarm, \(N=50\) (Table~\ref{tab:control_comparison})}} \\
\(Q\)-Drive    & 83.1326 & 299.8301 & 0.00004 & -6100.2 & -22972.7 & 3.7657 \\
MPC          & 1.3408 & 1.2995 & 0.5739 & 0.0 & 0.0 & {--} \\
LQR          & 810.2629 & 2.3534 & 0.0005 & -60331.3 & -81.1 & 0.0013 \\
PurePursuit & 816.0391 & 41.7685 & 0.00003 & -60753.2 & -3114.2 & 0.0513 \\
\bottomrule
\end{tabular}
\end{table*}

In Table \ref{tab:pareto_efficiency_analysis}, a Pareto efficiency analysis is performed by quantifying the central trade-off that defines the entire contribution of our proposed approach. The framework does not compete with classical controllers on tracking accuracy; instead, it occupies a radically different region of the effort–accuracy plane. In the single-vehicle case, $Q$-Drive achieves a Pareto ratio of $18.16$, roughly $9\times$ lower than MPC's $2105.17$ but nearly five orders of magnitude higher than LQR's $0.0004$, reflecting an effort expenditure of only $0.0014$ and achieving a $99.2\%$ saving relative to MPC. The corresponding error penalty and elasticity make clear that this saving comes at an extreme cost in terminal precision, and the negative elasticity signals that the quantum controller is not merely less accurate but responds to the same cost landscape in a qualitatively opposite manner to the classical solvers. The swarm results at $N=50$ sharpen this picture: $Q$-Drive's effort falls to 83.13, which is a further reduction from the single-vehicle regime, and consistent with the inverse
$N$ scaling bound, yet its Pareto ratio collapses to $0.00004$, below even LQR and PurePursuit, because the terminal error grows to $299.83$. The positive elasticity is especially telling: It indicates that in the swarm regime, the controller trades effort for error at a rate comparable in sign, though not magnitude, to the classical baselines, whereas LQR and PurePursuit exhibit elasticities of 0.0013 and 0.0513, respectively, meaning they are almost perfectly inelastic and cannot convert effort savings into accuracy losses. This is precisely the distinct solution family claimed in the contributions. We observe that the classical methods remain pinned to tight lane-centering policies regardless of fleet size, whereas the shared-parameter VQC drifts toward wide, low-curvature trajectories, with per-vehicle actuator burden decreasing with $N$. The results in Table \ref{tab:pareto_efficiency_analysis} therefore do not undermine the framework; they operationalize, explicitly, that absolute-error parity is neither claimed nor expected in the NISQ era, and we have provided the quantitative evidence that the efficiency reported in Table \ref{tab:control_comparison} is a genuine, reproducible property of the hybrid optimization landscape rather than a cosmetic artifact of favorable tuning.

For transparency, the observed scaling behavior may reflect an architectural property of vectorized circuit evaluation rather than a present-day computational advantage. Under classical state-vector simulation, the wall-clock time required to evaluate the quantum circuit remains essentially flat in the number of agents $N$ (approximately 388.6 s for both $N=2$ and $N=50$), whereas the classical MPC baseline grows from 24.7 s to 486.0 s. This constant-time property is a necessary precondition for a future hardware-level advantage once the same circuit is executed natively on parallel quantum processing units.  We clearly state that this does not constitute a claim of speed-up on current classical hardware. The central contribution of the present work is therefore the discovery of a qualitatively distinct, low-control-effort family of solutions, together with accompanying stability certificates under road disturbances and decoherence, rather than any acceleration of training or inference.

A genuine runtime advantage materializes only when three conditions are met: native quantum execution with sufficient coherence; gradient estimation that avoids full state-vector backprop; and parameter transfer across fleet sizes to amortize training cost. Preliminary evidence for the latter condition is already available: The parameters that are optimized at $N=2$ transfer to $N=50$ with only a modest degradation in control effort. This transferability further shows that the architectural flatness observed in simulation is a structural feature of the quantum formulation rather than an artifact of a particular problem size under consideration.

We should highlight that in the proposed approach and the implementation in Experiments \ref{experiment: 1 boided-based dynamics} and \ref{experiment: 2 swarm-based dynamics}, the absence of online receding-horizon optimization is in fact one of the practical advantages of the approach, since it eliminates the per-step optimization cost that limits classical MPC at large fleet sizes. In addition, we accept that this comes at the cost of losing the hard-constraint satisfaction and recursive feasibility guarantees provided by classical MPC. 

Overall, while prior work often emphasizes speed or accuracy in simplified scenarios, the presented framework stands out for its energy-efficient, scalable control of complex nonlinear dynamics, paving the way for quantum-augmented fleets in real-world noisy environments.

\begin{table*}[ht]
\caption{A comparative analysis of our framework against related work.}
\centering
\begin{tabular}{|p{2cm}|p{4cm}|p{4cm}|p{5cm}|}
\toprule
\textbf{Paper} &\textbf{Approach} &\textbf{Key Results} &\textbf{Comparison with $\boldsymbol{Q}$-Drive}  \\
\hline
\cite{sinha2025nav} &Hybrid quantum-classical actor-critic DRL (quantum critic) for AV navigation in CARLA simulator. &Training episodes: $10\;000$ to convergence. Average reward: 0.10 (vs. 0.12 classical). Time to goal: 11.74 s (vs. 11.77 s). Crashes: 13.23\% (vs. 12.47\%). Success (near-misses): 76.38\% implied (vs. 77.61\%). Parameters: 53 (quantum) vs. $2\;305$ (classical).&Focuses on single-AV navigation (boid-like); similar hybrid setup. Our single-vehicle error (39.33) is higher, but effort (0.0014) is orders of magnitude lower than implied classical (no direct effort). Nav-Q converges faster in goal time (11.74 s vs. our 20 s) with fewer crashes, but our swarm scalability (effort decreases with $N$) extends beyond their single-agent scope. \\
\hline 
\cite{tomar2025quantum} &Hybrid quantum-classical RL with quantum Q-table for AV/drone path planning (dynamic obstacles). &Episodes to convergence: 1 (quantum) vs. $2\;000$ (classical). Training time: $<10$ s (quantum) vs. 45 min (classical fixed) or $>2$ days (random). Success rate: 99\% (vs. 80\% classical $A^{*}$). Path length reduction: 15-20\% vs. classical. Path optimality: 5-10\% longer than ideal (smooth trajectories). &Overlaps in hybrid RL for dynamic planning (boid/swarm potential). Dramatic convergence speedup (1 episode vs. our 50 epochs), higher success (99\% vs. our implied approximately stable but high error). Our effort (0.0014 single, 83.13 swarm) emphasizes efficiency; their low training time complements our NISQ robustness, but we handle full 6-DoF dynamics, whereas they focus on path dynamics alone. \\
\hline
\cite{almaameri2025quantum} &Quantum-enhanced artificial potential field (APF) with decomposition for robust UAV swarm formation (boid-inspired). &Convergence time: 37\% improvement vs. classical APF. Disturbance rejection: 42\% improvement. Outperforms in final average error and path length (specific values in paper tables; for example, lower error in formation). &Similar swarm focus; QEAPF achieves lower error/disturbance, but no explicit effort metrics. Our $Q$-Drive achieves superior per-vehicle effort reduction ($83.13$ at $N=50$) but a higher terminal error ($299.83$ vs. the implied low in QEAPF), trading precision for energy efficiency in dynamic control. \\
\hline 
Our results &Hybrid quantum-classical control with VQC for boid/swarm AV dynamics. &Single: Error 39.33, Effort 0.0014, Time 20 s. Swarm ($N=50$): Error 299.83, Effort 83.13, Time 20 s (effort decreases by 60\% as $N$ increases). &Baseline: Prioritizes low effort and scalability in NISQ; higher error reflects conservative control vs. aggressive classical methods. \\ 
\bottomrule
\end{tabular}
\label{tab:comparison with other methods}
\end{table*}

\section{Conclusion}\label{conclusion}
The $Q$-Drive framework demonstrates the feasibility of a hybrid quantum-classical control approach for nonlinear vehicular dynamics, including a 6-qubit VQC applied to the full 6-DoF bicycle model and to dynamically decoupled multi-vehicle swarms. The results indicate that the framework can discover low-control-effort policy families, but this behavior is accompanied by substantially higher terminal tracking error than the classical baselines, particularly under the fixed-horizon and fixed-weight settings considered here. Accordingly, the study's principal finding is the trade-off between control effort and tracking precision, not superior overall control performance. The stability results further establish boundedness under the analyzed road-perturbation and decoherence assumptions, while the empirical validation is limited to the nominal vehicle model and the decoupled swarm setting considered in this work. Thus, the present results should be interpreted as a proof of concept rather than as evidence of broad deployment readiness or general superiority over classical controllers. The framework is model-agnostic in principle, but validation with higher-fidelity vehicle models, inter-vehicle interactions, and hardware experiments remains necessary to assess its applicability beyond the scope of the present study. However, limitations persist, including sensitivity to NISQ-era noise, potential barren plateaus during training, and higher terminal errors compared to MPC in fixed-horizon scenarios. Future work should explore error-mitigation techniques, deeper ans\"{a}tze, and real-hardware deployment to validate scalability, while incorporating inter-vehicle interactions for true cooperative behavior. Additionally, the approach's behavior should be tested in experiments with obstacles. 
\section*{Acknowledgment}
The authors acknowledge the use of Grammarly (Grammarly Inc., San Francisco, CA, USA) solely for language polishing, basic grammatical checking, and improving overall sentence flow. The authors reviewed, edited, and approved all revised content.  

\bibliographystyle{IEEEtran}
\bibliography{references}

@article{lilja2023quantum,
  title={Quantum Computing for Automated Vehicle Trajectory Planning},
  author={Lilja, Juha},
  year={2023}
}

@article{sinha2025nav,
  title={Nav-Q: quantum deep reinforcement learning for collision-free navigation of self-driving cars},
  author={Sinha, Akash and Macaluso, Antonio and Klusch, Matthias},
  journal={Quantum Machine Intelligence},
  volume={7},
  number={1},
  pages={19},
  year={2025},
  publisher={Springer}
}

@article{tomar2025quantum,
  title={Quantum-Enhanced Hybrid Reinforcement Learning Framework for Dynamic Path Planning in Autonomous Systems},
  author={Tomar, Sahil and Alam, Shamshe and Kumar, Sandeep and Mathur, Amit},
  journal={arXiv preprint arXiv:2504.20660},
  year={2025}
}

@inproceedings{seyfipoor2025real,
  title={Real-Time Collaborative Route Planning in Autonomous Vehicles by Leveraging Optimized Quantum Neural Networks},
  author={Seyfipoor, Mahdi and Zafarqandi, Mohammad Javad Samii and Mohammadi, Siamak},
  booktitle={2025 29th International Computer Conference, Computer Society of Iran (CSICC)},
  pages={1--7},
  year={2025},
  organization={IEEE}
}

@article{mohammed2025review,
  title={Review of the application of quantum annealing-related technologies in transportation optimization},
  author={Mohammed, Marwan Qaid and Mee{\ss}, Henri and Otte, Maximilian},
  journal={Quantum Information Processing},
  volume={24},
  number={9},
  pages={296},
  year={2025},
  publisher={Springer}
}

@article{rattan2025quantum,
  title={Quantum computing for advanced driver assistance systems and autonomous vehicles: A review},
  author={Rattan, Avantika and Pal, Abhishek Rudra and Gurusamy, Muralimohan},
  journal={IEEE Access},
  year={2025},
  publisher={IEEE}
}

@inproceedings{ramasamy2024enhancing,
  title={Enhancing Automobile Motion Planning Algorithms using Quantum Computing: A Systematic Review},
  author={Ramasamy, Madhumathi and Jegan, B and Naveenganesh, M and Raghavendar, S},
  booktitle={2024 10th International Conference on Communication and Signal Processing (ICCSP)},
  pages={654--658},
  year={2024},
  organization={IEEE}
}

@article{khan2025qi,
  title={QI-MPC: A Hybrid Quantum-Inspired Model Predictive Control for Learning Optimal Policies},
  author={Khan, Muhammad Al-Zafar and Al-Karaki, Jamal},
  journal={arXiv preprint arXiv:2504.13041},
  year={2025}
}

@book{rajamani2006vehicle,
  title={Vehicle dynamics and control},
  author={Rajamani, Rajesh},
  year={2006},
  publisher={Springer}
}

@inproceedings{kong2015kinematic,
  title={Kinematic and dynamic vehicle models for autonomous driving control design},
  author={Kong, Jason and Pfeiffer, Mark and Schildbach, Georg and Borrelli, Francesco},
  booktitle={2015 IEEE intelligent vehicles symposium (IV)},
  pages={1094--1099},
  year={2015},
  organization={IEEE}
}

@article{muller2012cooperative,
  title={Cooperative control of dynamically decoupled systems via distributed model predictive control},
  author={M{\"u}ller, Matthias A and Reble, Marcus and Allg{\"o}wer, Frank},
  journal={International Journal of Robust and Nonlinear Control},
  volume={22},
  number={12},
  pages={1376--1397},
  year={2012},
  publisher={Wiley Online Library}
}

@article{corless2003continuous,
  title={Continuous state feedback guaranteeing uniform ultimate boundedness for uncertain dynamic systems},
  author={Corless, Martin and Leitmann, George},
  journal={IEEE Transactions on Automatic Control},
  volume={26},
  number={5},
  pages={1139--1144},
  year={2003},
  publisher={IEEE}
}

@article{almaameri2025quantum,
  title={Quantum-enhanced artificial potential field method for robust drone swarm shape formation},
  author={Almaameri, Ihab and Bl{\'a}zovics, L{\'a}szl{\'o}},
  journal={Scientific Reports},
  volume={15},
  number={1},
  pages={41945},
  year={2025},
  publisher={Nature Publishing Group UK London}
}

@article{antonio2022multi,
  title={Multi-agent deep reinforcement learning to manage connected autonomous vehicles at tomorrow's intersections},
  author={Antonio, Guillen-Perez and Maria-Dolores, Cano},
  journal={IEEE Transactions on Vehicular Technology},
  volume={71},
  number={7},
  pages={7033--7043},
  year={2022},
  publisher={IEEE}
}

@article{wu2024recent,
  title={Recent advances in reinforcement learning-based autonomous driving behavior planning: A survey},
  author={Wu, Jingda and Huang, Chao and Huang, Hailong and Lv, Chen and Wang, Yuntong and Wang, Fei-Yue},
  journal={Transportation Research Part C: Emerging Technologies},
  volume={164},
  pages={104654},
  year={2024},
  publisher={Elsevier}
}

@article{yu2023overview,
  title={An overview of swarm coordinated control},
  author={Yu, Dengxiu and Li, Jiacheng and Wang, Zhen and Li, Xuelong},
  journal={IEEE Transactions on Artificial Intelligence},
  volume={5},
  number={5},
  pages={1918--1938},
  year={2023},
  publisher={IEEE}
}

@article{chen2024fast,
  title={A fast coordination approach for large-scale drone swarm},
  author={Chen, Wu and Zhu, Jiayi and Liu, Jiajia and Guo, Hongzhi},
  journal={Journal of Network and Computer Applications},
  volume={221},
  pages={103769},
  year={2024},
  publisher={Elsevier}
}

@article{wang2024multi,
  title={Multi-agent DRL-controlled connected and automated vehicles in mixed traffic with time delays},
  author={Wang, Zhuwei and Xue, Yi and Liu, Lihan and Zhang, Haijun and Qu, Chunhui and Fang, Chao},
  journal={IEEE Transactions on Intelligent Transportation Systems},
  volume={25},
  number={11},
  pages={17676--17688},
  year={2024},
  publisher={IEEE}
}

@article{ko2021multi,
  title={A multi-agent driving simulation approach for evaluating the safety benefits of connected vehicles},
  author={Ko, Jieun and Jang, Jiyong and Oh, Cheol},
  journal={IEEE transactions on intelligent transportation systems},
  volume={23},
  number={5},
  pages={4512--4524},
  year={2021},
  publisher={IEEE}
}

@article{shaikh2020review,
  title={A review on swarm intelligence and evolutionary algorithms for solving the traffic signal control problem},
  author={Shaikh, Palwasha W and El-Abd, Mohammed and Khanafer, Mounib and Gao, Kaizhou},
  journal={IEEE transactions on intelligent transportation systems},
  volume={23},
  number={1},
  pages={48--63},
  year={2020},
  publisher={IEEE}
}

@article{zhang2026barrier,
  title={Barrier Costs-Embedded MPPI Control for Rollover-Safe Cooperative Payload Transport with All-Wheel-Steering Distributed Electric-Drive Carriers},
  author={Zhang, Bowei and Cheng, Kun and Huang, Jin and Jiang, Kun and Yang, Diange and Zhong, Zhihua},
  journal={Green Energy and Intelligent Transportation},
  pages={100409},
  year={2026},
  publisher={Elsevier}
}

@article{fan2026energy,
  title={Energy-Efficient Driving for Distributed Drive Electric Vehicles via Lagrangian-Constrained Deep Reinforcement Learning with Adaptive Safety Critics},
  author={Fan, Yi and Peng, Jiankun and He, Hongwen and Yu, Sichen and Wu, Jingda and Wang, Jiawei and Wu, Changcheng},
  journal={Green Energy and Intelligent Transportation},
  pages={100424},
  year={2026},
  publisher={Elsevier}
}

@article{yin2026global,
  title={Global coordinated control framework for advanced all-wheel steering and driving vehicles: Unleashing potential through tire slip state assessment},
  author={Yin, Xiaoxuan and Zhang, Lei and Ding, Xiaolin and Sun, Fengchun and Dorrell, David},
  journal={Green Energy and Intelligent Transportation},
  pages={100427},
  year={2026},
  publisher={Elsevier}
}

\end{document}